%% file: main.tex
\documentclass{article}

\usepackage{microtype}
\usepackage{graphicx}
\usepackage{subcaption}
\usepackage{multirow}
\usepackage{multicol}
\usepackage{pifont}

\usepackage{makecell}
\usepackage{booktabs}

\usepackage{hyperref}

\usepackage[accepted]{icml2026}

\usepackage{amsmath}
\usepackage{amssymb}
\usepackage{mathtools}
\usepackage{amsthm}
\usepackage[table]{xcolor}
\usepackage{threeparttable}

\usepackage{xcolor}
\usepackage{tcolorbox}
\usepackage{listings}
\usepackage{enumitem}
\usepackage{amssymb}
\usepackage{booktabs}
\usepackage{dblfloatfix}
\usepackage{threeparttablex}
\usepackage{rotating}

\definecolor{codegreen}{rgb}{0,0.6,0}
\definecolor{codegray}{rgb}{0.5,0.5,0.5}
\definecolor{codepurple}{rgb}{0.58,0,0.82}
\definecolor{hlgreen}{rgb}{0.7,1,0.7}
\definecolor{hlyellow}{rgb}{1,1,0.6}

\lstdefinestyle{pythonstyle}{
    commentstyle=\color{codegreen},
    keywordstyle=\color{magenta},
    numberstyle=\tiny\color{codegray},
    stringstyle=\color{codepurple},
    basicstyle=\ttfamily\footnotesize,
    breakatwhitespace=false,
    breaklines=true,
    captionpos=b,
    keepspaces=true,
    numbers=none,
    numbersep=5pt,
    showspaces=false,
    showstringspaces=false,
    showtabs=false,
    tabsize=2,
    language=Python,
    escapechar=|,
}

\definecolor{thinkingbg}{RGB}{255, 248, 220}
\definecolor{thinkingframe}{RGB}{218, 165, 32}
\definecolor{toolcallbg}{RGB}{230, 240, 255}
\definecolor{toolcallframe}{RGB}{70, 130, 180}
\definecolor{toolresultbg}{RGB}{232, 245, 233}
\definecolor{toolresultframe}{RGB}{76, 175, 80}
\definecolor{iterationbg}{RGB}{245, 245, 245}
\definecolor{codebg}{RGB}{248, 248, 248}

\tcbset{
    base/.style={
        boxrule=0.5pt,
        left=6pt,
        right=6pt,
        top=4pt,
        bottom=4pt,
        fonttitle=\bfseries\small,
        before skip=6pt,
        after skip=6pt,
    }
}

\newtcolorbox{agentthinking}[1][]{
    base,
    colback=thinkingbg,
    colframe=thinkingframe,
    title={\textbf{$\triangleright$} Agent Reasoning},
    #1
}

\newtcolorbox{toolcall}[1][]{
    base,
    colback=toolcallbg,
    colframe=toolcallframe,
    title={\textbf{$\blacktriangleright$} Tool Call},
    #1
}

\newtcolorbox{toolresult}[1][]{
    base,
    colback=toolresultbg,
    colframe=toolresultframe,
    title={\textbf{$\checkmark$} Tool Result},
    #1
}

\newtcolorbox{iteration}[1][]{
    base,
    colback=iterationbg,
    colframe=black!60,
    fonttitle=\bfseries,
    #1
}

\newtcolorbox{promptbox}[1][]{
    base,
    colback=codebg,
    colframe=black!40,
    fonttitle=\bfseries\ttfamily\small,
    #1
}
\usepackage[capitalize,noabbrev]{cleveref}
\newcommand{\cmark}{\textcolor{green!60!black}{\ding{51}}}
\newcommand{\xmark}{\textcolor{red}{\ding{55}}}

\theoremstyle{plain}

\theoremstyle{definition}

\theoremstyle{remark}

\usepackage[textsize=tiny]{todonotes}

\icmltitlerunning{KernelCraft: Benchmarking for Agentic Close-to-Metal Kernel Generation on Emerging Hardware}

\begin{document}

\twocolumn[
\icmltitle{
\hspace{-0.5em}%
\raisebox{-2.4em}[0pt][0pt]{%
\includegraphics[height=4em]{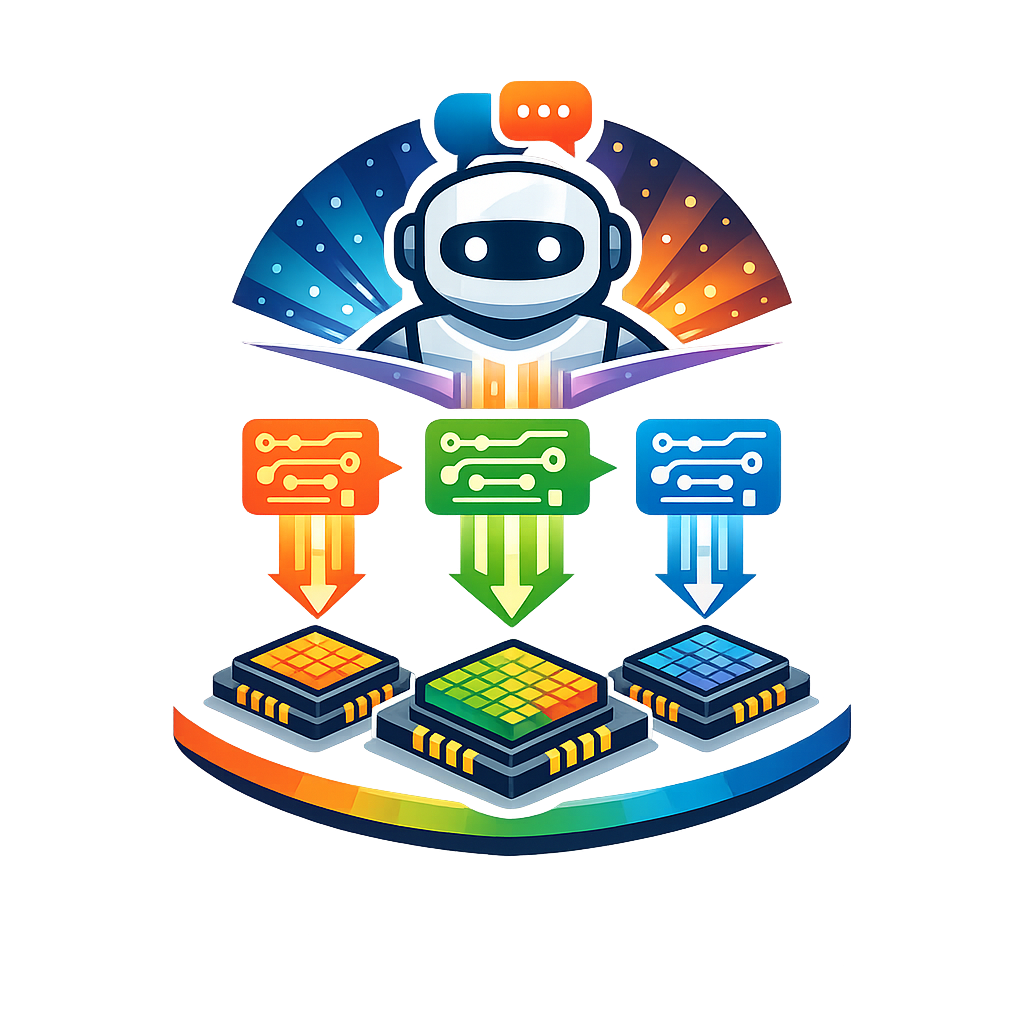}%
}%
\hspace{0.05em}%
KernelCraft: Benchmarking for Agentic Close-to-Metal\\
Kernel Generation on Emerging Hardware
}
\icmlsetsymbol{equal}{*}
\begin{icmlauthorlist}
\icmlauthor{Jiayi Nie}{equal,cam}
\icmlauthor{Haoran Wu}{equal,cam}
\icmlauthor{Yao Lai}{cam}
\icmlauthor{Zeyu Cao}{cam}
\icmlauthor{Cheng Zhang}{imperial}
\icmlauthor{Binglei Lou}{imperial}\\
\icmlauthor{Erwei Wang}{amd}
\icmlauthor{Jianyi Cheng}{edin}
\icmlauthor{Timothy M. Jones}{cam}
\icmlauthor{Robert Mullins}{cam}
\icmlauthor{Rika Antonova}{cam}
\icmlauthor{Yiren Zhao}{imperial}
\end{icmlauthorlist}

\icmlaffiliation{cam}{Department of Computer Science and Technology, University of Cambridge, Cambridge, United Kingdom}
\icmlaffiliation{imperial}{Department of Electrical and Electronic Engineering, Imperial College London, London, United Kingdom}
\icmlaffiliation{edin}{School of Informatics, University of Edinburgh, Edinburgh, United Kingdom}
\icmlaffiliation{amd}{AMD, San Jose, USA}

\icmlcorrespondingauthor{Jiayi Nie}{jn517@cam.ac.uk}
\icmlcorrespondingauthor{Haoran Wu}{hw691@cam.ac.uk}
\icmlcorrespondingauthor{Yiren Zhao}{a.zhao@imperial.ac.uk}
\icmlkeywords{Machine Learning, ICML}

\vskip 0.3in
]

\printAffiliationsAndNotice{\icmlEqualContribution}

\begin{abstract}
  \input{text/abstract}

\end{abstract}

\input{text/intro}
\input{text/env}

\input{text/benchmark}
\input{text/exper}
\input{text/discussion}
\input{text/analysis}

\input{text/conclusion}

\bibliography{refs}
\bibliographystyle{icml2026}

\newpage
\appendix

\input{text/related}
\input{text/appendix}

\end{document}

%% file: text/abstract.tex

New AI accelerators with novel instruction set architectures (ISAs) often require developers to manually craft low-level kernels --- a time-consuming and error-prone process that does not scale across hardware targets. This delays emerging hardware platforms from reaching the market. While prior LLM-based code generation has shown promise in mature GPU ecosystems, it remains unclear whether agentic LLM systems can quickly produce valid and efficient kernels for emerging hardware with new ISAs. We present KernelCraft: the first benchmark for evaluating an LLM agent's ability to generate and optimize low-level kernels for customized accelerators through a function-calling, feedback-driven workflow. We evaluate agent performance across three emerging accelerators on more than 20 machine-learning tasks, each with five diverse task configurations. Across four leading reasoning models, the strongest agents generate functionally correct kernels for unseen ISAs within a few refinement steps, and produce optimized kernels that match or outperform compiler baselines. These results demonstrate KernelCraft's potential to accelerate the accelerator chip development cycle. KernelCraft is available at \url{https://kernelcraft-cam.github.io/}.

%% file: text/intro.tex
\newpage
\section{Introduction}

\begin{figure}[!t]
    \centering
    \includegraphics[width=\columnwidth]{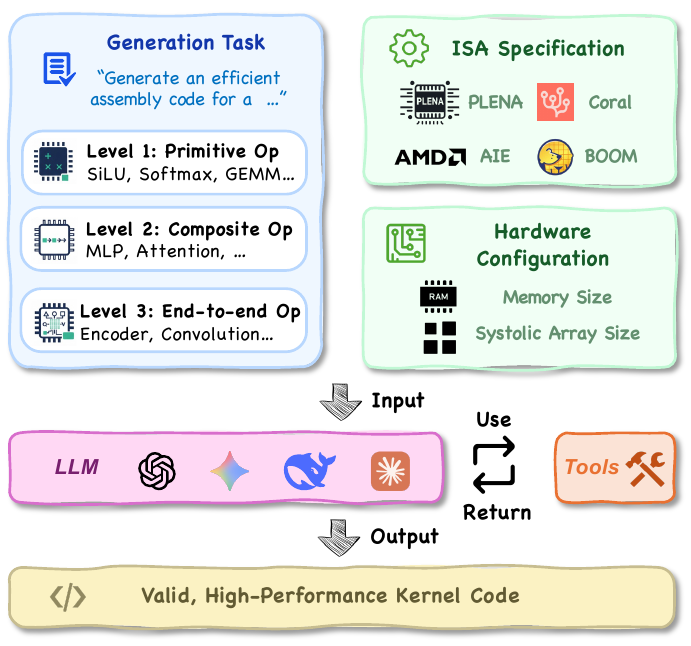}
    \caption{\textbf{Overview of KernelCraft.} KernelCraft tasks span three workload levels: primitive operations, composite operations, and end-to-end systems. Given the task description, ISA specification, and hardware configuration, an LLM-based agent generates kernels while using the provided tools for debugging and iterative refinement.}
    \label{fig:system-overview}
    \vspace{-15pt}
\end{figure}

\begin{table*}[!h]
\centering
\small
\caption{Comparison of LLM-based Hardware Kernel Generation Benchmarks.}
\label{tab:benchmark-comparison}
\resizebox{\linewidth}{!}{%
\begin{threeparttable}
\begin{tabular}{l|cccccc}
\toprule
\multirow{2}{*}{\textbf{Feature}}
& \textbf{KernelBench} 
& \textbf{TritonBench} 
& \textbf{NPUEval} 
& \textbf{BackendBench} 
& \textbf{MultiKernelBench} 
& \multirow{2}{*}{\textbf{KernelCraft (Ours)}} \\
 & \cite{ouyang2025kernelbench} 
 & \cite{li2025tritonbench} 
 & \cite{kalade2025npueval} 
 & \cite{saroufim2025backendbench} 
 & \cite{wen2025multikernelbench} 
 & \\
\midrule
Target Language & CUDA & Triton & C++ (AIE kernel) & PyTorch Backend & CUDA/AscendC/Pallas & Assembly \\
Task Variations\tnote{*} & \xmark  & \xmark & \cmark  & \xmark & \xmark & \cmark \\
Evaluation Metrics\tnote{\dag} & Correct. + Perf. & Code Sim. + Correct. + Perf. & Correct. + Perf. & Correct. &  Correct. + Perf. & Correct. + Perf. \\
Tool-Use &\xmark  & \xmark  &  \xmark& \xmark & \xmark & \cmark \\
Multi-turn Regeneration
 & \cmark & \xmark & \cmark & \xmark& \xmark  & \cmark \\
\bottomrule
\end{tabular}
\begin{tablenotes}[flushleft]
\footnotesize
\item[*] Includes variations over batch size, hidden dimension, and quantization configurations.
\item[\dag] Correct.\ = correctness; Perf.\ = performance; Code Sim.\ = code similarity (to a reference implementation).
\end{tablenotes}
\end{threeparttable}
}
\vspace{-10pt}
\end{table*}

The rapid evolution of large language models (LLMs)~\cite{gpt4, llama} has necessitated a paradigm shift in AI accelerator design to accommodate increasingly complex computational patterns and memory bottlenecks~\cite{llmcompass}. To maximize throughput and energy efficiency, emerging accelerators increasingly adopt specialized and heterogeneous instruction set architectures (ISAs), in contrast to the more general-purpose designs of conventional CPUs and GPUs.

Unlike standard ISAs, these customized architectures~\cite{wu2025combating,flightllm} expose low-level hardware abstractions, including computation patterns, data movement, and memory hierarchy, directly to the programmer. While this provides fine-grained control, it creates a ``programmability wall": most emerging accelerators lack the mature compiler toolchains required to automatically map high-level tensor programs to optimized bare-metal kernels. Building and maintaining a robust compiler stack for a customized accelerator with a new ISA~\cite{ml_compiler} requires substantial engineering effort. Therefore, this lack of a mature compiler creates a significant barrier to entry, often leaving innovative hardware underutilized or obsolete before it can be adopted by the AI community.

Existing machine learning compilers such as Apache TVM~\cite{chen2018tvm} aim to bridge this gap through automated kernel generation and optimization. However, adapting these frameworks to novel accelerators remains a ``cold-start" problem, requiring engineers to manually encode hardware-specific constraints and memory hierarchies into the backend. As a result, translating high-level operator intent into close-to-metal kernel code continues to rely heavily on human expertise and careful reasoning about target-specific details, balancing functional correctness under hardware constraints with high performance through deep, hardware-specific tuning. This dependence makes kernel development time-consuming, error-prone, and difficult to scale, demanding deep expertise in both the target architecture and the computational workload. Consequently, even mathematically simple operators—such as linear layers, normalization, and attention—can require significant manual effort to implement efficiently, hindering the rapid iteration cycles for software-hardware co-design that is required by evolving AI workloads.

The rise of agentic systems~\cite{browser_use2024, command_agent} presents a solution to this human-centric implementation bottleneck. Recent work has explored leveraging LLMs to automate kernel generation with promising results for domain-specific languages, including Triton~\cite{li2025autotriton} and CUDA kernel optimization~\cite{chen2025cuda, lange2025towards}. However, these efforts largely target mature ecosystems with abundant training data and well-established programming patterns. As shown in~\cref{tab:benchmark-comparison}, these benchmarks fail to address the unique challenges of ``zero-shot" kernel generation for emerging accelerators: In these scenarios, an effective agentic system must operate without prior programming examples,  relying instead on the long-tail feedback from verification on hardware simulators and formal architectural specifications.

Hence, a critical question remains unanswered: \textit{Can agentic LLM systems quickly generate correct and close-to-metal kernel code for emerging hardware with novel instruction sets and architectural designs?} 

To address this gap, we propose \textbf{KernelCraft}, 
a tool-using agentic system that rapidly generates and optimizes high-performance low-level hardware kernels for customized accelerators with novel ISAs.\footnote{Kernel correctness is non-trivial for emerging hardware. KernelCraft therefore integrates tool-based verification (e.g., compilation checks, simulation, and test-based validation) to detect functional errors during generation and optimization.} As illustrated in \Cref{fig:system-overview}, KernelCraft integrates an agentic execution loop with a systematic benchmark of workload specifications and hardware descriptions. The framework enables controlled evaluation and comparison of LLM agent capabilities across multiple emerging hardware platforms, providing a unified testbed for assessing functional correctness, optimization quality, and generalization to previously unseen ISAs.

\begin{figure*}[!t]
    \centering
    \includegraphics[width=\textwidth]{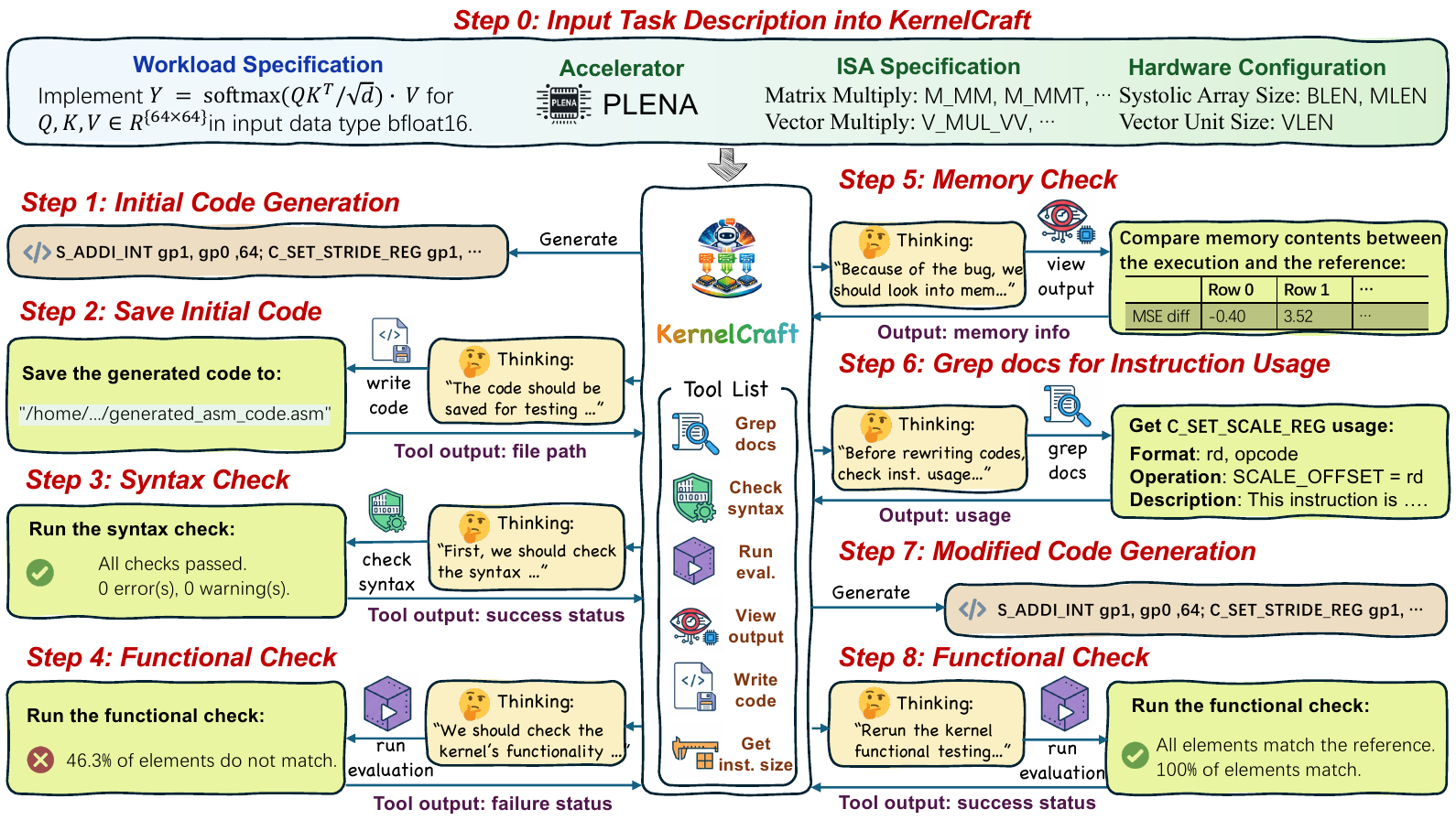}
\caption{KernelCraft benchmarks LLM agents for accelerator assembly-kernel generation through a diagnosis-and-repair loop. Given workload, ISA, and hardware specifications, the agent writes an assembly kernel that KernelCraft automatically saves and verifies using syntax and reference-based functional checks. On mismatches, KernelCraft performs memory-level diff diagnostics and feeds the results back for iterative patching until the kernel satisfies correctness criteria.}    
    \label{fig:benchmark_flow}
    \vspace{-10pt}
\end{figure*}

Our contributions are as follows:
\begin{itemize}[itemsep=1pt, topsep=1pt, parsep=0pt, partopsep=0pt]
    \item We introduce KernelCraft, the first benchmark for evaluating LLM agents on close-to-metal assembly kernel generation for emerging accelerators with novel ISAs, covering 33 tasks (23 ML, 10 CPU), six native tool interfaces, and diagnostic feedback across three emerging NPU platforms.
    \item We demonstrate that frontier LLM agents can autonomously generate functionally correct bare-metal kernels for new instruction sets through limited iterative refinement, achieving an up to 56\% success on primitive operations.
    \item We show that beyond correctness, agents can discover hardware-aware optimizations autonomously, generating kernels with competitive or superior performance. 
    \item We illustrate through case studies that KernelCraft extends naturally to broader hardware development workflows, from optimizing compiler templates to co-designing new instructions for emerging workloads.
\end{itemize}

\vspace{-12pt}

\paragraph{Conflict of Interest Disclosure.} The author Erwei Wang is employed by AMD, whose NPU is among the hardware platforms evaluated in this paper. AMD also provides research funding and equipment to the group of author Yiren Zhao at Imperial College London.

%% file: text/env.tex
\section{Preliminaries}
\subsection{Hardware kernel}

A hardware kernel is a low-level program that implements a specific computational task on a given hardware target, such as a customized AI accelerator or a general-purpose CPU.
In practice, hardware kernels may be expressed at different abstraction levels, ranging from high-level kernel languages or domain-specific languages (e.g., CUDA or Triton) to low-level representations such as intermediate code or assembly.
Regardless of the abstraction level, a hardware kernel ultimately specifies the computation, data movement, and control flow executed on the target hardware.

In this work, we focus on \emph{ISA-level hardware kernels}, which are written directly in a hardware-specific instruction set architecture (ISA).
As a result, kernel correctness and performance are tightly coupled to the underlying ISA semantics and hardware constraints.
Such kernels consist of low-level scalar, vector, and memory-access instructions, rather than compiler intermediate representations.


\subsection{Tool-use-based LLM agents}

Tool use~\cite{qu2025tool, li2025review, yuan2025easytool} (also referred to as function calling) enables large language models to interact with external systems through predefined and structured interfaces (e.g., invoking external APIs or executing programs).
This paradigm has been widely adopted by modern LLMs, including DeepSeek~\cite{liu2025deepseek} and Qwen~\cite{yang2025qwen3}, where the model selects appropriate functions based on user queries and outputs structured invocation arguments that are executed by the host application.
The execution results are then fed back to the model to support subsequent reasoning and response generation. By decoupling natural language reasoning from computation and external knowledge access, tool-use-based agents overcome several inherent limitations of standalone LLMs and serve as a foundation for contemporary agentic systems.

%% file: text/benchmark.tex
\begin{table*}[t]
\centering
\caption{\textbf{Tools Available in KernelCraft.} Workload as input specifies the target task (e.g., linear, ffn, attention).}
\label{tab:tools}
\resizebox{\linewidth}{!}{
\begin{tabular}{llll}
\toprule
\textbf{Tool} & \textbf{Input} & \textbf{Output} & \textbf{Description} \\
\midrule
\texttt{write\_code} & assembly code, workload & success, line count & Save assembly code as a file \\
\texttt{check\_syntax} & workload & success, errors, instruction count & Validate syntax and compile \\
\texttt{run\_evaluation} & workload & match rate, latency & Evaluate correctness and performance \\
\texttt{view\_output} & workload & value comparison statistics & Compare actual vs expected output \\
\texttt{get\_instruction\_size} & workload & instruction counts by category & Count instructions by category \\
\texttt{grep\_docs} & query & matched documentation & Search ISA and hardware documentation \\
\bottomrule
\end{tabular}
}
\vspace{-10pt}
\end{table*}

\section{KernelCraft}

KernelCraft is an open-source benchmark designed to evaluate large language models for low-level assembly kernel generation on domain-specific accelerators.
It provides a unified framework for assessing model performance in generating correct and efficient bare-metal assembly code across a diverse set of accelerator architectures and workload complexities in an end-to-end workflow.

Each task in KernelCraft includes multiple variations, resulting in a large and diverse test space that stresses both correctness and performance across different execution contexts.
In addition, the benchmark natively supports multiple tools and execution backends, enabling agents to flexibly select and combine tools during kernel generation. \Cref{fig:benchmark_flow} shows one example agent flow through this iterative loop, from the initial specifications to a correct kernel.

\subsection{Hardware targets and kernel tasks}

\begin{table}[!t]
\centering
\small
\caption{Target Hardware Platforms for Kernel Generation.}
\label{tab:accelerators}
\resizebox{\columnwidth}{!}{
\begin{tabular}{lllll}
\toprule
\textbf{Hardware} & \textbf{ISA} & \textbf{Compiler}  & \textbf{Backend}\\
\midrule
PLENA & Custom ISA & PLENA compiler & PLENA Simulator\\
AMD NPU & Custom ISA & Peano & NPU Hardware\\
Coral NPU & RISC-V + RVV & RISC-V Compiler & Verilator\\
BOOM & RISC-V & RISC-V Compiler & Verilator\\
\bottomrule
\end{tabular}
}
\end{table}

\paragraph{Hardware targets.}
We consider two classes of hardware targets in KernelCraft (\Cref{tab:accelerators}): domain-specific AI accelerators (PLENA~\cite{wu2025combating}, AMD NPU~\cite{amd-aie}, Coral NPU~\cite{google2025_coral_npu}) and CPUs (e.g., the RISC-V--based Sonic BOOM core~\cite{zhaosonicboom}).
We select these systems to cover a diverse set of kernel development settings, ranging from fully customized accelerator ISAs with dedicated toolchains (PLENA and AMD NPU) to open-ISA targets built on RISC-V (Coral NPU and Sonic BOOM), which provide accessible documentation and widely used simulation frameworks. Each hardware target is described by its own ISA documentation and hardware specifications, which form the core of the agent's system prompt, along with platform-specific tools (\Cref{tab:tools}). Full prompt structure is detailed in \Cref{appendix:agent-prompt}.

\begin{table}[!ht]
  \centering
  \small
  \caption{Benchmark tasks for AI accelerator kernel generation. $\odot$: element-wise multiplication; $\sigma$: sigmoid; $\Phi$: Gaussian CDF; $\mu$, $\sigma^2$: mean and variance; }
  \label{tab:tasks}
  \resizebox{\columnwidth}{!}{%
  \begin{tabular}{rlll}
  \toprule
  \textbf{ID} & \textbf{Category} & \textbf{Task} & \textbf{Description} \\
  \midrule
  \rowcolor{gray!15}
  \multicolumn{4}{c}{\textbf{\textit{Level 1: Primitive Operations}}} \\
  \midrule
  1 & \multirow{3}{*}{Activation} & SiLU & $x \cdot \sigma(x)$ \\
  2 &  & ReLU & $\max(0, x)$ \\
  3 &  & GELU & $x \cdot \Phi(x)$ \\
  \cmidrule{1-4}
  4 & \multirow{3}{*}{Normalization} & Softmax & $e^{x_i} / \sum_j e^{x_j}$ \\
  5 &  & LayerNorm & $(x - \mu) / \sqrt{\sigma^2 + \epsilon}$ \\
  6 &  & RMSNorm & $x / \text{RMS}(x)$  \\
  \cmidrule{1-4}
  7 & \multirow{4}{*}{Matrix} & GEMV & $y = Ax$ \\
  8 &  & GEMM & $Y = XW$ \\
  9 &  & BatchMatMul & $Y_i = X_i W_i$ for batch $i$ \\
  10 &  & Linear & $Y = XW + b$ \\
  \cmidrule{1-4}
  11 & \multirow{2}{*}{Spatial} & Conv2D & 2D convolution \\
  12 &  & DepthwiseConv & Per-channel 2D convolution \\
  \midrule
  \rowcolor{gray!15}
  \multicolumn{4}{c}{\textbf{\textit{Level 2: Composite Operations}}} \\
  \midrule
  13 & Encoding & RoPE & Rotary position encoding \\
  \cmidrule{1-4}
  14 & \multirow{2}{*}{MLP} & FFN & Linear, SiLU, Linear \\
  15 &  & SwiGLU & (Linear $\odot$ SiLU(Linear)), Linear \\
  \cmidrule{1-4}
  16 & \multirow{2}{*}{\shortstack[l]{Attention\\(core)}} & ScaledDotProduct & $\text{softmax}(QK^\top\!/\!\sqrt{d})V$ \\
  17 &  & FlashAttention & Tiled attention with online softmax \\
  \cmidrule{1-4}
  18 & \multirow{3}{*}{\shortstack[l]{Attention\\(+proj)}} & MHA & Multi-head attention with projections \\
  19 &  & GQA & Grouped K/V heads \\
  20 &  & MQA & Single shared K/V head \\
  \midrule
  \rowcolor{gray!15}
  \multicolumn{4}{c}{\textbf{\textit{Level 3: End-to-End System}}} \\
  \midrule
 21 & CNN & ConvBlock & Conv2D, BatchNorm, ReLU \\
    22 & \multirow{2}{*}{Transformer} & DecoderBlock \newline {\scriptsize (LLaMA-style)} & RMSNorm, self-attn, SwiGLU \\
23 &  & DecoderBlock \newline {\scriptsize (T5-style)} & RMSNorm, self-attn, cross-attn, FFN \\
  \bottomrule
  \end{tabular}
  }
\end{table}

\begin{table*}[!th]
\centering
\small
\caption{Success rates for kernel generation across models and accelerators. Each task is evaluated on 5 configurations; cells show successful/total within the iteration budget per level.}
\label{tab:success-rate}
\resizebox{\linewidth}{!}{%
\begin{tabular}{ll|cccc|cccc|cccc}
\toprule
& & \multicolumn{4}{c|}{\textbf{PLENA}} 
& \multicolumn{4}{c|}{\textbf{AMD NPU}} 
& \multicolumn{4}{c}{\textbf{Coral NPU}} \\
\textbf{ID} & \textbf{Task} 
& GPT-5.2 & Gemini-3-flash & Sonnet 4 & DeepSeek R1
& GPT-5.2 & Gemini-3-flash & Sonnet 4 & DeepSeek R1
& GPT-5.2 & Gemini-3-flash & Sonnet 4 & DeepSeek R1 \\
\midrule
\rowcolor{gray!15}
\multicolumn{14}{c}{\textbf{\textit{Level 1: Primitive Operations}} (Max 15 iterations)} \\
\midrule
1  & SiLU          & \textbf{5/5} & \textbf{5/5} & \textbf{2/5} & \textcolor{gray}{0/5} & \textbf{1/5} & \textbf{1/5} & \textcolor{gray}{0/5} & \textcolor{gray}{0/5} & \textbf{3/5} & \textbf{3/5} & \textcolor{gray}{0/5} & \textcolor{gray}{0/5} \\
2  & ReLU          & \textbf{2/5} & \textcolor{gray}{0/5} & \textbf{1/5} & \textcolor{gray}{0/5} & \textbf{2/5} & \textbf{1/5} & \textcolor{gray}{0/5} & \textcolor{gray}{0/5} & \textbf{5/5} & \textbf{4/5} & \textbf{1/5} & \textcolor{gray}{0/5} \\
3  & GELU          & \textbf{4/5} & \textbf{4/5} & \textbf{1/5} & \textcolor{gray}{0/5} & \textbf{1/5} & \textbf{2/5} & \textcolor{gray}{0/5} & \textcolor{gray}{0/5} & \textbf{5/5} & \textbf{5/5} & \textcolor{gray}{0/5} & \textcolor{gray}{0/5} \\
4  & Softmax       & \textbf{5/5} & \textbf{3/5} & \textbf{4/5} & \textcolor{gray}{0/5} & \textcolor{gray}{0/5} & \textcolor{gray}{0/5} & \textcolor{gray}{0/5} & \textcolor{gray}{0/5} & \textbf{4/5} & \textbf{2/5} & \textcolor{gray}{0/5} & \textcolor{gray}{0/5} \\
5  & LayerNorm     & \textbf{3/5} & \textbf{5/5} & \textbf{2/5} & \textcolor{gray}{0/5} & \textbf{2/5} & \textbf{1/5} & \textcolor{gray}{0/5} & \textcolor{gray}{0/5} & \textbf{1/5} & \textcolor{gray}{0/5} & \textcolor{gray}{0/5} & \textcolor{gray}{0/5} \\
6  & RMSNorm       & \textbf{3/5} & \textbf{5/5} & \textbf{1/5} & \textbf{1/5} & \textbf{1/5} & \textcolor{gray}{0/5} & \textcolor{gray}{0/5} & \textcolor{gray}{0/5} & \textbf{1/5} & \textbf{1/5} & \textcolor{gray}{0/5} & \textcolor{gray}{0/5}  \\
7  & GEMV          & \textbf{5/5} & \textbf{2/5} & \textbf{1/5} & \textcolor{gray}{0/5} & \textbf{2/5} & \textbf{1/5} & \textcolor{gray}{0/5} & \textcolor{gray}{0/5} & \textbf{4/5} & \textbf{5/5} & \textcolor{gray}{0/5} & \textcolor{gray}{0/5} \\
8  & GEMM          & \textbf{4/5} & \textbf{2/5} & \textcolor{gray}{0/5} & \textcolor{gray}{0/5} & \textbf{4/5} & \textbf{3/5} & \textbf{2/5} & \textbf{1/5} & \textbf{2/5} & \textbf{4/5} & \textbf{1/5} & \textbf{1/5} \\
9  & BatchMatMul   & \textbf{2/5} & \textbf{2/5} & \textcolor{gray}{0/5} & \textcolor{gray}{0/5} & \textcolor{gray}{0/5} & \textcolor{gray}{0/5} & \textcolor{gray}{0/5} & \textcolor{gray}{0/5} & \textcolor{gray}{0/5} & \textcolor{gray}{0/5} & \textcolor{gray}{0/5} & \textcolor{gray}{0/5} \\
10 & Linear        & \textbf{4/5} & \textbf{2/5} & \textcolor{gray}{0/5} & \textcolor{gray}{0/5} & \textbf{3/5} & \textbf{2/5} & \textbf{1/5} & \textcolor{gray}{0/5} & \textbf{2/5} & \textcolor{gray}{0/5} & \textcolor{gray}{0/5} & \textcolor{gray}{0/5} \\
11 & Conv2D        & \multicolumn{4}{c|}{--$^\ddagger$} & \textcolor{gray}{0/5} & \textcolor{gray}{0/5} & \textcolor{gray}{0/5} & \textcolor{gray}{0/5} & \textbf{2/5} & \textbf{1/5} & \textcolor{gray}{0/5} & \textcolor{gray}{0/5} \\
12 & DepthwiseConv & \multicolumn{4}{c|}{--$^\ddagger$} & \textcolor{gray}{0/5} & \textcolor{gray}{0/5} & \textcolor{gray}{0/5} & \textcolor{gray}{0/5} & \textbf{5/5} & \textbf{3/5} & \textcolor{gray}{0/5} & \textcolor{gray}{0/5} \\
\midrule
\rowcolor{blue!8}
\multicolumn{2}{l|}{\textit{Level 1 Subtotal}} & \textit{37/50} & \textit{30/50} & \textit{12/50} & \textit{1/50} & \textit{16/60} & \textit{11/60} & \textit{3/60} & \textit{1/60} & \textit{34/60} & \textit{28/60} & \textit{2/60} & \textit{1/60} \\
\midrule
\rowcolor{gray!15}
\multicolumn{14}{c}{\textbf{\textit{Level 2: Composite Operations}} (Max 20 iterations)} \\
\midrule
13 & RoPE             & \textbf{1/5} & \textcolor{gray}{0/5} & \textcolor{gray}{0/5} & \textcolor{gray}{0/5} & \textcolor{gray}{0/5} & \textcolor{gray}{0/5} & \textcolor{gray}{0/5} & \textcolor{gray}{0/5} & \multicolumn{4}{c}{--$^\dagger$} \\
14 & FFN              & \textbf{3/5} & \textbf{2/5} & \textcolor{gray}{0/5} & \textcolor{gray}{0/5} & \textbf{2/5} & \textbf{1/5} & \textcolor{gray}{0/5} & \textcolor{gray}{0/5} & \textbf{1/5} & \textcolor{gray}{0/5} & \textcolor{gray}{0/5} & \textcolor{gray}{0/5} \\
15 & SwiGLU           & \textbf{4/5} & \textcolor{gray}{0/5} & \textcolor{gray}{0/5} & \textcolor{gray}{0/5} & \textcolor{gray}{0/5} & \textcolor{gray}{0/5} & \textcolor{gray}{0/5} & \textcolor{gray}{0/5} & \textcolor{gray}{0/5} & \textcolor{gray}{0/5} & \textcolor{gray}{0/5} & \textcolor{gray}{0/5} \\
16 & ScaledDotProduct & \textbf{3/5} & \textbf{2/5} & \textcolor{gray}{0/5} & \textcolor{gray}{0/5} & \textbf{1/5} & \textcolor{gray}{0/5} & \textcolor{gray}{0/5} & \textcolor{gray}{0/5} & \multicolumn{4}{c}{--$^\dagger$} \\
17 & FlashAttention   & \textbf{3/5} & \textbf{1/5} & \textcolor{gray}{0/5} & \textcolor{gray}{0/5} & \multicolumn{4}{c|}{--$^\S$} & \multicolumn{4}{c}{--$^\dagger$} \\
18 & MHA              & \textbf{3/5} & \textcolor{gray}{0/5} & \textcolor{gray}{0/5} & \textcolor{gray}{0/5} & \textcolor{gray}{0/5} & \textcolor{gray}{0/5} & \textcolor{gray}{0/5} & \textcolor{gray}{0/5} & \multicolumn{4}{c}{--$^\dagger$} \\
19 & GQA              & \textbf{1/5} & \textcolor{gray}{0/5} & \textcolor{gray}{0/5} & \textcolor{gray}{0/5} & \textcolor{gray}{0/5} & \textcolor{gray}{0/5} & \textcolor{gray}{0/5} & \textcolor{gray}{0/5} & \multicolumn{4}{c}{--$^\dagger$} \\
20 & MQA              & \textbf{1/5} & \textcolor{gray}{0/5} & \textcolor{gray}{0/5} & \textcolor{gray}{0/5} & \textcolor{gray}{0/5} & \textcolor{gray}{0/5} & \textcolor{gray}{0/5} & \textcolor{gray}{0/5} & \multicolumn{4}{c}{--$^\dagger$} \\
\midrule
\rowcolor{blue!8}
\multicolumn{2}{l|}{\textit{Level 2 Subtotal}} & \textit{19/40} & \textit{5/40} & \textit{0/40} & \textit{0/40} & \textit{3/35} & \textit{1/35} & \textit{0/35} & \textit{0/35} & \textit{1/10} & \textit{0/10} & \textit{0/10} & \textit{0/10} \\
\midrule
\rowcolor{gray!15}
\multicolumn{14}{c}{\textbf{\textit{Level 3: End-to-End System}} (Max 25 iterations)} \\
\midrule
21 & ConvBlock              & \multicolumn{4}{c|}{--$^\ddagger$} & \textcolor{gray}{0/5} & \textcolor{gray}{0/5} & \textcolor{gray}{0/5} & \textcolor{gray}{0/5}  & \textcolor{gray}{0/5} & \textbf{1/5} & \textcolor{gray}{0/5} & \textcolor{gray}{0/5} \\
22 & DecoderBlock (LLaMA)   & \textcolor{gray}{0/5} & \textcolor{gray}{0/5} & \textcolor{gray}{0/5} & \textcolor{gray}{0/5} & \textcolor{gray}{0/5} & \textcolor{gray}{0/5} & \textcolor{gray}{0/5} & \textcolor{gray}{0/5}  & \multicolumn{4}{c}{--$^\dagger$} \\
23 & DecoderBlock (T5)      & \textcolor{gray}{0/5} & \textcolor{gray}{0/5} & \textcolor{gray}{0/5} & \textcolor{gray}{0/5} & \textcolor{gray}{0/5} & \textcolor{gray}{0/5} & \textcolor{gray}{0/5} & \textcolor{gray}{0/5}  & \multicolumn{4}{c}{--$^\dagger$} \\
\midrule
\rowcolor{blue!8}
\multicolumn{2}{l|}{\textit{Level 3 Subtotal}} & \textit{0/10} & \textit{0/10} & \textit{0/10} & \textit{0/10} & \textit{0/15} & \textit{0/15} & \textit{0/15} & \textit{0/15} & \textit{0/5} & \textit{1/5} & \textit{0/5} & \textit{0/5} \\
\midrule
\rowcolor{yellow!15}
\multicolumn{2}{l|}{\textbf{Total}} & \textbf{56/100} & \textbf{35/100} & \textbf{12/100} & \textbf{1/100} & \textbf{19/110} & \textbf{12/110} & \textbf{3/110} & \textbf{1/110} & \textbf{35/75} & \textbf{29/75} & \textbf{2/75} & \textbf{1/75} \\
\bottomrule
\end{tabular}
}
\vspace{0.3em}
{\small $^\dagger$ Not officially supported by Coral NPU. \; $^\ddagger$ Not officially supported by PLENA ISA. \; $^\S$ Not officially supported by AMD NPU compiler.}
\end{table*}

\paragraph{Kernel tasks.}

We group ML- and CPU-related kernel generation tasks into three difficulty levels, based on increasing computational complexity and system integration, as summarized in \Cref{tab:tasks} and \Cref{tab:cpu-task-results}. All tasks target \emph{bare-metal kernels}: low-level implementations written directly against the target ISA or minimal compiler interfaces, where the kernel handles instruction selection, data movement, scheduling, and parallelism without high-level libraries or runtime abstractions. Level~1 covers primitive building blocks, including element-wise activations, normalization, linear algebra, spatial operators, and embedding lookups for AI accelerators, as well as arithmetic, memory, and basic linear algebra kernels for CPUs. Level~2 includes composite or algorithmic operations that combine primitives or require non-trivial control flow, such as MLPs, attention mechanisms, positional encoding, and loss functions on AI accelerators, and sorting or recursive algorithms on CPUs. Level~3 covers end-to-end workloads that integrate multiple operators, including complete Transformer and CNN blocks, along with synthetic and real-world applications and system-level CPU benchmarks.

\subsection{Evaluation}

Our evaluation consists of two critical metrics on the generated kernels, namely \emph{success rate} and \emph{kernel performance}.

For the success rate, we consider a generation successful if the produced assembly kernel is functionally correct.
Correctness is verified by comparing kernel outputs against platform-specific ground-truth implementations under the target hardware semantics.
Because differences in instruction ordering, quantization, and accumulation can introduce numerical variation in assembly-level execution, exact equivalence with reference results is not always achievable.
We therefore apply tolerance-based validation, with platform-specific tolerance settings described in detail in \Cref{app:tolerances}.
A workload is deemed functionally valid if the generated output matches the reference within the specified tolerances.
We report success rates across five test cases with varying task configurations in \Cref{tab:success-rate}.


For performance evaluation, we measure execution cycles via cycle-accurate simulation or on-device execution, depending on platform availability. Generated kernels are compared against compiler baselines from each platform’s standard toolchain (\Cref{app: Kernel_Gen_Baseline}). We report speedup as $\frac{B}{G}$, where $B$ and $G$ denote the baseline and generated-kernel execution cycles or latency, respectively.

%% file: text/exper.tex
\section{Experiments}
We evaluate KernelCraft across 3 accelerators and 4 frontier models with 5 configurations each, totaling over 1,100 experiments. We report success rates and kernel-level performance below; detailed experimental settings are provided in Appendix~\ref{app:Evaluation_Settings}.

\subsection{Task success rate}

Table~\ref{tab:success-rate} shows a clear performance gap across models. GPT-5.2~\cite{gpt52_2026} performs best on PLENA, AMD NPU, and Coral NPU, with success rates of 56\%, 17\%, and 47\%, respectively. Gemini-3-Flash~\cite{gemini3flash_2026} ranks second, achieving 35\%, 11\%, and 39\%. Sonnet 4~\cite{claudeSonnet4_2026} and DeepSeek R1~\cite{guo2025deepseek-r1} perform substantially worse; DeepSeek R1 generates only one successful kernel per platform.

Performance drops sharply as task complexity increases. Level~1 primitives achieve reasonable success rates, reaching 74\% for GPT-5.2 on PLENA. However, Level~2 composite operations are substantially harder: GPT-5.2 drops to 45\% on PLENA and below 10\% on the other platforms. Level~3 end-to-end blocks remain largely unsolved, with only one successful ConvBlock kernel across all model--accelerator combinations, generated by Gemini-3-Flash on Coral NPU.

Success rates also differ significantly across accelerators. PLENA achieves the highest overall rates, while AMD NPU is the most challenging, remaining below 20\% even for the best model. We partly attribute this gap to documentation quality: as shown in \Cref{tab:system-prompt-tokens}, PLENA’s system prompt is nearly 3$\times$ longer than AMD NPU’s, mainly due to more extensive ISA documentation. We provide per-task and per-model token-usage breakdowns in Appendix~\ref{app:token_usage_analysis}. Since Coral NPU does not officially support attention-based workloads, we evaluate it on a reduced task set; nevertheless, the same trend holds: higher task complexity strongly correlates with lower success rates across models and accelerators.

Overall, these results show that KernelCraft captures a broad complexity spectrum in kernel generation. Due to cost constraints, we cap refinement at 15, 20, and 25 iterations for Levels~1--3, respectively. Despite these limits, the clear gradient from primitive to composite to end-to-end tasks suggests opportunities for improving agentic capabilities, such as in-context learning (Section~\ref{sec:ablation}).

\begin{table}[!t]
\centering
\caption{KernelCraft System Prompt and Tools Description Token Count Breakdown by Accelerator (Claude Sonnet 4 Tokenizer)}
\label{tab:system-prompt-tokens}
\resizebox{\columnwidth}{!}{
\begin{tabular}{lccc}
\toprule
\textbf{Component} & \textbf{PLENA} & \textbf{AMD NPU} & \textbf{Coral NPU} \\
\midrule
ISA Documentation & 9.8k & 2.8k & 3.7k \\
Memory Layout & 4.3k & 2.6k & 0.5k \\
Hardware Description & 1.0k & 0.1k & 1.1k \\
Total System Prompt & 15.1k & 5.5k & 5.3k \\
\midrule
Tools Description & 1.8k & 1.2k & 1.5k \\
\bottomrule
\end{tabular}
}
\end{table}

\begin{figure*}[!t]
    \centering
    \includegraphics[width=\textwidth]{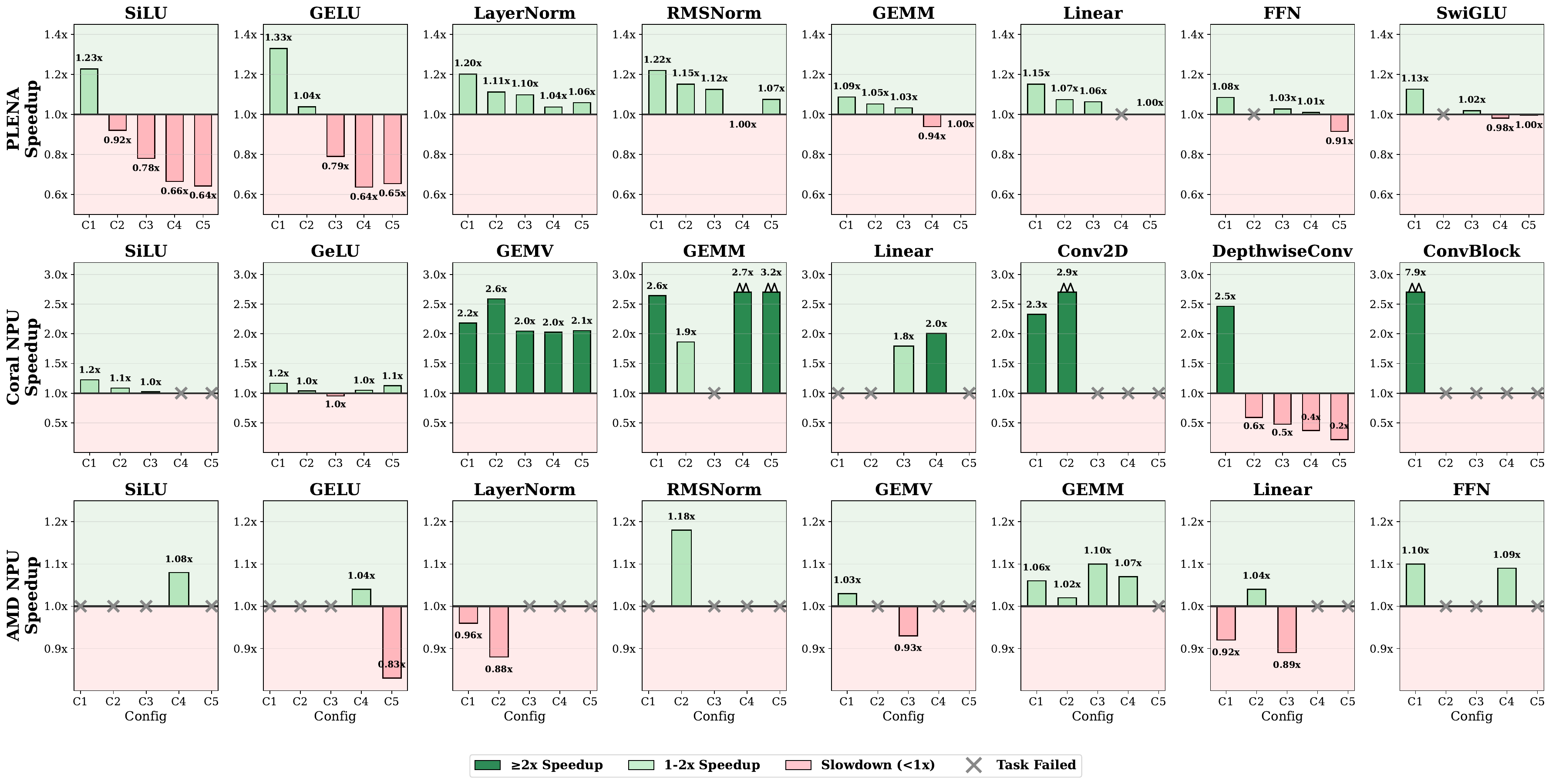}
    \caption{Speedup of best KernelCraft agent's kernels over compiler baselines on representative workloads of varying complexity across three accelerator platforms (PLENA: native compiler, Coral: RVV -O2, AMD: Peano)}
    \label{fig:perf_results}
    \vspace{-10pt}
\end{figure*}

\subsection{Kernel performance}
\Cref{fig:perf_results} presents a representative visualization of kernel-level speedups achieved by the best-performing KernelCraft agent on each task against each platform compiler. On PLENA, normalization tasks achieve consistent 1.06--1.22$\times$ speedups across all configurations, while elementwise tasks degrade at larger scales. Coral NPU exhibits the largest gains, with speedups reaching 2--8$\times$ on GEMV, GEMM, and ConvBlock. AMD NPU results cluster tightly around the baseline (0.89--1.18$\times$), with the most reliable improvements observed on GEMM and composite operations. Despite GPT-5.2 and Gemini-3-Flash demonstrating the highest task completion rates (Table~\ref{tab:success-rate}), Gemini-3-Flash produces the most aggressively optimized kernels on Coral NPU—including the 7.93$\times$ speedup on ConvBlock—suggesting that success rate and optimization quality are not perfectly correlated. These results indicate that \textit{frontier models can generate both correct and performance-competitive kernels for novel ISAs}, though optimization quality varies significantly across operations, task complexity, and scale. Complete per-model cycle counts are provided in Tables~\ref{tab:cycles-plena}, \ref{tab:cycles-coral}, and \ref{tab:cycles-amd} (Appendix~\ref{app:perf_sections}).

%% file: text/discussion.tex
\subsection{Discussion}
\label{sec:ablation}
In this section, we outline a set of ablations intended to better understand the contribution of individual design choices and present key findings on agent-generated kernel behavior.

\paragraph{Extended reasoning is essential for hard kernel generation tasks.} Having identified the importance of extended thinking for successful kernel generation. We conducted an ablation study to quantify the impact of ``thinking" tokens on task success, comparing standard inference against configurations with extended reasoning enabled. As shown in Table~\ref{tab:reasoning-ablation}, without extended thinking, GPT-5.2 fails to produce functional kernels within the iteration budget (0/5 success). Enabling reasoning tokens allows the agent to reason about hardware design and ISA implementation for Level 2 kernels, sustaining more iterations rather than terminating early. We selected moderate reasoning levels for each model as defaults across the main experiments to balance capability with cost; settings are documented in~\cref{tab:lm_models_settings}.

\begin{table}[!t]
\centering
\caption{Ablation study on extended reasoning for GPT-5.2 on Level 2 tasks (5 runs each). Avg Iter: average iterations across 5 runs; Avg Tok: average tokens per iteration across 5 runs; Succ: success rate. Without thinking tokens, the model terminates early with 0\% success. Medium thinking increases token usage but enables more iterations, achieving 60--80\% success.}
\label{tab:reasoning-ablation}
\small
\resizebox{\columnwidth}{!}{%
\begin{tabular}{lccc ccc}
\toprule
& \multicolumn{3}{c}{\textbf{No Thinking}} & \multicolumn{3}{c}{\textbf{Med. Thinking}} \\
\cmidrule(lr){2-4} \cmidrule(lr){5-7}
\textbf{Task} & Avg Iter & Avg Tok & Succ & Avg Iter & Avg Tok & Succ \\
\midrule
Attention & 5.6 & 817 & 0/5 & 19.3 & 2135 & 3/5 \\
MHA       & 6.8 & 611 & 0/5 & 18.4 & 2671 & 3/5 \\
FFN       & 11.8 & 867 & 0/5 & 18.8 & 2270 & 3/5 \\
SwiGLU    & 9.4 & 997 & 0/5 & 18.0 & 2157 & 4/5 \\
\bottomrule
\end{tabular}
}
\end{table}

\begin{table}[!t]
\centering
\caption{In-context learning ablation for level 2 tasks (5 runs each). One-shot example: Scaled Dot-Product for PLENA, MatMul for AMD NPU.}
\label{tab:icl-ablation}
\small
\resizebox{\columnwidth}{!}{%
\begin{tabular}{llcccc}
\toprule
& & \multicolumn{2}{c}{\textbf{GPT-5.2}} & \multicolumn{2}{c}{\textbf{Gemini 3 Flash}} \\
\cmidrule(lr){3-4} \cmidrule(lr){5-6}
\textbf{Hardware} & \textbf{Task} & Zero-shot & One-shot & Zero-shot & One-shot \\
\midrule
\multirow{3}{*}{PLENA} 
                       & MHA  & 3/5 & 4/5 & 0/5 & 1/5 \\
                       & MQA  & 1/5 & 4/5 & 0/5 & 0/5 \\
\midrule
\multirow{2}{*}{AMD NPU}   
                    & FFN  & 0/5 & 2/5 & 0/5 & 1/5\\
                       & SDPA & 0/5 & 1/5 & 0/5 & 0/5 \\
\bottomrule
\end{tabular}
}
\end{table}

\paragraph{In-context learning is critical when ISA documentation is scarce.}
We started out deliberately evaluating agents under a zero-shot setting across all accelerator platforms to assess their true ability to leverage tools and iteratively refine code on novel ISA specifications. Under this setting, agents struggle to produce correct kernels for complex, multi-operation tasks on PLENA and Coral. For the AMD NPU, where documentation is more limited, we provide a GEMM demonstration in the system prompt for the main experiments~(\Cref{tab:success-rate}).

To further quantify the impact of in-context learning, we conducted an ablation study (\Cref{tab:icl-ablation}). Providing demonstrations of related workloads significantly improves success rates: GPT-5.2 improves from 1/5 to 4/5 on MQA kernels for PLENA with a Scaled Dot-Product example, and from 0/5 to 2/5 on AMD NPU FFN with a GEMM example. However, effectiveness remains model-dependent—Gemini 3 Flash shows minimal improvement across both platforms. These results suggest that curating reference implementations of foundational operations provides a strong basis for agentic kernel generation, enabling rapid adaptation to novel hardware and ever-increasingly complex ML workloads.

\begin{table}[!t]
\centering
\caption{Tool-use efficiency across models for Level 1 tasks (success rate as $n$/5 over 5 configs). Error types: \textbf{S} = Syntax errors, \textbf{F} = Functionality incorrect, \textbf{T} = Tool-use failures. Averages are rounded to the nearest integer.}
\label{tab:tool-efficiency}
\resizebox{\columnwidth}{!}{%
\small
\begin{tabular}{lccccc}
\toprule
\textbf{Task} & \textbf{GPT-5.2} & \textbf{GPT-5.2C} & \textbf{Qwen} & \textbf{DS-R1} & \textbf{DS-V3.2} \\
\midrule
SiLU & 5/5 & 0/5 & 0/5 & 0/5 & 0/5 \\
\quad \textit{Errors} & F & T & S, F & S, F & T \\
\quad \textit{Avg Tool calls} & 14 & 3 & 16 & 13 & 3 \\
\quad \textit{Avg Evaluations} & 4 & 1 & 4 & 4 & 1 \\
\midrule
GEMM & 4/5 & 0/5 & 0/5 & 0/5 & 0/5 \\
\quad \textit{Errors} & S, F & T & S, F & S, F & T \\
\quad \textit{Avg Tool calls} & 18 & 2 & 16 & 12 & 2 \\
\quad \textit{Avg Evaluations} & 6 & 1 & 4 & 4 & 0 \\
\bottomrule
\end{tabular}
}
\vspace{-10pt}
\end{table}

\vspace{-4pt}
\paragraph{Tool-use efficiency.}
As summarized in Table~\ref{tab:benchmark-comparison}, existing work on agents for kernel generation typically does not leverage native tool use, instead employing external orchestration loops where the agent serves solely as a code generator while a separate system manages compilation, error checking, and feedback. Such designs require extensive prompt engineering, conflating model capabilities with surrounding infrastructure and making it difficult to evaluate inherent agentic abilities.
KernelCraft addresses this gap by relying solely on native API tool-use capabilities, directly measuring a model's ability to autonomously handle coding writing through tool-calling interfaces.

We evaluate GPT-5.2, GPT-5.2 Chat \cite{gpt52_chat_2026}, Qwen3-Coder-Flash, DeepSeek-R1, and DeepSeek-V3.2. Notably, despite producing optimized CUDA kernels on KernelBench~\cite{ouyang2025kernelbench}, DeepSeek-V3.2 and DeepSeek-R1 completely failed on KernelCraft. 
Analysis of failure modes reveals fundamental deficiencies in tool use: DeepSeek-V3.2 returns \texttt{finish\_reason="stop"} after the first \texttt{write\_code} call, terminating the agentic loop before ever invoking \texttt{run\_evaluation}. 
We additionally compare GPT-5.2 against its chatbot variant, GPT-5.2 Chat, which fails to use tools entirely -- outputting assembly code as plain text in its response rather than calling the \texttt{write\_code} tool, leaving the code never saved or evaluated. 
We demonstrate that tool-calling is a fundamental capability needed in KernelCraft for achieving good performance. More importantly, using tool-call enabled LLMs eliminate the need for carefully designed hand-crafted agentic loops like the ones in KernelBench.

\vspace{-8pt}

\paragraph{Close-to-metal kernel generation enables optimizations absent from immature toolchains.}
Agent-generated kernels can outperform compiler baselines by applying cross-operator optimizations that the target platform's toolchain does not yet support.
For example, on the Coral NPU ConvBlock workload, the agent achieves up to a $7\times$ speedup.
The ConvBlock consists of a Conv2D followed by BatchNorm and ReLU, where activations and weights are quantized to int8 while BatchNorm parameters remain in floating point, reflecting realistic quantized inference settings.
In this setting, the compiler implementation incurs expensive float--int conversions inside the inner loop.
In contrast, the agent discovers BatchNorm folding by precomputing normalization parameters as fixed-point constants and fusing BatchNorm into integer arithmetic.
This eliminates all floating-point operations from the inner loop and yields a $3.34\times$ speedup over the compiler baseline.
While such fusion passes are often absent from the immature toolchains of emerging accelerators, the agent can discover and apply them directly at the assembly level.

\paragraph{Documentation quality and harness design measurably contribute to kernel generation success.}
Assembly generation for novel ISAs differs from standard code generation and requires task-specific support (Appendix~\ref{app:harness-ablations}). We quantify the impact of documentation, tooling, and prompting. Ablating the ISA documentation at three levels of detail (\Cref{tab:isa-depth-ablation,fig:isa-ablation-visualization}), we find that denser documentation consistently improves success, and that practical programming guidance is the most useful context for agents. Among the supporting files (\Cref{tab:hw-mem-ablation}), the memory-layout document is by far the most critical across all four platforms. On BOOM, even removing the ISA specification has no effect on success rate (\Cref{tab:boom-ablation}), consistent with RISC-V being well represented in pretraining data. Tooling and prompting contribute comparably. \texttt{grep\_docs} is the load-bearing tool (\Cref{tab:tool-ablation}): successful runs rely on it more as task complexity grows, whereas failed runs do not. Removing the KernelCraft prompts sharply degrades performance (\Cref{tab:template-ablation}), and the two prompt components show opposite platform sensitivities.


\paragraph{Error Analysis.}
We classify each failed run over the PLENA
(\Cref{fig:plena-error-model,fig:plena-error-task-level}) and Coral NPU
(\Cref{fig:coral-error-model,fig:coral-error-task-level}) logs into four
execution-failure categories (Syntax Error, Simulator Runtime Error, Timeout,
Tool Orchestration Error) and two functional-correctness categories (Low Accuracy,
Zero Match), and observe qualitatively different failure modes across platforms and
models. Syntax errors mainly arise from hallucinated opcodes unsupported by the target ISA. Timeouts concentrate on Coral at L2,
where models fall back to scalar operations instead of the RISC-V vector extension and
blow the cycle budget. Tool-orchestration errors are dominated by DeepSeek-R1, which
exhausts its output budget on planning without invoking \texttt{write\_code}. These
patterns align with our tool-usage analysis
(\Cref{fig:coral-tool-usage-by-level,fig:plena-gpt-tool-usage-by-level,fig:plena-gpt-tool-use-by-task}),
where successful runs invoke supporting tools more heavily as complexity grows. We refer
readers to Appendix~\ref{app:additional-figures}, along with per-platform scaling behavior
(\Cref{fig:scaling-analysis-plena,fig:scaling-analysis-coral,fig:scaling-analysis-amd})
and per-workload ISA coverage
(\Cref{tab:coral-isa-coverage,tab:plena-isa-coverage}), for full details.

%% file: text/analysis.tex
\vspace{-3pt}
\section{Case Studies on KernelCraft Abilities}
\label{sec:casestudies}

\subsection{Improving compiler templates}
\label{sec:casestudy-template}

Beyond generating kernels from scratch, KernelCraft also demonstrates versatility in optimizing existing compiler templates. By incorporating hand-designed kernel templates into the system prompt, agents directly modify template logic while using the same tool-calling interface to evaluate assembly and iteratively refine performance. On the PLENA compiler’s FFN kernel template, the agent identifies that fully unrolled projection loops incur excessive scalar pointer arithmetic. It then leverages hardware loop instructions from the ISA specification and precomputes loop-invariant pointers to reduce overhead. For configuration C2 in \Cref{tab:plena-workload-configs}, the optimized template achieves a 94.5\% reduction in instruction count and a 2.9\% latency improvement by replacing repeated unrolled code with hardware loops and optimizing the tiling order. Appendix~\ref{appendix:ffn-template} presents both the original and KernelCraft-optimized templates. These results show that KernelCraft can naturally extend beyond kernel generation to compiler development and optimization tasks.

\subsection{Extending KernelCraft to the well-established ISA} 
\label{sec:casestudy-cpu}
\begin{table}[t]
\centering
\small
\caption{Evaluation of CPU kernel generation results using GPT-5.2. \emph{Iter.} denotes the iteration in which the agent first produces a functionally correct kernel. Speedup is reported relative to the RISC-V compiler baseline compiled with \texttt{-O2}. These results demonstrate that KernelCraft is able to generate correct assembly kernels within a small number of iterations.}
\label{tab:cpu-task-results}
\resizebox{\columnwidth}{!}{%
\begin{tabular}{llccc}
\toprule
\textbf{Task} & \textbf{Description} & \textbf{Succ.} & \textbf{Iter.} & \textbf{Speedup} \\
\midrule
\rowcolor{gray!15}
\multicolumn{5}{c}{\textbf{\textit{Level 1: Primitive Operations}} (max 10 iterations)} \\
\midrule
multiply & Shift-and-add multiplication & \cmark & 3  & 1.04$\times$ \\
vvadd    & Vector-vector add            & \cmark & 5  & 1.10$\times$ \\
copy     & Copy memory block            & \cmark & 8  & 1.50$\times$ \\
median   & 3-element median filter      & \cmark & 10  & 1.03$\times$ \\
dotprod  & Dot product                  & \cmark & 8  & 1.01$\times$ \\
\midrule
\rowcolor{blue!8}
\multicolumn{2}{l}{\textit{Level 1 Subtotal}} & \textit{5/5} &  & \\
\midrule
\rowcolor{gray!15}
\multicolumn{5}{c}{\textbf{\textit{Level 2: Algorithmic Operations}} (max 10 iterations)} \\
\midrule
qsort    & Quicksort algorithm          & \cmark & 9  & 1.63$\times$ \\
rsort    & Radix sort algorithm         & \cmark & 6 & 0.12$\times$ \\
towers   & Towers of Hanoi              & \cmark & 8  & 1.18$\times$ \\
\midrule
\rowcolor{blue!8}
\multicolumn{2}{l}{\textit{Level 2 Subtotal}} & \textit{3/3} &  & \\
\midrule
\rowcolor{gray!15}
\multicolumn{5}{c}{\textbf{\textit{Level 3: End-to-End System}} (max 15 iterations)} \\
\midrule
dhrystone & Mixed integer operations    & \cmark & 12 & 0.93$\times$ \\
pmp       & Physical Memory Protection  & \cmark & 8 & 1.68$\times$ \\
\midrule
\rowcolor{blue!8}
\multicolumn{2}{l}{\textit{Level 3 Subtotal}} & \textit{2/2} &  & \\
\midrule
\rowcolor{yellow!15}
\multicolumn{2}{l}{\textbf{Total}} & \textbf{10/10} & &  \\
\bottomrule
\end{tabular}
}
\end{table}

To evaluate the generality of KernelCraft beyond accelerator-oriented kernels, we extend our study to a diverse set of CPU workloads.
These benchmarks span primitive operations, algorithmic kernels, and end-to-end system workloads, with results summarized in \Cref{tab:cpu-task-results}.
Our results show that KernelCraft is able to generate kernels whose performance is comparable to that of well-established compiler toolchains in most cases.

\subsection{Extending KernelCraft to GPUs}

We further evaluate the generality of KernelCraft on NVIDIA's Streaming Assembly (SASS). Since NVIDIA SASS is not publicly documented and varies across GPU generations, we leverage CuAssembler~\cite{cuassembler}, an open-source PTX-to-SASS workflow, to extract instruction information, convert it into concise documentation, and integrate it into KernelCraft. We add a preliminary NVIDIA SASS experiment, shown in~\Cref{tab:sass} (Appendix ~\ref{app:sass_appendix}), demonstrating that most models struggle to generate correct SASS under limited documentation.

\subsection{Competing with human expert-tuned kernels}
We compare the assembly code generated by KernelCraft with hand-written kernels for the Coral NPU development team, with results summarized in \Cref{tab:case_study_human}.
The reference kernels are fully optimized for CNNs, with computation pipelines carefully scheduled and all execution stages explicitly fixed by human experts.
Our experimental results show that, while KernelCraft is able to generate a functionally correct kernel for this complex task within 15 iterations, its performance remains significantly lower than that of the fully optimized hand-tuned code.
This performance gap highlights both the difficulty of matching expert-level manual optimization and the opportunity for further improvement in automated kernel generation.

\begin{table}[t]
\centering
\small
\caption{Comparison of KernelCraft against human expert--optimized and compiler-generated kernels for depthwise convolution. All values report cycle counts and speedup relative to the compiler baseline is shown in parentheses. The selected C1, C2, and C3 correspond to those defined in \Cref{tab:coral-workload-configs}.}
\label{tab:case_study_human}
\begin{tabular}{crrr}
\toprule
\textbf{Config} & \textbf{Human Expert} & \textbf{KernelCraft} & \textbf{Compiler} \\
\midrule
C1 & 33327 (20$\times$) & 269239 (2.5$\times$) & 673098 \\
C2 & 75732 (18.2$\times$) & 2300040 (0.6$\times$) & 1380024 \\
C3 & 87470 (31.5$\times$) & 5510952 (0.5$\times$) &  2755476 \\
\bottomrule
\end{tabular}
\vspace{-15pt}
\end{table}

\subsection{Co-designing ISA for emerging ML workloads}
\label{sec:casestudy-isa}
During our experiments, we observed an intriguing behavior: when agents encounter inefficiencies in the current ISA specification, they attempt to propose new instructions to address the gap. This motivated a case study to evaluate LLM agents as collaborators in ISA co-design for emerging workloads, specifically diffusion language models (dLLMs) \cite{nie2025large, ye2025dream} on the PLENA accelerator.

Unlike autoregressive models, dLLMs perform iterative parallel denoising: in a typical setting, at each decoding step, the model predicts tokens across all masked positions, scores them by confidence, and selectively unmasks the highest-confidence predictions. This sampling pattern is not efficiently captured by the baseline PLENA ISA. To test the agent's ability to independently identify architectural gaps, we intentionally withheld specialized instructions proposed in prior work~\cite{lou2026beyond}. By extending KernelCraft with a conversational mode for iterative human feedback, the agent independently identified the deficiency and proposed instructions that closely mirrored the human-expert ISA design. Once the human in the loop provided the held-out instruction descriptions in the conversation context, the agent successfully implemented the sampling kernel. These findings highlight the potential for an agent-assisted co-design flow, where agents not only implement kernels but also proactively identify and resolve the hardware bottlenecks. Appendix~\ref{appendix:dllm} presents the entire case study.

%% file: text/conclusion.tex
\section{Conclusion}
We presented KernelCraft, the first benchmark for evaluating LLM agents' ability to generate low-level assembly kernels for emerging AI accelerators with newly introduced ISAs. 
Covering three complexity levels and multiple accelerator platforms, KernelCraft provides standardized tasks, tool interfaces, and diagnostic feedback to support fair comparisons of agent capabilities.
Evaluation of four frontier LLMs shows promising capabilities for generating effective kernels, but also reveals large performance gaps and several limitations. For example, some frontier models still struggle to complete end-to-end blocks, and exhibit a significant drop in performance for cases with insufficient documentation. Nonetheless, agent-generated kernels can match, and in some cases outperform, hand-optimized C++ implementations on the same hardware, while achieving up to $2\times$ speedup over compiler-generated baselines.

\section{Future Work}

\paragraph{Extension with formal verification.}
Kernel correctness verification remains crucial yet unstandardized, for both LLM-generated code and existing toolchains. KernelCraft currently ensures functional correctness only through numerical validation. Future work could explore agents generating formal correctness proofs.

\vspace{-5pt}
\paragraph{Fusing ISA knowledge into models.}
Our BOOM case study shows that internalized ISA knowledge leads to notably strong performance, suggesting that future work could embed ISA specifications directly into model training or fine-tuning.

\vspace{-5pt}
\paragraph{Agent system capabilities.}
KernelCraft could be extended with memory mechanisms that reduce per-iteration token usage, more specialized skill designs, and multi-agent settings.

\vspace{-2pt}
\section*{Impact Statement}

Our work aims to facilitate the development of new AI accelerators. The direct impact is that it will speed up deployment. While the downstream technologies will then have other societal impacts, which should be considered carefully, our work will have only an indirect (enabling) effect. Hence, there is nothing else to highlight in this statement.

\section*{Acknowledgment}
This work was supported by the Advanced Research and Invention Agency (ARIA) under the \href{https://aria.org.uk/opportunity-spaces/nature-computes-better/scaling-compute/}{Scaling Compute Programme}.

We thank David Gao from the Google Coral NPU team for providing expert hand-optimized kernels, which serve as human-expert baselines for evaluating KernelCraft-generated code on the Coral platform. This work was supported in part by the AMD University Program.

%% file: text/related.tex
\onecolumn

\section{Related Work}
\subsection{Benchmarks for Compute Kernel Generation}

In high-performance computing and deep learning systems, a compute kernel refers to a routine compiled for high-throughput accelerators (e.g., GPUs and NPUs) to perform specific mathematical operations. Unlike general-purpose software, such kernels require careful management of memory hierarchies and fine-grained thread parallelism.
Table~\ref{tab:benchmark-comparison} summarizes existing benchmarks designed to evaluate LLMs on compute kernel generation tasks. KernelBench~\cite{ouyang2025kernelbench} and TritonBench~\cite{li2025tritonbench} assess LLMs’ capabilities in automatically generating GPU kernel code using CUDA and Triton, respectively. NPUEval~\cite{kalade2025npueval} extends this evaluation to kernel generation for NPUs, while MultiKernelBench~\cite{wen2025multikernelbench} further broadens the scope by covering multiple hardware backends, including GPUs, NPUs, and TPUs.
In contrast, BackendBench~\cite{saroufim2025backendbench} focuses on evaluating PyTorch backend and kernel development, with a strong emphasis on production-level correctness and performance validation through PyTorch’s native testing infrastructure, rather than on standalone kernel code generation for specific accelerator programming models.
However, none of the above benchmarks are designed for custom accelerators. Our KernelCraft bridges this gap by benchmarking LLMs on generating low-level assembly code for kernel construction on custom accelerator architectures.

\subsection{LLM for Code Generation}

Large Language Models (LLMs) have demonstrated remarkable proficiency in code generation tasks, having been trained on large public and private code corpora. This is exemplified by state-of-the-art models such as OpenAI's GPT-5, Anthropic's Claude-4.5-Opus, and open-source alternatives like Qwen-3~\cite{yang2025qwen3} and DeepSeek-V3.2~\cite{liu2025deepseek}.

In academia, researchers are seeking ways to further improve the performance of code generation with LLMs. 
However, relying solely on single-pass generation often fails to address complex programming challenges. 
Consequently, recent studies have shifted focus towards Code Agents~\cite{dong2025survey}, systems that empower LLMs with the ability to plan, debug, and interact with execution environments iteratively to solve intricate software engineering tasks. 
A common paradigm is multi-turn refinement, where agents iteratively generate, execute, and repair code based on execution feedback.
In this paradigm, the agent generates code, executes it against a set of unit tests or a compiler, and utilizes the resulting error logs or execution traces to repair the code in subsequent turns. Prominent examples include Reflexion~\cite{shinn2023reflexion}, which employs verbal reinforcement to help agents reflect on feedback and correct their reasoning, and Self-Debug~\cite{chen2024teaching}, which enables LLMs to autonomously identify and fix bugs by analyzing execution results and explaining the code line-by-line. 
Building upon these iterative strategies, AlphaEvolve~\cite{novikov2025alphaevolve} has explored evolutionary approaches to optimize the generation process further. 
It introduces a framework where code generated by the LLM is rigorously assessed, and based on this feedback, the system utilizes evolutionary algorithms to iteratively refine the prompts.

Beyond internal reasoning and prompt optimization, recent advancements also equip single agents with external tool-use capabilities to overcome the limitations of static parametric knowledge.
By integrating with compilers, interpreters, and retrieval systems, agents can verify intermediate logic and access up-to-date documentation.
For instance, Toolformer~\cite{schick2023toolformer} demonstrates the efficacy of self-supervised API usage, while frameworks such as Parsel~\cite{zelikman2023parsel} enable agents to decompose complex algorithmic tasks into hierarchical function calls.
Complementary to these approaches, Easytool~\cite{yuan2025easytool} addresses a practical bottleneck in tool-augmented agents by transforming diverse and verbose tool documentations into concise and unified tool instructions, significantly reducing token consumption and improving tool-use performance.
Scaling tool-use capabilities to real-world software development, CodeAgent~\cite{zhang2024codeagent} tackles repository-level code generation by integrating a comprehensive suite of tools spanning information retrieval, code navigation, and testing, enabling agents to manage complex dependencies and external documentation that single-pass models often overlook.
Pushing this paradigm further into dynamic environments, Voyager~\cite{wang2024voyager} introduces lifelong learning, where an agent continuously writes and executes code to explore a game world.
It leverages execution feedback not only for immediate debugging, but also to curate a library of reusable code skills.
This paradigm transforms the LLM from a passive text generator into an active problem solver that interacts with its environment to validate hypotheses and accumulate practical experience.

While these single-agent frameworks equipped with refinement loops and external tools have shown promise, they often struggle with complex software engineering tasks due to limited context retention and the lack of diverse perspectives. To address these challenges, recent research has shifted towards Multi-Agent Systems (MAS), which simulate human development teams by assigning distinct roles to specialized agents. ChatDev~\cite{qian2024chatdev} pioneers this approach by modeling the software development lifecycle as a communicative chain, where agents act as CEOs, CTOs, and programmers to collaborate through a waterfall model. Building on this, MetaGPT~\cite{hong2023metagpt} incorporates Standardized Operating Procedures (SOPs) into the collaboration, requiring agents to generate structured outputs like Product Requirement Documents (PRDs) and UML diagrams before coding, thereby reducing hallucination and enhancing architectural consistency. Furthermore, AgentCoder~\cite{huang2023agentcoder} refines the verification process by introducing a multi-agent loop specifically designed for competitive programming; it employs separate agents for coding and test-case generation, allowing for rigorous self-verification against synthesized tests. Collectively, these frameworks demonstrate that decomposing complex coding tasks into specialized, collaborative sub-tasks yields superior robustness and code quality compared to monolithic generation approaches.

\subsection{Automatic Compute Kernel Generation}
Due to the excellent code generation ability of LLMs, recent work has explored using LLMs and agentic frameworks to automate compute kernel generation and optimization for hardware accelerators. 
A number of approaches focus on GPU kernels written in CUDA, where LLMs are combined with verification, search, or evolutionary strategies to iteratively discover high-performance implementations. 
For example, the AI CUDA Engineer~\cite{lange2025towards} and EvoEngineer~\cite{guo2025evoengineer} frameworks automate CUDA kernel generation and optimization, demonstrating that LLM-guided evolution can achieve substantial speedups over baseline PyTorch or hand-written kernels while maintaining correctness. 
In a similar spirit, inference-time scaling is effective for kernel generation: NVIDIA engineers~\cite{chen2025automating} demonstrate that DeepSeek-R1, coupled with a verifier-driven closed-loop workflow, can automatically generate and refine optimized GPU attention kernels that outperform expert-designed implementations in several cases.
Beyond CUDA-centric pipelines, some work emphasizes portability and operator coverage across accelerator platforms. 
TritorX~\cite{hammond2025agentic} presents an agentic system for generating Triton-based PyTorch ATen kernels at scale, prioritizing correctness and generality across diverse operators, data types, and shapes, thereby enabling rapid construction of backends for emerging ML accelerators. 
Complementary to direct kernel generation, AlphaEvolve~\cite{novikov2025alphaevolve} demonstrates that LLM-based agents can optimize hardware accelerator kernels across multiple abstraction levels, ranging from kernel-level tiling heuristics to direct optimization of compiler-generated intermediate representations that encapsulate kernels, such as FlashAttention.

Recent agentic approaches frame kernel generation and optimization as a closed-loop process rather than one-shot code synthesis: models iteratively propose kernels, invoke tools such as compilation/execution/profiling, and use automated verification to reject incorrect candidates and guide further refinements~\cite{lange2025towards, chen2025automating, wei2025astra, wang2025geak}. 
Robust-kbench~\cite{lange2025towards} couples such scaffolding with verifier-guided evolutionary refinement to translate PyTorch modules into faster CUDA implementations under a more rigorous evaluation setup. 
NVIDIA engineers similarly demonstrate inference-time scaling in a closed-loop workflow that repeatedly generates and refines CUDA attention kernels~\cite{chen2025automating}. 
Astra~\cite{wei2025astra} explores multi-agent optimization starting from existing CUDA kernels, coordinating iterative edits with testing and profiling to improve performance while maintaining correctness. 
Beyond CUDA as a programming model, GEAK~\cite{wang2025geak} targets Triton and uses Reflexion-style feedback to produce efficient kernels evaluated on a dedicated benchmark suite. 
Overall, these agentic systems remain largely GPU-centric; NPUEval~\cite{kalade2025npueval} is a notable exception that targets non-GPU devices by benchmarking NPU-oriented, domain-specific C++ kernels with compiler/hardware feedback.

While these approaches demonstrate the promise of LLMs for automatic kernel generation and optimization, they predominantly target established programming models (e.g., CUDA, Triton, or compiler IRs) and mainstream accelerator architectures. 
In contrast, our KernelCraft focuses on low-level compute kernel generation for custom accelerators, pushing automatic kernel generation below CUDA and compiler IRs by benchmarking LLMs on directly producing assembly code.

%% file: text/appendix.tex
\onecolumn

\section{Token Usage Analysis}
\label{app:token_usage_analysis}
\begin{figure*}[!h]
    \centering

    \includegraphics[width=0.75\textwidth]{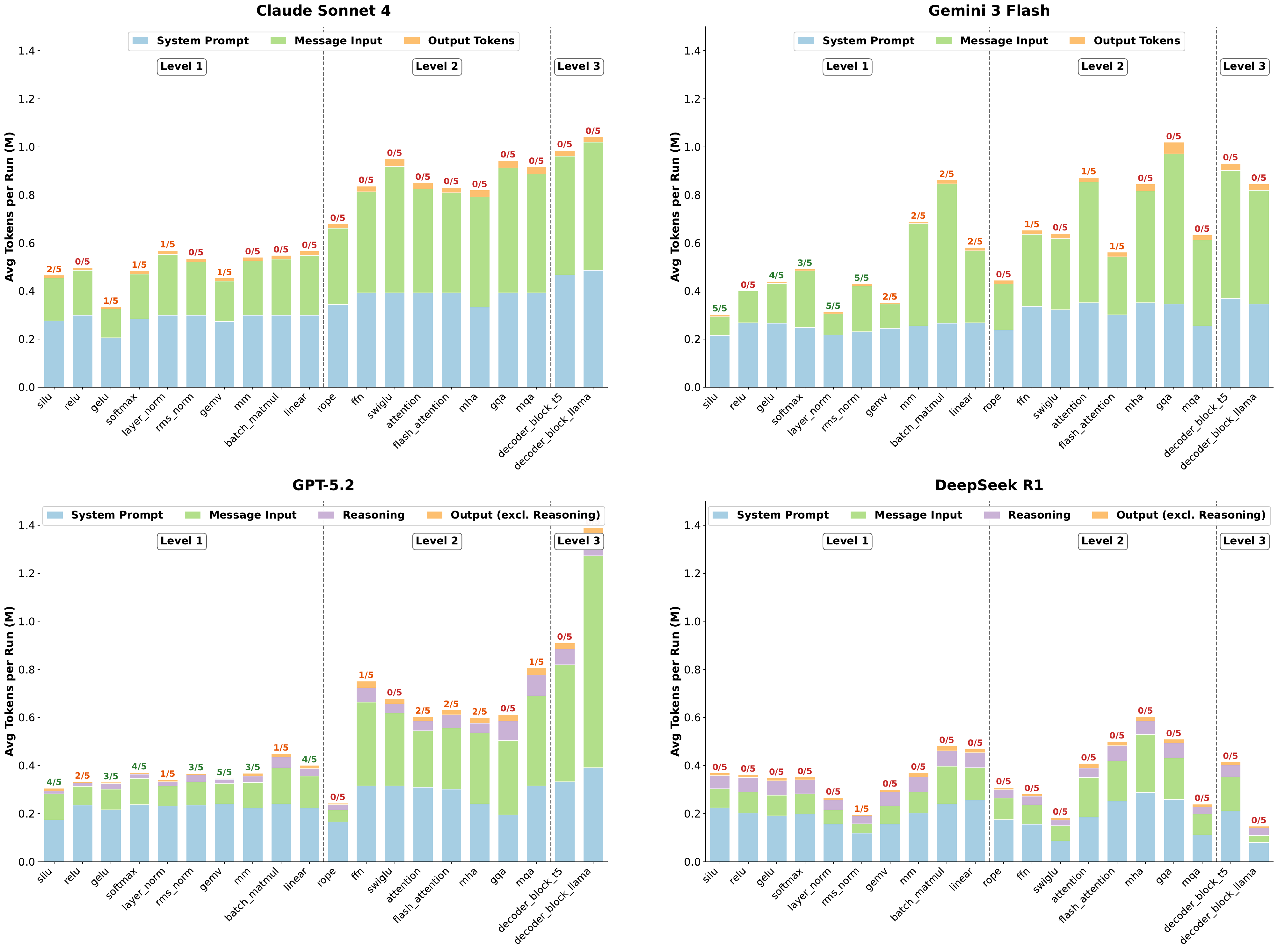}
\caption{Average token usage per workload across four LLMs on PLENA (5 runs each). Bars show per-run averages decomposed into system prompt, input, reasoning (GPT-5.2 and DeepSeek R1 only), and output tokens. Claude Sonnet 4 and Gemini 3 Flash include reasoning tokens within the output token count. Success rates are shown above each bar.}
    \label{fig:plena_tokens}
\end{figure*}

\begin{figure*}[!h]
    \centering
    \includegraphics[width=0.7\textwidth]{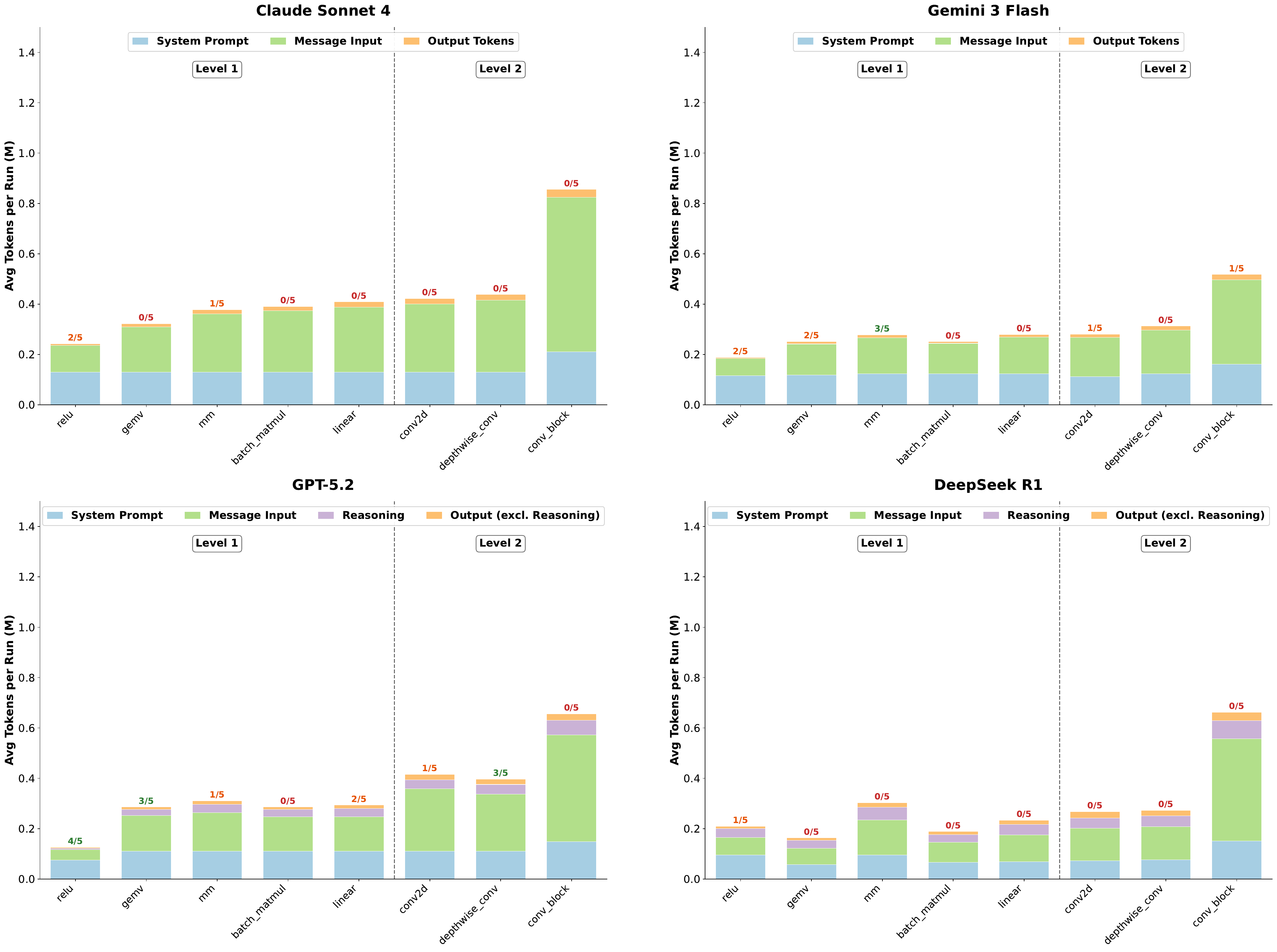}
\caption{Average token usage per workload across four LLMs on Coral NPU (5 runs each). Bars show per-run averages decomposed into system prompt, input, reasoning (GPT-5.2 and DeepSeek R1 only), and output tokens. Claude Sonnet 4 and Gemini 3 Flash include reasoning tokens within the output token count. Success rates are shown above each bar.}
    \label{fig:coral_tokens}
\end{figure*}

\newpage
\section{Evaluation Settings }
\setcounter{footnote}{0}
\label{app:Evaluation_Settings}

Our evaluation consists of two stages: functional correctness checking and performance evaluation of generated assembly kernels.
For correctness checking, we verify whether the generated assembly code produces numerically correct results under the target hardware semantics.
This process includes random input data generation, preparation of execution environments, execution via simulation or on physical hardware, and result validation.

Random data generation is used to create input stimuli for each kernel, enabling systematic correctness verification.
For each target platform, we compile the generated assembly kernel together with the corresponding input data to produce an executable artifact suitable for simulation or hardware execution.
Kernels are then executed either using cycle-accurate or RTL-based simulators, or directly on physical hardware, depending on platform availability.

Result validation is performed by comparing the execution outputs against standard reference results.
For machine learning workloads, we use PyTorch as the reference implementation, feeding identical input stimuli and checking numerical equivalence between the reference outputs and the kernel outputs.

After functional correctness is established, we conduct a performance evaluation.
We measure simulated cycle counts when using simulators, or record actual execution time when running on physical hardware.
Performance is reported by comparing the generated assembly kernels against compiler-generated baselines or other ground-truth implementations under identical configurations.

Overall, we implement a unified and automated evaluation pipeline across all evaluated accelerators and CPUs.
The hardware-specific execution environments are detailed below.

\subsection{Hardware Setup}

\begin{itemize}
    \item \textbf{PLENA.}\footnote{\url{https://github.com/AICrossSim/PLENA_Simulator.git}}
    PLENA provides a dedicated transactional simulator that executes compiled assembly code together with an explicit memory layout specification.
    Given an assembly kernel, the simulator models instruction execution and memory accesses to produce numerical outputs.
    Kernel correctness is verified by comparing the simulated outputs against reference results produced by PyTorch using identical inputs.

    \item \textbf{AMD NPU.}\footnote{\url{https://github.com/Xilinx/mlir-aie.git}}
    The AMD NPU does not provide a publicly available cycle-accurate simulator.
    Instead, we evaluate kernels on physical AMD NPU hardware accessed via the AMD Strix cluster on the Ryzen AI Cloud.
    We use the Peano compiler as a backend for AMD NPU code generation. High-level PyTorch programs are first lowered through the MLIR-AIR compiler stack to produce low-level representations, which are then compiled by Peano into executable binaries. In addition, Peano can directly compile low-level assembly code into executable binaries. In our experiments, KernelCraft-generated assembly kernels are compiled using Peano, deployed to the AMD NPU, and executed on-device. Correctness is validated by comparing hardware execution outputs against standard reference results.

    \item \textbf{Coral NPU.}\footnote{\url{https://github.com/google-coral/coralnpu.git}}
    The Coral NPU provides an open-source hardware and software stack, including complete RTL descriptions and a cocotb- and Verilator-based RTL simulation framework.
    We compile kernels to the Coral NPU ISA and evaluate them using RTL simulation, enabling both functional verification and cycle-level performance measurement.

    \item \textbf{BOOM.}\footnote{\url{https://github.com/riscv-boom/riscv-boom.git}}
    For CPU evaluation, we use the open-source RISC-V--based Sonic BOOM core.
    C++ kernels are compiled using a RISC-V toolchain into assembly and executable binaries.
    Execution and correctness are evaluated using the Verilator-based RTL simulation framework provided with BOOM.
\end{itemize}

\subsection{Functional Correctness Checking Methods and Their Robustness}
\label{app:tolerances}

Correctness is verified by comparing kernel outputs against reference implementations elementwise. We follow KernelBench \cite{ouyang2025kernelbench} in using \texttt{numpy.allclose} with tolerance parameters. Our tolerances are platform-specific because each target performs computation in a different arithmetic format:

\begin{itemize}
    \item \textbf{PLENA.} Activations use BF16 in VRAM; weights use MXFP8 (E4M3, 8-bit shared scale per block of eight) in HBM. Matrix multiplications accumulate in FP32, but intermediate results are quantized to BF16 between stages, introducing rounding errors. For multi-stage operators (e.g., FFNs), accumulated rounding can deviate up to one mantissa step ($2^{-6} \approx 0.015$). We use $\epsilon_{\text{abs}} = 0.012$, $\epsilon_{\text{rel}} = 0.01$. Ground-truth outputs are computed in PyTorch with MX quantization simulation.
    
    \item \textbf{AMD NPU.} Operating in full BF16 precision, we use \texttt{torch.allclose} defaults. Ground-truth outputs are computed in C++ with full precision.
    
    \item \textbf{Coral NPU.} Supporting only integer arithmetic at the time of writing with the latest Coral NPU repo (Coral NPU is an developing project), we require exact equivalence ($\epsilon = 0$). Workloads are implemented in C++ with int8 quantization.
\end{itemize}

To contextualise our tolerance choices against industry practice, we examine existing production-level toolchains. NVIDIA CUTLASS uses a symmetric reference for NVFP/MXFP error tolerance:
\begin{verbatim}
# For MXFP in CUTLASS, eps = 0.1, non-zero floor = 1.526e-5
|a - b|  <  eps * (|a| + |b|)        (general case)
|a - b|  <  eps * nonzero_floor      (near-zero case)
\end{verbatim}
Triton adopts \texttt{torch.allclose} with \texttt{atol=1e-3} and \texttt{rtol=1e-3}
for MXFP GEMM kernels. We further test these two error tolerance mechanisms on KernelCraft's PLENA GEMM kernel experiments and find that the generated kernels pass with negligible errors. As shown in \Cref{tab:tolerance}, the generated kernels pass under all three tolerance mechanisms (KernelCraft, CUTLASS, and Triton). We integrate these industry-standard metrics from CUTLASS and Triton into the KernelCraft evaluation system, giving future users the flexibility to select tolerance settings appropriate to their hardware platform.

\begin{table}[!h]
  \centering
  \caption{Tolerance comparison on four functionally correct GEMM kernels
  (GPT-5.2). Match rate = percentage of element-wise output values within tolerance to Ground Truth.}
  \label{tab:tolerance}
  \footnotesize
  \begin{tabular}{l r ccc}
  \toprule
  \rowcolor{gray!10}
  \textbf{Config} & \textbf{Size $(m,k,n)$ } & \textbf{KernelCraft} & \textbf{Cutlass} &
  \textbf{Triton}\\
  \midrule
  C1 & $4{\times}64{\times}64$     & 100\% & 99.6\% & 100\%   \\
  C2 & $8{\times}128{\times}128$   & 100\% & 100\%  & 98.6\% \\
  C3 & $16{\times}128{\times}256$  & 100\% & 99.8\% & 99.0\% \\
  C5 & $64{\times}256{\times}512$  & 100\% & 99.8\% & 97.3\% \\
  \bottomrule
  \end{tabular}

  \vspace{0.3em}
  \end{table}

Additionally, we conducted a systematic input robustness evaluation (\cref{tab:robustness}) to directly address boundary and adversarial inputs. We tested three functionally correct PLENA kernels across 13 conditions spanning four categories: random seeds (5 seeds), operand magnitude scaling (±0.003, ±0.03, ±0.3), edge cases (all zeros, all ones, all negative), and extreme values (1\% outliers at 20× magnitude, asymmetric operand scaling). All three kernels pass across all conditions, confirming robustness to input distribution shifts.

\begin{table}[!h]
  \centering
   \caption{Input robustness evaluation on three functionally correct PLENA kernels
  (GPT-5.2). Operand magnitude scales all tensors (activations and weights) uniformly.}
  \label{tab:robustness}
  \footnotesize
  \begin{tabular}{lp{4.8cm} ccc}
  \toprule
  \rowcolor{gray!10}
  \textbf{Category} & \textbf{Conditions} & \textbf{SiLU} & \textbf{Linear} &
  \textbf{FFN} \\
  \midrule
  Random seeds
    & 5 random seeds
    & 5/5 & 5/5 & 5/5 \\
  Operand magnitude
    & $\pm$0.003, $\pm$0.03, $\pm$0.3
    & 3/3 & 3/3 & 3/3 \\
  Edge cases
    & All zeros, all ones, all negative
    & 3/3 & 3/3 & 3/3 \\
  Extreme values
    & 1\% input extreme value,
      asymmetric operand magnitude
    & 2/2 & 2/2 & 2/2 \\
  \midrule
  \rowcolor{blue!8}
  \textbf{Total} & \textbf{13 conditions}
    & \textbf{13/13} & \textbf{13/13} & \textbf{13/13} \\
  \bottomrule
  \end{tabular}

  \vspace{0.3em}
  \end{table}

\subsection{Kernel Generation Baseline}
\label{app: Kernel_Gen_Baseline}
We use platform-specific reference implementations to establish ground-truth correctness:

\begin{itemize}
    \item \textbf{PLENA.} Ground-truth outputs are computed in PyTorch with MX quantization.
    Generated kernels are compiled using PLENA’s template-based compiler and executed on the PLENA transactional simulator.

    \item \textbf{Coral NPU.} Workloads are implemented in C++ with int8 quantization.
    Generated kernels are compiled and evaluated using a Verilator-based RTL simulation framework.

    \item \textbf{AMD NPU.} Ground-truth outputs are computed in C++ using full-precision arithmetic.
    Generated kernels are compiled using AMD’s NPU compiler toolchain and executed on AMD NPU hardware.
\end{itemize}

\subsection{Models}
\begin{table*}[!h]
\small
\caption{LLM Models Compared in This Study}
\centering
\label{tab:lm_models_settings}
\begin{tabular}{llll}
\toprule
\textbf{Provider} & \textbf{Model} & \textbf{Thinking Configuration} & \textbf{API Client} \\
\midrule
Anthropic & claude-sonnet-4-20250514 & budget\_tokens: 10K & Anthropic SDK \\
\midrule
OpenAI & GPT-5.2 & reasoning\_effort: medium & OpenAI SDK \\
\midrule
DeepSeek & DeepSeek-R1-0528  & Built-in CoT (not configurable)  &  OpenAI SDK (OpenRouter API) \\
\midrule
Google & gemini-3-flash-preview  & thinking\_level: medium & Google GenAI SDK \\
\bottomrule
\end{tabular}
\end{table*}

\section{Extending KernelCraft to GPUs}
\label{app:sass_appendix}
\begin{table*}[!h]
\centering
\small
\caption{Success rates for SASS kernel generation on NVIDIA GPUs. Each task is evaluated on five configurations; each cell reports successful cases over total cases within the iteration budget for each level.}
\label{tab:sass}
\begin{tabular}{ll|cccc}
\toprule
& & \multicolumn{4}{c}{\textbf{Nvidia SASS}} \\
\textbf{ID} & \textbf{Task} 
& GPT-5.2 & Gemini-3-flash & Sonnet 4 & DeepSeek R1 \\
\midrule
\rowcolor{gray!15}
\multicolumn{6}{c}{\textbf{\textit{Level 1: Primitive Operations}} (Max 15 iterations)} \\
\midrule
1  & SiLU          & \textbf{1/5} & \textcolor{gray}{0/5} & \textcolor{gray}{0/5} & \textcolor{gray}{0/5} \\
2  & ReLU          & \textcolor{gray}{0/5} & \textcolor{gray}{0/5} & \textcolor{gray}{0/5} & \textcolor{gray}{0/5} \\
3  & GELU          & \textcolor{gray}{0/5} & \textcolor{gray}{0/5} & \textcolor{gray}{0/5} & \textcolor{gray}{0/5} \\
4  & Softmax       & \textcolor{gray}{0/5} & \textcolor{gray}{0/5} & \textcolor{gray}{0/5} & \textcolor{gray}{0/5} \\
5  & LayerNorm     & \textcolor{gray}{0/5} & \textcolor{gray}{0/5} & \textcolor{gray}{0/5} & \textcolor{gray}{0/5} \\
6  & RMSNorm       & \textcolor{gray}{0/5} & \textcolor{gray}{0/5} & \textcolor{gray}{0/5} & \textcolor{gray}{0/5} \\
7  & GEMV          & \textcolor{gray}{0/5} & \textcolor{gray}{0/5} & \textcolor{gray}{0/5} & \textcolor{gray}{0/5} \\
8  & GEMM          & \textbf{1/5} & \textcolor{gray}{0/5} & \textcolor{gray}{0/5} & \textcolor{gray}{0/5} \\
9  & BatchMatMul   & \textcolor{gray}{0/5} & \textcolor{gray}{0/5} & \textcolor{gray}{0/5} & \textcolor{gray}{0/5} \\
10 & Linear        & \textbf{1/5} & \textcolor{gray}{0/5} & \textcolor{gray}{0/5} & \textcolor{gray}{0/5} \\
\bottomrule
\end{tabular}
\vspace{0.3em}
\end{table*}

\newpage
\input{text/appendix/configs_perf_tables}

\newpage
\input{text/appendix/ablations}

\newpage
\input{text/appendix/analysis}

\newpage
\input{text/appendix/template_case_study}

\newpage
\input{text/appendix/co_design_dllm}

\newpage
\input{text/appendix/system_prompt}

\newpage
\input{text/appendix/example_trace}

%% file: text/appendix/configs_perf_tables.tex
\section{Workload Configurations and Performance Table}
\label{app:perf_sections}
\begin{table}[!ht]
\small
\centering
\caption{PLENA Workload Configurations. Each workload uses 5 configurations of increasing complexity.}
\label{tab:plena-workload-configs}

\resizebox{\linewidth}{!}{%
\begin{tabular}{rl ccccc}
\toprule
\textbf{ID} & \textbf{Task (Parameters)} & \textbf{C1} & \textbf{C2} & \textbf{C3} & \textbf{C4} & \textbf{C5} \\
\midrule

1--3   & SiLU, ReLU, GELU ($n$)                & 256           & 1024          & 2048          & 8192          & 16384         \\
4--6   & Softmax, LayerNorm, RMSNorm ($n$)     & 256           & 1024          & 2048          & 8192          & 16384         \\
7      & GEMV ($M, N$)                         & (64,512)      & (128,512)     & (256,512)     & (512,512)     & (512,1024)    \\
8      & GEMM ($M, K, N$)                      & (4,64,64)     & (8,128,128)   & (16,128,256)  & (32,256,256)  & (64,256,512)  \\
9      & BatchMatMul ($b, M, K, N$)            & (2,4,64,64)   & (4,8,128,128) & (8,16,128,256)& (16,32,256,256)& (32,64,256,512)\\
10     & Linear ($n, h_\text{in}, h_\text{out}$)& (4,64,64)    & (8,128,128)   & (16,128,256)  & (32,256,256)  & (64,256,512)  \\
\midrule

13--14 & FFN, SwiGLU ($n, h, h_\text{i}$)      & (4,64,128)    & (8,128,256)   & (16,128,512)  & (32,256,512)  & (64,256,1024) \\
15--16 & Attention, FlashAttention ($s, h$)    & (32,64)       & (64,128)      & (64,256)      & (128,128)     & (128,256)     \\
17     & MHA ($s, h, n_\text{h}$)              & (32,64,1)     & (64,128,2)    & (64,256,4)    & (128,256,4)   & (128,512,8)   \\
18     & GQA ($s, h, n_\text{h}, n_\text{kv}$) & (32,128,2,1)  & (64,256,4,2)  & (64,512,8,4)  & (128,512,8,4) & (128,512,8,2) \\
19     & MQA ($s, h, n_\text{h}$)              & (32,128,2)    & (64,256,4)    & (64,512,8)    & (128,512,8)   & (128,512,8)   \\
20     & RoPE ($s, d_\text{head}$)             & (32,64)       & (64,128)      & (128,64)      & (128,128)     & (256,128)     \\
\midrule

21--22 & DecoderBlock \newline {\scriptsize (T5-style)}, DecoderBlock \newline {\scriptsize (LLaMA-style)} ($s, h, n_\text{h}, h_\text{i}$) 
       & (32,64,1,128) & (64,128,2,256) & (64,256,4,512) & (128,256,4,512) & (128,512,8,1024) \\

\bottomrule
\end{tabular}
}

\vspace{0.5em}
\footnotesize{%
  \textbf{Notation:} 
  $n$ = input tokens ($n = s \times b$),
  $s$ = sequence length,
  $h$ = \texttt{hidden\_size},
  $h_\text{i}$ = \texttt{intermediate\_size},
  $n_\text{h}$ = \texttt{num\_attention\_heads},
  $n_\text{kv}$ = \texttt{num\_key\_value\_heads},
  $d_\text{head}$ = \texttt{head\_dim}.
  Attention-based workloads use batch size $b=1$.
}
\end{table}

\begin{table}[!ht]                                                              
    \small                                                                        
    \centering                                                                    
    \caption{AMD NPU Workload Configurations. Each workload uses 5 configurations of
   increasing complexity.}                                                        
    \label{tab:aie-workload-configs}                                                  
    \resizebox{\linewidth}{!}{%
    \begin{tabular}{rl ccccc}                                                     
      \toprule                                                                    
      \textbf{ID} & \textbf{Task (parameters)} & \textbf{C1} & \textbf{C2} &      
  \textbf{C3} & \textbf{C4} & \textbf{C5} \\                                      
      \midrule                                                                    
      1--3  & SiLU, ReLU, GELU ($n$)                                              
            & 256           & 1024          & 2048          & 8192          & 16384 \\                                   
      4--6  & Softmax, LayerNorm, RMSNorm ($n$)                                   
            & 256           & 1024          & 2048          & 8192          & 16384 \\     
      7     & GEMV ($M, N$)                                                       
            & (64,64) & (128,256) & (256,256) & (512,256) & (512,1024) \\           
      8     & GEMM ($M, K, N$)                                                    
            & (8,64,64) & (16,256,128) & (32,256,256) & (64,512,256) &           
  (128,512,1024) \\                                                                
      9     & BatchMatMul ($b, M, K, N$)                                          
            & (8,16,64,64) & (16,32,256,128) & (16,64,256,256) & (32,64,512,256) &    
  (32,64,256,1024) \\                                                             
      10    & Linear ($n, h_\text{in}, h_\text{out}$)                             
            & (8,64,64) & (16,256,128) & (32,256,256) & (64,256,512) &            
  (128,1024,512) \\                                                                
      \midrule                                                                    
    13--14 & FFN, SwiGLU ($n, h, h_\text{i}$)                                                              
        & (8,64,128) & (16,128,256) & (32,256,512) & (64,256,512) &                                      
  (128,512,1024) \\
      15--16 & (SDPT)Attention, FlashAttention ($s, h$)                     
            & (8,16) & (16,32) & (32,64) & (64,64) & (128,64) \\        
      17    & MHA ($s, h, n_\text{h}$)                                            
            & (8,64,1) & (16,128,2) & (32,128,8) & (64,256,4) & (128,256,4) \\    
      18    & GQA ($s, h, n_\text{h}, n_\text{kv}$)                               
            & (8,64,4,1) & (16,128,2,1) & (32,128,8,2) & (64,256,4,2) &           
  (128,256,4,2) \\                                                                
      19    & MQA ($s, h, n_\text{h}$)                                            
            & (8,64,4) & (16,128,2) & (32,128,8) & (64,256,4) & (128,512,8) \\    
      20    & RoPE ($s, d_\text{head}$)                                           
            & (8,16) & (16,64) & (32,16) & (64,64) & (128,64) \\                  
      \midrule                                                                    
      21--22 & DecoderBlock \newline {\scriptsize (T5-style)}, DecoderBlock \newline {\scriptsize (LLaMA-style)} ($s, h, n_\text{h}, h_\text{i}$)        
            & (8,64,4,256) & (16,128,2,512) & (32,128,8,512) & (64,256,4,1024) &  
  (128,256,4,1024) \\                                                             
      \bottomrule                                                                 
    \end{tabular}                                                                 
    }                                                                             
    \vspace{0.5em}                                                                
                                                                                  
    \footnotesize{Notation conventions: $n$=input tokens ($n = s \times b$),      
  $s$=sequence length, $h$=\texttt{hidden\_size},                                 
  $h_\text{i}$=\texttt{intermediate\_size},                                       
  $n_\text{h}$=\texttt{num\_attention\_heads},                                    
  $n_\text{kv}$=\texttt{num\_key\_value\_heads},                                  
  $d_\text{head}$=\texttt{head\_dim}. Attention-based workloads use batch size    
  $b=1$ throughout.}                                                              
\end{table}

\begin{table}[!ht]
\small
\centering
\caption{Coral NPU Workload Configurations. Each workload uses 5 configurations of increasing complexity. All configs fit in 32KB DTCM and run under 2.5 minutes on Verilator.}
\label{tab:coral-workload-configs}

\resizebox{\linewidth}{!}{%
\begin{tabular}{rl ccccc}
\toprule
\textbf{ID} & \textbf{Task (Parameters)} & \textbf{C1} & \textbf{C2} & \textbf{C3} & \textbf{C4} & \textbf{C5} \\
\midrule
2  & ReLU ($n$)                                      & 512          & 2048         & 4096         & 8192         & 14336        \\
7  & GEMV ($M, N$)                                   & (64,32)      & (64,64)      & (128,64)     & (64,256)     & (256,64)     \\
8  & GEMM ($M, K, N$)                                & (32,32,32)   & (32,64,64)   & (64,64,64)   & (128,32,64)  & (32,64,128)  \\
9  & BatchMatMul ($b, M, K, N$)                      & (2,32,32,32) & (4,32,32,32) & (2,64,64,32) & (4,32,64,32) & (2,32,64,64) \\
10 & Linear ($b, h_\text{in}, h_\text{out}$)         & (4,32,64)    & (8,64,64)    & (4,256,64)   & (8,64,256)   & (12,128,128) \\
11 & Conv2D ($H, c_\text{in}, c_\text{out}, k$)      & (8,4,8,1)    & (8,8,8,3)    & (8,8,16,3)   & (8,8,32,3)   & (8,32,8,3)   \\
12 & DepthwiseConv ($H, c, k$)                       & (8,16,3)     & (16,8,3)     & (16,16,3)    & (16,16,5)    & (16,32,5)    \\
\midrule
23 & ConvBlock ($H, c_\text{in}, c_\text{out}, k$)   & (8,4,8,1)    & (8,8,8,3)    & (8,8,16,3)   & (8,8,32,3)   & (8,32,8,3)   \\
\bottomrule
\end{tabular}
}

\vspace{0.5em}
\footnotesize{%
  \textbf{Notation:}
  $n$ = number of elements,
  $M, K, N$ = matrix dimensions,
  $b$ = batch size,
  $h_\text{in}, h_\text{out}$ = input/output features,
  $H$ = spatial dimension (height = width),
  $c_\text{in}, c_\text{out}$ = input/output channels,
  $c$ = channels (depthwise),
  $k$ = kernel size.
  All operations use int8 quantization.
}
\end{table}

\definecolor{spdYellow}{RGB}{255, 243, 205}
\definecolor{spdGreen}{RGB}{212, 237, 218}
\definecolor{spdDarkGreen}{RGB}{170, 220, 190}
\definecolor{spdRed}{RGB}{248, 215, 218}

\begin{sidewaystable*}[p]
\centering
\scriptsize
\caption{Cycle Counts and Speedup on PLENA. Each KernelCraft cell reports cycles and speedup vs. the Compiler baseline (in parentheses). \colorbox{spdGreen}{Green} indicates a speedup ($\ge 1.0\times$), and \colorbox{spdRed}{Red} indicates a slowdown ($<1.0\times$). Lower cycles / higher speedup is better.}
\label{tab:cycles-plena}

\resizebox{\textheight}{!}{%
\begin{tabular}{ll|rrrrrrrrrr|rrrrrrrr|rr}
\toprule
& & \multicolumn{10}{c|}{\textbf{Level 1}} & \multicolumn{8}{c|}{\textbf{Level 2}} & \multicolumn{2}{c}{\textbf{Level 3}} \\
\cmidrule(lr){3-12} \cmidrule(lr){13-20} \cmidrule(lr){21-22}
\textbf{Cfg} & \textbf{Method}
& SiLU & ReLU & GELU & Softmax & LayerNorm & RMSNorm & GEMV & GEMM & BatchMatMul & Linear
& FFN & SwiGLU & Attn & FlashAttn & MHA & GQA & MQA & RoPE
&\makecell{DecoderBlock\\{\scriptsize(LLaMA-style)}}  & \makecell{DecoderBlock\\{\scriptsize(T5-style)}}\\
\midrule

\multirow{5}{*}{C1}
  & Compiler                      & 92 & --$^\dagger$ & 97 & --$^\dagger$ & 185 & 128 & --$^\dagger$ & 1421 & --$^\dagger$ & 1591 & 5480 & 8423 & --$^\dagger$ & --$^\dagger$ & --$^\dagger$ & --$^\dagger$ & --$^\dagger$ & --$^\dagger$ & --$^\dagger$ & --$^\dagger$ \\
  & KernelCraft [GPT-5.2]         & \cellcolor{spdGreen}75 (1.23$\times$) & 95 & \cellcolor{spdGreen}81 (1.20$\times$) & 117 & -- & -- & 2251 & \cellcolor{spdGreen}1308 (1.09$\times$) & 2602 & \cellcolor{spdGreen}1382 (1.15$\times$) & \cellcolor{spdGreen}5055 (1.08$\times$) & \cellcolor{spdGreen}7484 (1.13$\times$) & -- & -- & 42334 & 97240 & 100255 & -- & -- & -- \\
  & KernelCraft [Gemini-3-flash]  & \cellcolor{spdGreen}78 (1.18$\times$) & -- & \cellcolor{spdGreen}73 (1.33$\times$) & 129 & \cellcolor{spdGreen}154 (1.20$\times$) & \cellcolor{spdGreen}105 (1.22$\times$) & 10572 & \cellcolor{spdGreen}1321 (1.08$\times$) & 2644 & \cellcolor{spdGreen}1400 (1.14$\times$) & \cellcolor{spdGreen}5183 (1.06$\times$) & -- & -- & -- & -- & -- & -- & -- & -- & -- \\
  & KernelCraft [Sonnet 4]        & \cellcolor{spdGreen}80 (1.15$\times$) & 96 & \cellcolor{spdGreen}97 (1.00$\times$) & -- & \cellcolor{spdGreen}181 (1.02$\times$) & \cellcolor{spdRed}133 (0.96$\times$) & -- & -- & -- & -- & -- & -- & -- & -- & -- & -- & -- & -- & -- & -- \\
  & KernelCraft [DeepSeek R1]     & -- & -- & -- & -- & -- & -- & -- & -- & -- & -- & -- & -- & -- & -- & -- & -- & -- & -- & -- & -- \\
\midrule

\multirow{5}{*}{C2}
  & Compiler                      & 255 & --$^\dagger$ & 272 & --$^\dagger$ & 596 & 403 & --$^\dagger$ & 9504 & --$^\dagger$ & 10012 & 37864 & 57285 & --$^\dagger$ & --$^\dagger$ & --$^\dagger$ & --$^\dagger$ & --$^\dagger$ & --$^\dagger$ & --$^\dagger$ & --$^\dagger$ \\
  & KernelCraft [GPT-5.2]         & \cellcolor{spdRed}277 (0.92$\times$) & 269 & \cellcolor{spdRed}293 (0.93$\times$) & 429 & \cellcolor{spdGreen}536 (1.11$\times$) & \cellcolor{spdGreen}361 (1.12$\times$) & 3366 & \cellcolor{spdGreen}9034 (1.05$\times$) & -- & \cellcolor{spdGreen}9325 (1.07$\times$) & -- & -- & 72192 & 72317 & -- & -- & -- & -- & -- & -- \\
  & KernelCraft [Gemini-3-flash]  & \cellcolor{spdRed}302 (0.84$\times$) & -- & \cellcolor{spdGreen}262 (1.04$\times$) & -- & \cellcolor{spdGreen}539 (1.11$\times$) & \cellcolor{spdGreen}350 (1.15$\times$) & 19900 & -- & -- & \cellcolor{spdGreen}9409 (1.06$\times$) & -- & -- & 73542 & 71720 & -- & -- & -- & -- & -- & -- \\
  & KernelCraft [Sonnet 4]        & \cellcolor{spdRed}343 (0.74$\times$) & -- & -- & 506 & -- & -- & -- & -- & -- & -- & -- & -- & -- & -- & -- & -- & -- & -- & -- & -- \\
  & KernelCraft [DeepSeek R1]     & -- & -- & -- & -- & -- & -- & -- & -- & -- & -- & -- & -- & -- & -- & -- & -- & -- & -- & -- & -- \\
\midrule

\multirow{5}{*}{C3}
  & Compiler                      & 415 & --$^\dagger$ & 448 & --$^\dagger$ & 1159 & 774 & --$^\dagger$ & 36378 & --$^\dagger$ & 37453 & 145332 & 219531 & --$^\dagger$ & --$^\dagger$ & --$^\dagger$ & --$^\dagger$ & --$^\dagger$ & --$^\dagger$ & --$^\dagger$ & --$^\dagger$ \\
  & KernelCraft [GPT-5.2]         & \cellcolor{spdRed}532 (0.78$\times$) & -- & -- & 862 & \cellcolor{spdGreen}1056 (1.10$\times$) & \cellcolor{spdGreen}688 (1.13$\times$) & 7249 & \cellcolor{spdGreen}35248 (1.03$\times$) & -- & \cellcolor{spdGreen}35251 (1.06$\times$) & \cellcolor{spdGreen}141556 (1.03$\times$) & \cellcolor{spdGreen}215500 (1.02$\times$) & -- & 141865 & 981297 & -- & -- & -- & -- & -- \\
  & KernelCraft [Gemini-3-flash]  & \cellcolor{spdRed}533 (0.78$\times$) & -- & \cellcolor{spdRed}567 (0.79$\times$) & 882 & \cellcolor{spdGreen}1059 (1.09$\times$) & \cellcolor{spdGreen}704 (1.10$\times$) & -- & -- & -- & -- & -- & -- & -- & -- & -- & -- & -- & -- & -- & -- \\
  & KernelCraft [Sonnet 4]        & -- & -- & -- & 895 & \cellcolor{spdGreen}1103 (1.05$\times$) & -- & -- & -- & -- & -- & -- & -- & -- & -- & -- & -- & -- & -- & -- & -- \\
  & KernelCraft [DeepSeek R1]     & -- & -- & -- & -- & -- & \cellcolor{spdGreen}704 (1.10$\times$) & -- & -- & -- & -- & -- & -- & -- & -- & -- & -- & -- & -- & -- & -- \\
\midrule

\multirow{5}{*}{C4}
  & Compiler                      & 1429 & --$^\dagger$ & 1558 & --$^\dagger$ & 4202 & 2793 & --$^\dagger$ & 140604 & --$^\dagger$ & 143885 & 563297 & 844781 & --$^\dagger$ & --$^\dagger$ & --$^\dagger$ & --$^\dagger$ & --$^\dagger$ & --$^\dagger$ & --$^\dagger$ & --$^\dagger$ \\
  & KernelCraft [GPT-5.2]         & \cellcolor{spdRed}2181 (0.66$\times$) & -- & \cellcolor{spdRed}2448 (0.64$\times$) & 3379 & \cellcolor{spdRed}4297 (0.98$\times$) & -- & 15529 & -- & 2255773 & -- & \cellcolor{spdGreen}557836 (1.01$\times$) & \cellcolor{spdRed}861279 (0.98$\times$) & 279841 & 282072 & 2223983 & -- & -- & 5835 & -- & -- \\
  & KernelCraft [Gemini-3-flash]  & \cellcolor{spdRed}2151 (0.66$\times$) & -- & \cellcolor{spdRed}2576 (0.60$\times$) & 3293 & \cellcolor{spdGreen}4051 (1.04$\times$) & \cellcolor{spdGreen}2801 (1.00$\times$) & -- & \cellcolor{spdRed}149776 (0.94$\times$) & 2294499 & -- & -- & -- & 323515 & -- & -- & -- & -- & -- & -- & -- \\
  & KernelCraft [Sonnet 4]        & -- & -- & -- & 4041 & -- & -- & 15780 & -- & -- & -- & -- & -- & -- & -- & -- & -- & -- & -- & -- & -- \\
  & KernelCraft [DeepSeek R1]     & -- & -- & -- & -- & -- & -- & -- & -- & -- & -- & -- & -- & -- & -- & -- & -- & -- & -- & -- & -- \\
\midrule

\multirow{5}{*}{C5}
  & Compiler                      & 2709 & --$^\dagger$ & 2966 & --$^\dagger$ & 8322 & 5505 & --$^\dagger$ & 556366 & --$^\dagger$ & 564030 & 2224214 & 3342678 & --$^\dagger$ & --$^\dagger$ & --$^\dagger$ & --$^\dagger$ & --$^\dagger$ & --$^\dagger$ & --$^\dagger$ & --$^\dagger$ \\
  & KernelCraft [GPT-5.2]         & \cellcolor{spdRed}4217 (0.64$\times$) & -- & \cellcolor{spdRed}4532 (0.65$\times$) & 4998 & -- & \cellcolor{spdGreen}5167 (1.07$\times$) & 31201 & \cellcolor{spdGreen}556524 (1.00$\times$) & -- & \cellcolor{spdGreen}562768 (1.00$\times$) & -- & \cellcolor{spdGreen}3353934 (1.00$\times$) & 557952 & -- & -- & -- & -- & -- & -- & -- \\
  & KernelCraft [Gemini-3-flash]  & \cellcolor{spdRed}4547 (0.60$\times$) & -- & -- & -- & \cellcolor{spdGreen}7865 (1.06$\times$) & \cellcolor{spdGreen}5122 (1.07$\times$) & -- & -- & -- & -- & \cellcolor{spdRed}2432184 (0.91$\times$) & -- & -- & -- & -- & -- & -- & -- & -- & -- \\
  & KernelCraft [Sonnet 4]        & -- & -- & -- & 7221 & -- & -- & -- & -- & -- & -- & -- & -- & -- & -- & -- & -- & -- & -- & -- & -- \\
  & KernelCraft [DeepSeek R1]     & -- & -- & -- & -- & -- & -- & -- & -- & -- & -- & -- & -- & -- & -- & -- & -- & -- & -- & -- & -- \\
\bottomrule
\end{tabular}
}

\vspace{0.5em}
{\small $^\dagger$ Not officially supported by PLENA compiler.}
\end{sidewaystable*}

\begin{sidewaystable*}[p]
\centering
\scriptsize
\caption{Speedup on AMD NPU relative to the \texttt{C++$\rightarrow$Peano} compiler baseline. \colorbox{spdGreen}{Green} indicates speedup ($\ge 1.0\times$), \colorbox{spdRed}{Red} indicates slowdown ($<1.0\times$), and -- indicates no correct kernel generated.}
\label{tab:cycles-amd}

\resizebox{\textheight}{!}{%
\begin{tabular}{ll|rrrrrrrrrrrr|rrrrrrr|rrr}
\toprule
& & \multicolumn{12}{c|}{\textbf{Level 1}} & \multicolumn{7}{c|}{\textbf{Level 2}} & \multicolumn{3}{c}{\textbf{Level 3}} \\
\cmidrule(lr){3-14} \cmidrule(lr){15-21} \cmidrule(lr){22-24}
\textbf{Cfg} & \textbf{Method}
& SiLU & ReLU & GELU & Softmax & LayerNorm & RMSNorm & GEMV & GEMM & BatchMatMul & Linear & Conv2D & DepthwiseConv
& FFN & SwiGLU & ScaledDotProduct & MHA & GQA & MQA & RoPE
& ConvBlock & \makecell{DecoderBlock\\{\scriptsize(LLaMA-style)}} & \makecell{DecoderBlock\\{\scriptsize(T5-style)}} \\
\midrule

\multirow{4}{*}{C1}
  & GPT-5.2
  & -- & -- & -- & --
  & \cellcolor{spdRed}0.96$\times$
  & --
  & \cellcolor{spdRed}0.87$\times$
  & \cellcolor{spdGreen}1.05$\times$
  & --
  & \cellcolor{spdRed}0.82$\times$
  & -- & --
  & \cellcolor{spdGreen}1.10$\times$
  & -- & -- & -- & -- & -- & --
  & -- & -- & -- \\

  & Gemini-3-flash
  & -- & -- & -- & --
  & \cellcolor{spdRed}0.90$\times$
  & --
  & \cellcolor{spdGreen}1.03$\times$
  & \cellcolor{spdRed}0.84$\times$
  & --
  & \cellcolor{spdRed}0.92$\times$
  & -- & --
  & \cellcolor{spdRed}0.89$\times$
  & -- & -- & -- & -- & -- & --
  & -- & -- & -- \\

  & Sonnet 4
  & -- & -- & -- & --
  & -- & -- & --
  & \cellcolor{spdGreen}1.06$\times$
  & --
  & \cellcolor{spdRed}0.62$\times$
  & -- & --
  & -- & -- & -- & -- & -- & -- & --
  & -- & -- & -- \\

  & DeepSeek R1
  & -- & -- & -- & -- & -- & -- & --
  & \cellcolor{spdRed}0.87$\times$
  & -- & -- & -- & --
  & -- & -- & -- & -- & -- & -- & --
  & -- & -- & -- \\

\midrule

\multirow{4}{*}{C2}
  & GPT-5.2
  & -- & -- & -- & -- & \cellcolor{spdRed}0.88$\times$
  & \cellcolor{spdGreen}1.18$\times$
  & --
  & \cellcolor{spdGreen}1.02$\times$
  & --
  & \cellcolor{spdGreen}1.04$\times$
  & -- & --
  & -- & -- & -- & -- & -- & -- & --
  & -- & -- & -- \\

  & Gemini-3-flash
  & -- & -- & -- & -- & -- & -- & --
  & \cellcolor{spdRed}0.99$\times$
  & --
  & \cellcolor{spdRed}0.58$\times$
  & -- & --
  & -- & -- & -- & -- & -- & -- & --
  & -- & -- & -- \\

  & Sonnet 4
  & -- & -- & -- & -- & -- & -- & -- & -- & -- & -- & -- & --
  & -- & -- & -- & -- & -- & -- & --
  & -- & -- & -- \\

  & DeepSeek R1
  & -- & -- & -- & -- & -- & -- & -- & -- & -- & -- & -- & --
  & -- & -- & -- & -- & -- & -- & --
  & -- & -- & -- \\

\midrule

\multirow{4}{*}{C3}
  & GPT-5.2
  & -- & -- & -- & -- & -- & --
  & \cellcolor{spdRed}0.93$\times$
  & \cellcolor{spdGreen}1.09$\times$
  & --
  & \cellcolor{spdRed}0.89$\times$
  & -- & --
  & -- & -- & \cellcolor{spdRed}0.69$\times$ & -- & -- & -- & --
  & -- & -- & -- \\

  & Gemini-3-flash
  & -- & -- & -- & -- & -- & -- & --
  & \cellcolor{spdGreen}1.10$\times$
  & -- & -- & -- & --
  & -- & -- & -- & -- & -- & -- & --
  & -- & -- & -- \\

  & Sonnet 4
  & -- & -- & -- & -- & -- & -- & -- & -- & -- & -- & -- & --
  & -- & -- & -- & -- & -- & -- & --
  & -- & -- & -- \\

  & DeepSeek R1
  & -- & -- & -- & -- & -- & -- & -- & -- & -- & -- & -- & --
  & -- & -- & -- & -- & -- & -- & --
  & -- & -- & -- \\

\midrule

\multirow{4}{*}{C4}
  & GPT-5.2
  & \cellcolor{spdGreen}1.08$\times$
  & \cellcolor{spdRed}0.99$\times$
  & \cellcolor{spdRed}0.98$\times$
  & -- & -- & -- & --
  & \cellcolor{spdGreen}1.07$\times$
  & -- & -- & -- & --
  & \cellcolor{spdGreen}1.09$\times$
  & -- & -- & -- & -- & -- & --
  & -- & -- & -- \\

  & Gemini-3-flash
  & \cellcolor{spdRed}0.98$\times$
  & \cellcolor{spdRed}0.99$\times$
  & \cellcolor{spdGreen}1.04$\times$
  & -- & -- & -- & -- & -- & -- & -- & -- & --
  & -- & -- & -- & -- & -- & -- & --
  & -- & -- & -- \\

  & Sonnet 4
  & -- & -- & -- & -- & -- & -- & -- & -- & -- & -- & -- & --
  & -- & -- & -- & -- & -- & -- & --
  & -- & -- & -- \\

  & DeepSeek R1
  & -- & -- & -- & -- & -- & -- & -- & -- & -- & -- & -- & --
  & -- & -- & -- & -- & -- & -- & --
  & -- & -- & -- \\

\midrule

\multirow{4}{*}{C5}
  & GPT-5.2
  & -- & \cellcolor{spdGreen}1.04$\times$ & -- & -- & -- & -- & -- & -- & -- & -- & -- & --
  & -- & -- & -- & -- & -- & -- & --
  & -- & -- & -- \\

  & Gemini-3-flash
  & -- & -- & \cellcolor{spdRed}0.83$\times$ & -- & -- & -- & -- & -- & -- & -- & -- & --
  & -- & -- & -- & -- & -- & -- & --
  & -- & -- & -- \\

  & Sonnet 4
  & -- & -- & -- & -- & -- & -- & -- & -- & -- & -- & -- & --
  & -- & -- & -- & -- & -- & -- & --
  & -- & -- & -- \\

  & DeepSeek R1
  & -- & -- & -- & -- & -- & -- & -- & -- & -- & -- & -- & --
  & -- & -- & -- & -- & -- & -- & --
  & -- & -- & -- \\

\bottomrule
\end{tabular}
}

\vspace{0.5em}
\end{sidewaystable*}

\begin{sidewaystable*}[p]
\centering
\scriptsize
\caption{Cycle Counts and Speedup on Coral NPU. Each KernelCraft cell reports cycles and speedup vs.\ the RVV Intrinsics -O2 baseline (in parentheses). \colorbox{spdGreen}{Green} indicates a speedup ($\ge 1.0\times$), and \colorbox{spdRed}{Red} indicates a slowdown ($<1.0\times$). Lower cycles / higher speedup is better.}
\label{tab:cycles-coral}

\resizebox{\textheight}{!}{%
\begin{tabular}{ll|rrrrrrrrrrrr|rr|r}
\toprule
& & \multicolumn{12}{c|}{\textbf{Level 1}} & \multicolumn{2}{c|}{\textbf{Level 2}} & \multicolumn{1}{c}{\textbf{Level 3}} \\
\cmidrule(lr){3-14} \cmidrule(lr){15-16} \cmidrule(lr){17-17}
\textbf{Cfg} & \textbf{Method}
& SiLU & ReLU & GELU & Softmax & LayerNorm & RMSNorm & GEMV & GEMM & BatchMatMul & Linear & Conv2D & DepthwiseConv
& FFN & SwiGLU
& ConvBlock \\
\midrule

\multirow{6}{*}{C1}
  & RVV Intrinsics -O2   & 595,299 & 1,566 & 600,040 & 76,657 & 2,742 & 2,393 & 7,602 & 31,518 & 4,764 & 18,542 & 8,860 & 21,498 & -- & -- & 33,371 \\
  & RVV Intrinsics -O3   & 595,815 & 1,549 & 599,017 & 76.650 & 2,739 & 2,396 & 7,609 & 31,535 & 3,978 & 18,540 & 7,081 & 29,261 & -- & -- & 27,705 \\
  & KernelCraft [GPT-5.2]
    & \cellcolor{spdGreen}486,781 (1.22$\times$)
    & \cellcolor{spdRed}1,744 (0.90$\times$)
    & \cellcolor{spdGreen}513,798 (1.17$\times$)
    & \cellcolor{spdRed}82,446 (0.93$\times$) & \cellcolor{spdRed}3,974(0.69$\times$) & \cellcolor{spdGreen}1,932 (1.24$\times$)
    & --
    & \cellcolor{spdGreen}19,689 (1.60$\times$)
    & -- & --
    & \cellcolor{spdRed}10,353 (0.86$\times$)
    & \cellcolor{spdGreen}12,195 (1.76$\times$)
    & -- & -- & -- \\
  & KernelCraft [Gemini-3-flash]
    & \cellcolor{spdGreen}591,325 (1.01$\times$)
    & \cellcolor{spdRed}1,597 (0.98$\times$)
    & \cellcolor{spdGreen}561,494 (1.07$\times$)
    & \cellcolor{spdRed}90,205 (0.85$\times$) & -- & --
    & \cellcolor{spdGreen}3,485 (2.18$\times$)
    & \cellcolor{spdGreen}11,927 (2.64$\times$)
    & -- & --
    & \cellcolor{spdGreen}3,807 (2.33$\times$)
    & \cellcolor{spdGreen}8,733 (2.46$\times$)
    & -- & --
    & \cellcolor{spdGreen}\textbf{4,206 (7.93$\times$)} \\
  & KernelCraft [Sonnet 4]
    & -- & -- & -- & -- & -- & -- & --
    & \cellcolor{spdGreen}21,137 (1.49$\times$)
    & -- & -- & -- & -- & -- & -- & -- \\
  & KernelCraft [DeepSeek R1]
    & -- & -- & -- & -- & -- & -- & -- & -- & -- & -- & -- & -- & -- & -- & -- \\
\midrule

\multirow{6}{*}{C2}
  & RVV Intrinsics -O2   & 595,349 & 5,541 & 600,083 & 152,212 & 5,101 & 3,842 & 11,524 & 118,005 & 9,198 & 61,523 & 100,994 & 17,529 & -- & -- & 127,591 \\
  & RVV Intrinsics -O3   & 595,865 & 5,521 & 599,065 & 152,215 & 5,103 & 3,841 & 11,526 & 117,978 & 7,422 & 61,529 & 73,320 & 25,729 & -- & -- & -- \\
  & KernelCraft [GPT-5.2]
    & \cellcolor{spdRed}642,732 (0.93$\times$)
    & \cellcolor{spdGreen}3,371 (1.64$\times$)
    & \cellcolor{spdGreen}577,369 (1.04$\times$)
    & \cellcolor{spdGreen}137,127 (1.11$\times$) & -- & --
    & \cellcolor{spdGreen}5,672 (2.03$\times$)
    & -- & -- & --
    & \cellcolor{spdGreen}34,609 (2.92$\times$)
    & \cellcolor{spdRed}29,658 (0.59$\times$)
    & -- & -- & -- \\
  & KernelCraft [Gemini-3-flash]
    & \cellcolor{spdGreen}548,798 (1.08$\times$)
    & \cellcolor{spdRed}5,791 (0.96$\times$)
    & \cellcolor{spdRed}736,913 (0.81$\times$)
    & \cellcolor{spdRed}155,318 (0.98$\times$) & -- & \cellcolor{spdRed}2,420 (0.99$\times$)
    & \cellcolor{spdGreen}4,457 (2.59$\times$)
    & \cellcolor{spdGreen}63,400 (1.86$\times$)
    & -- & -- & --
    & \cellcolor{spdRed}65,433 (0.27$\times$)
    & -- & -- & -- \\
  & KernelCraft [Sonnet 4]
    & --
    & \cellcolor{spdRed}6,490 (0.85$\times$)
    & -- & -- & -- & -- & -- & -- & -- & -- & -- & -- & -- & -- & -- \\
  & KernelCraft [DeepSeek R1]
    & -- & -- & -- & -- & -- & -- & -- & -- & -- & -- & -- & -- & -- & -- & -- \\
\midrule

\multirow{6}{*}{C3}
  & RVV Intrinsics -O2   & 595,457 & 10,838 & 600,192 & 606,726 & 17,979 & 11,260 & 22,522 & 231,383 & 4,773 & 125,401 & 133,403 & 21,504 & -- & -- & 186,820 \\
  & RVV Intrinsics -O3   & 595,968 & 10,823 & 599,171 & 606,725 & 17,977 & 11,263 & 22,524 & 231,386 & 3,983 & 125,398 & 105,839 & 29,272 & -- & -- & -- \\
  & KernelCraft [GPT-5.2]
    & \cellcolor{spdGreen}581,712 (1.02$\times$)
    & \cellcolor{spdGreen}9,903 (1.09$\times$)
    & \cellcolor{spdRed}641,771 (0.94$\times$)
    & \cellcolor{spdGreen}577,834 (1.05$\times$) & -- & --
    & \cellcolor{spdGreen}18,786 (1.20$\times$)
    & -- & --
    & \cellcolor{spdGreen}69,950 (1.79$\times$)
    & --
    & \cellcolor{spdRed}44,970 (0.48$\times$)
    & -- & -- & -- \\
  & KernelCraft [Gemini-3-flash]
    & \cellcolor{spdRed}612,103 (0.97$\times$)
    & \cellcolor{spdGreen}8,919 (1.22$\times$)
    & \cellcolor{spdRed}629,134 (0.95$\times$)
    & -- & -- & --
    & \cellcolor{spdGreen}11,020 (2.04$\times$)
    & -- & -- & -- & -- & --
    & -- & -- & -- \\
  & KernelCraft [Sonnet 4]
    & -- & -- & -- & -- & -- & -- & -- & -- & -- & -- & -- & -- & -- & -- & -- \\
  & KernelCraft [DeepSeek R1]
    & -- & -- & -- & -- & -- & -- & -- & -- & -- & -- & -- & -- & -- & -- & -- \\
\midrule

\multirow{6}{*}{C4}
  & RVV Intrinsics -O2   & 595,670 & 21,441 & 600,398 & 1,131,907 & 35,324 & 21,351 & 37,714 & 239,176 & 9,182 & 243,484 & 220,148 & 32,460 & -- & -- & 317,393 \\
  & RVV Intrinsics -O3   & 596,183 & 21,412 & 599,385 & 1,129,130 & 35,328 & 21,353 & 37,686 & 239,037 & 7,423 & 243,484 & 187,560 & 31,391 & -- & -- & 364,219 \\
  & KernelCraft [GPT-5.2]
    & --
    & \cellcolor{spdGreen}19,692 (1.09$\times$)
    & \cellcolor{spdGreen}573,128 (1.05$\times$)
    & \cellcolor{spdRed}1,432,793 (0.79$\times$) & -- & --
    & \cellcolor{spdGreen}23,472 (1.61$\times$)
    & \cellcolor{spdGreen}192,507 (1.24$\times$)
    & --
    & \cellcolor{spdGreen}121,398 (2.01$\times$)
    & --
    & \cellcolor{spdRed}102,621 (0.32$\times$)
    & -- & -- & -- \\
  & KernelCraft [Gemini-3-flash]
    & --
    & \cellcolor{spdRed}22,769 (0.94$\times$)
    & \cellcolor{spdGreen}591,670 (1.01$\times$)
    & -- & -- & --
    & \cellcolor{spdGreen}18,616 (2.03$\times$)
    & \cellcolor{spdGreen}88,311 (2.71$\times$)
    & -- & -- & --
    & \cellcolor{spdRed}87,622 (0.37$\times$)
    & -- & -- & -- \\
  & KernelCraft [Sonnet 4]
    & -- & -- & -- & -- & -- & -- & -- & -- & -- & -- & -- & -- & -- & -- & -- \\
  & KernelCraft [DeepSeek R1]
    & -- & -- & -- & -- & --  & -- & --
    & \cellcolor{spdGreen}230,753 (1.04$\times$)
    & -- & -- & -- & -- & -- & -- & -- \\
\midrule

\multirow{6}{*}{C5}
  & RVV Intrinsics -O2   & 596,101 & 37,320 & 595,670 & -- & 68,615 & 40,090 & 44,570 & 233,399 & 4,774 & 343,134 & 344,584 & 52,185 & -- & -- & 372,400 \\
  & RVV Intrinsics -O3   & 596,620 & 37,303 & 599,818 & -- & 68,615 & 40,086 & 44,576 & 233,406 & 3,980 & 343,134 & 343,935 & 51,833 & -- & -- & 353,201 \\
  & KernelCraft [GPT-5.2]
    & --
    & \cellcolor{spdRed}41,020 (0.91$\times$)
    & \cellcolor{spdGreen}529,600 (1.12$\times$)
    & -- & -- & --
    & \cellcolor{spdGreen}33,929 (1.31$\times$)
    & -- & -- & -- & --
    & \cellcolor{spdRed}238,855 (0.22$\times$)
    & -- & -- & -- \\
  & KernelCraft [Gemini-3-flash]
    & -- & --
    & \cellcolor{spdGreen}571,999 (1.04$\times$)
    & -- & -- & --
    & \cellcolor{spdGreen}21,702 (2.05$\times$)
    & \cellcolor{spdGreen}72,038 (3.24$\times$)
    & -- & -- & -- & --
    & -- & -- & -- \\
  & KernelCraft [Sonnet 4]
    & -- & -- & -- & -- & -- & -- & -- & -- & -- & -- & -- & -- & -- & -- & -- \\
  & KernelCraft [DeepSeek R1]
    & -- & -- & -- & -- & -- & -- & -- & -- & -- & -- & -- & -- & -- & -- & -- \\
\bottomrule
\end{tabular}
}

\vspace{0.5em}
\end{sidewaystable*}

%% file: text/appendix/ablations.tex
\section{Harness Ablation}
\label{app:harness-ablations}

This appendix section collects the full set of harness ablation results referenced in \Cref{sec:ablation}.

\begin{table*}[!h]
\centering
\caption{Cross-platform ISA documentation ablation with GPT-5.2. Documentation
levels: D1, instruction syntax only; D2, adds per-instruction usage examples and
latencies; D3, adds full behavioral descriptions (PLENA only). Fractions denote the
pass rate over tested configurations.}
\label{tab:isa-depth-ablation}
\footnotesize
\setlength{\tabcolsep}{4pt}
\begin{tabular}{@{}l ccc @{\hspace{1.5em}} l cc @{\hspace{1.5em}} l cc@{}}
\toprule
\multicolumn{4}{c}{\textbf{PLENA}} &
\multicolumn{3}{c}{\textbf{Coral NPU}} &
\multicolumn{3}{c}{\textbf{AMD NPU}} \\
\cmidrule{1-10}
\textbf{Task} & \textbf{D1} & \textbf{D2} & \textbf{D3} &
\textbf{Task} & \textbf{D1} & \textbf{D2} &
\textbf{Task} & \textbf{D1} & \textbf{D2} \\
\midrule
SiLU             & 4/5 & 5/5 & \textbf{5/5} & ReLU          & 5/5 & \textbf{5/5} & ReLU             & 2/5 & \textbf{2/5} \\
Linear           & 0/5 & 1/5 & \textbf{4/5} & Linear        & 1/5 & \textbf{2/5} & Linear           & 2/5 & \textbf{3/5} \\
ScaledDotProduct & 1/5 & 3/5 & \textbf{3/5} & DepthwiseConv & 1/5 & \textbf{5/5} & FFN              & 0/5 & \textbf{2/5} \\
MHA              & 0/5 & 3/5 & \textbf{3/5} & FFN           & 0/5 & \textbf{1/5} & ScaledDotProduct & 0/5 & \textbf{1/5} \\
\midrule
\rowcolor{blue!8}
\textbf{Total} & \textbf{5/20} & \textbf{11/20} & \textbf{12/20} &
& \textbf{15/20} & \textbf{18/20} &
& \textbf{4/20} & \textbf{8/20} \\
\bottomrule
\end{tabular}
\end{table*}

\begin{figure}[!h]
    \centering
    \includegraphics[width=0.6\linewidth]{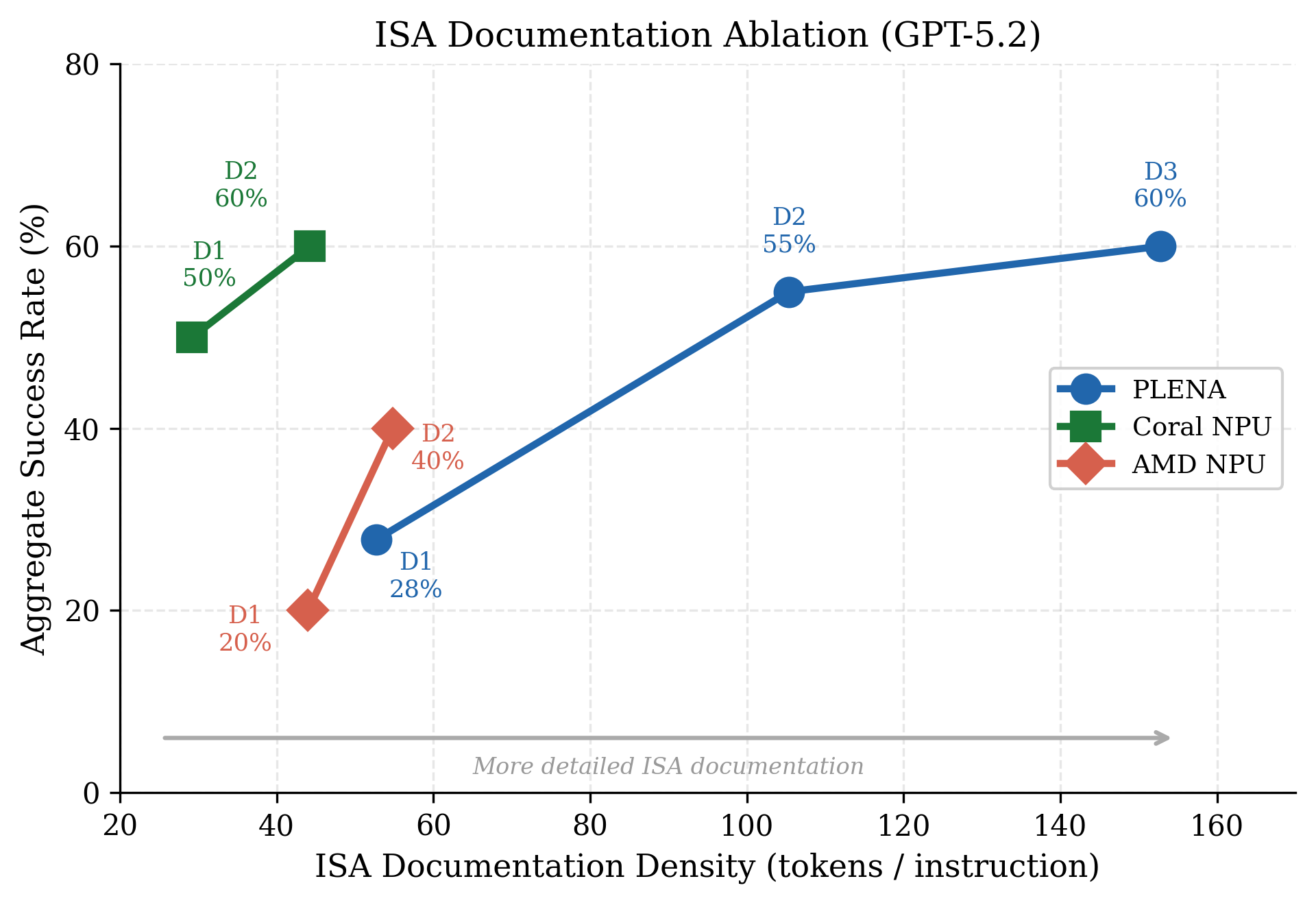}
    \caption{Visualization of success rate over ISA documentation density level. levels: D1, instruction syntax only; D2, adds per-instruction usage examples and latencies; D3, adds full behavioral descriptions (PLENA only).}
    \label{fig:isa-ablation-visualization}
\end{figure}

\begin{table*}[!h]
\centering
\caption{Tool ablation on PLENA with GPT-5.2. Each ablation removes one tool while
keeping all others. Fractions denote the pass rate over tested configurations. }
\label{tab:tool-ablation}
\footnotesize
\setlength{\tabcolsep}{4pt}
\begin{tabular}{@{}l cc @{\hspace{1.5em}} l cc @{\hspace{1.5em}} l cc@{}}
\toprule
\multicolumn{3}{c}{\textit{Without} \texttt{grep\_docs}} &
\multicolumn{3}{c}{\textit{Without} \texttt{view\_output}} &
\multicolumn{3}{c}{\textit{Without} \texttt{get\_instruction\_size}} \\
\cmidrule{1-9}
\textbf{Task} & \textbf{All} & \textbf{Abl.} &
\textbf{Task} & \textbf{All} & \textbf{Abl.} &
\textbf{Task} & \textbf{All} & \textbf{Abl.} \\
\midrule
ReLU      & 2/5 & 0/5 & SwiGLU    & 4/5 & 2/5 & GELU    & 4/5 & 4/5 \\
GEMV      & 5/5 & 2/5 & FlashAttn & 3/5 & 3/5 & Softmax & 5/5 & 4/5 \\
FlashAttn & 3/5 & 2/5 &           &     &     &         &     &     \\
MQA       & 1/5 & 0/5 &           &     &     &         &     &     \\
\midrule
\rowcolor{blue!8}
\textbf{Subtotal} & \textbf{11/20} & \textbf{4/20} &
\textbf{Subtotal} & \textbf{7/10} & \textbf{5/10} &
\textbf{Subtotal} & \textbf{9/10} & \textbf{8/10} \\
\bottomrule
\end{tabular}
\end{table*}

\begin{table*}[!h]                                                                 
  \centering                                                                        
\caption{Prompt engineering ablation with GPT-5.2. We ablate with the full prompt
versus no prompt, and ablate the two main components independently---guidance
(assembly skeletons, tiling rules) and coaching (debugging checklists, pitfalls).
Fractions denote the pass rate over tested configurations.}                                  
  \label{tab:template-ablation}                                                     
  \resizebox{\textwidth}{!}{%
  \begin{tabular}{ll cc cc ll cc cc}                                                
  \toprule                                                                          
  & & \multicolumn{4}{c}{\textbf{PLENA}} & & & \multicolumn{4}{c}{\textbf{Coral     
  NPU}} \\                                                                          
  \cmidrule(lr){3-6} \cmidrule(lr){9-12}                                            
  & & \multicolumn{2}{c}{\textit{full prompt / no prompt}} &                                
  \multicolumn{2}{c}{\textit{Which component?}}                                     
  & & & \multicolumn{2}{c}{\textit{full prompt / no prompt}} &                              
  \multicolumn{2}{c}{\textit{Which component?}} \\                                  
  \cmidrule(lr){3-4} \cmidrule(lr){5-6} \cmidrule(lr){9-10} \cmidrule(lr){11-12}
  \textbf{Lvl} & \textbf{Task} & Full Prompt & No prompt &                
  \shortstack{w/o\\Guidance} & \shortstack{w/o\\Coaching}                           
  & \textbf{Lvl} & \textbf{Task} &  Full Prompt & No Prompt &              
  \shortstack{w/o\\Guidance} & \shortstack{w/o\\Coaching} \\                        
  \midrule                                                  
  L1 & Linear & \textbf{4/5} & 2/5 & 3/5 & 1/5                                      
  & L1 & Linear & \textbf{2/5} & 0/5 & 0/5 & 4/5 \\                                 
  L2 & Attention & \textbf{3/5} & 2/5 & 3/5 & 2/5                                   
  & L1 & DW Conv & \textbf{5/5} & 0/5 & 0/5 & 1/5 \\                                
  \midrule                                                                          
  \rowcolor{blue!8}                                                                 
  \multicolumn{2}{l}{\textbf{Total}} & \textbf{7/10} & \textbf{4/10} & \textbf{6/10}
   & \textbf{3/10}                                                                  
  & \multicolumn{2}{l}{\textbf{Total}} & \textbf{7/10} & \textbf{0/10} &            
  \textbf{0/10} & \textbf{5/10} \\                                                  
  \bottomrule                                               
  \end{tabular}                                                                     
  }                                                         
  \vspace{0.3em}                                                                    
  \end{table*}

\begin{table*}[!h]   

\centering
\caption{Supporting documentation ablation across all accelerators with GPT-5.2.
Each condition removes one supporting documentation file while retaining the full ISA spec and accelerator-specific template.
\textbf{w/o mem} removes \texttt{memory\_layout.md} only;
\textbf{w/o hw} removes \texttt{hardware\_config.md} only.
\textbf{w/o gemm} removes GEMM-related documentation.
The \textit{Full} column uses data from \Cref{tab:success-rate}.}
\label{tab:hw-mem-ablation}
\resizebox{\textwidth}{!}{%
\begin{tabular}{ll ccc cccc ccc}
\toprule
& & \multicolumn{3}{c}{\textbf{PLENA}} 
& \multicolumn{4}{c}{\textbf{AMD NPU}} 
& \multicolumn{3}{c}{\textbf{Coral NPU}} \\
\cmidrule(lr){3-5} 
\cmidrule(lr){6-9} 
\cmidrule(lr){10-12}
\textbf{Lvl} & \textbf{Task} 
& Full & w/o mem & w/o hw 
& Full & w/o mem & w/o hw & w/o gemm 
& Full & w/o mem & w/o hw \\
\midrule
L1 & ReLU     
& \textbf{2/5} & 0/5 & 1/5 
& \textbf{2/5} & 1/5 & 2/5 & 0/5 
& \textbf{5/5} & 5/5 & 5/5 \\

L2 & FFN       
& \textbf{3/5} & 2/5 & 3/5 
& \textbf{2/5} & 0/5 & 1/5 & 0/5 
& \textbf{1/5} & 0/5 & 0/5 \\
\midrule
\rowcolor{blue!8}
\multicolumn{2}{l}{\textbf{Total}} 
& \textbf{5/10} & \textbf{2/10} & \textbf{4/10} 
& \textbf{4/10} & \textbf{1/10} & \textbf{3/10} & \textbf{0/10} 
& \textbf{6/10} & \textbf{5/10} & \textbf{5/10} \\
\bottomrule
\end{tabular}
}
\end{table*}

\begin{table}[!h]
\centering
\caption{Supporting documentation ablation for BOOM with GPT-5.2.
Each condition removes one supporting documentation file while retaining the full ISA spec and accelerator-specific template.
\textbf{w/o mem} removes \texttt{memory\_layout.md} only;
\textbf{w/o hw} removes \texttt{hardware\_config.md} only.}
\label{tab:boom-ablation}
\begin{tabular}{llcccc}
\toprule
\textbf{Lvl} & \textbf{Task} & Full & w/o mem & w/o hw & w/o isa\\
\midrule
L1 & multiply & \textbf{1/1} & 1/1 & 1/1 & 1/1 \\
L2 & qsort  & \textbf{1/1} & 0/1 & 1/1 & 1/1\\
\midrule
\rowcolor{blue!8}
\multicolumn{2}{l}{\textbf{Total}} 
& \textbf{2/2} & \textbf{1/2} & \textbf{2/2} & \textbf{2/2} \\
\bottomrule
\end{tabular}
\end{table}

\begin{table}[!h]
\centering
\caption{KernelCraft against several baselines.
(1) KernelBench~\cite{ouyang2025kernelbench}, an agentic CUDA/DSL kernel generation
pipeline, adapted to generate C++ kernels with the platform compiler, run in its
default setup with no KernelCraft documentation or tools; (2) an Iterative LLM
generating C++ kernels with KernelCraft's prompts and documentation; and (3) the full KernelCraft agent writing directly in assembly. The Human Expert column reports hand-optimized kernels where available. Each cell shows speedup against the Compiler -O2 baseline; \xmark{} represents an agent that never succeeds.}

\label{tab:kernelbench-ablation}
\footnotesize
\begin{tabular}{llr cccc}
\toprule
\rowcolor{gray!10}
\textbf{Workload} & \textbf{Cfg} & \textbf{Compiler -O2} & \textbf{KernelBench} & \textbf{Iterative Agent+Docs+Prompts} & \textbf{KernelCraft} & \textbf{Human Expert} \\
\midrule
GEMM           & C1 & 31{,}518  & \xmark & 1.57$\times$  & 1.60$\times$ & -- \\
GEMM           & C2 & 118{,}005 & \xmark & 1.97$\times$  & \xmark & -- \\
DepthwiseConv  & C1 & 21{,}498  & \xmark & 1.53$\times$ & 1.76$\times$ & 18.73$\times$ \\
DepthwiseConv  & C2 & 17{,}529  & \xmark & 0.44$\times$ & 0.59$\times$ & 19.00$\times$ \\
GELU & C1 & 600{,}040  & \xmark & \xmark & 1.17$\times$ & -- \\
GELU & C2 & 600{,}083   & \xmark & \xmark & 1.04$\times$ & -- \\
\midrule
\rowcolor{blue!8}
\multicolumn{3}{l}{\textbf{Success Rate}} & 0/6 & 4/6 & 5/6 & \\
\bottomrule
\end{tabular}
\vspace{0.3em}
\end{table}

\clearpage

%% file: text/appendix/analysis.tex
\section{Supplementary Analysis}
\label{app:additional-figures}

This appendix section collects additional analysis that complements the results in the main text, covering error breakdowns, tool-usage patterns, scaling behavior, and benchmark ISA coverage across platforms.

\subsection{Error Analysis}
\label{app:error-analysis}
We classify every failed run into four execution-failure categories (Syntax Error,
Simulator Runtime Error, Timeout, Tool Orchestration Error) and two functional-correctness
categories (Low Accuracy, Zero Match), and report the breakdown by model and by
task-complexity level for both PLENA and Coral NPU.

\begin{figure}[!h]
    \centering
    \includegraphics[width=0.8\linewidth]{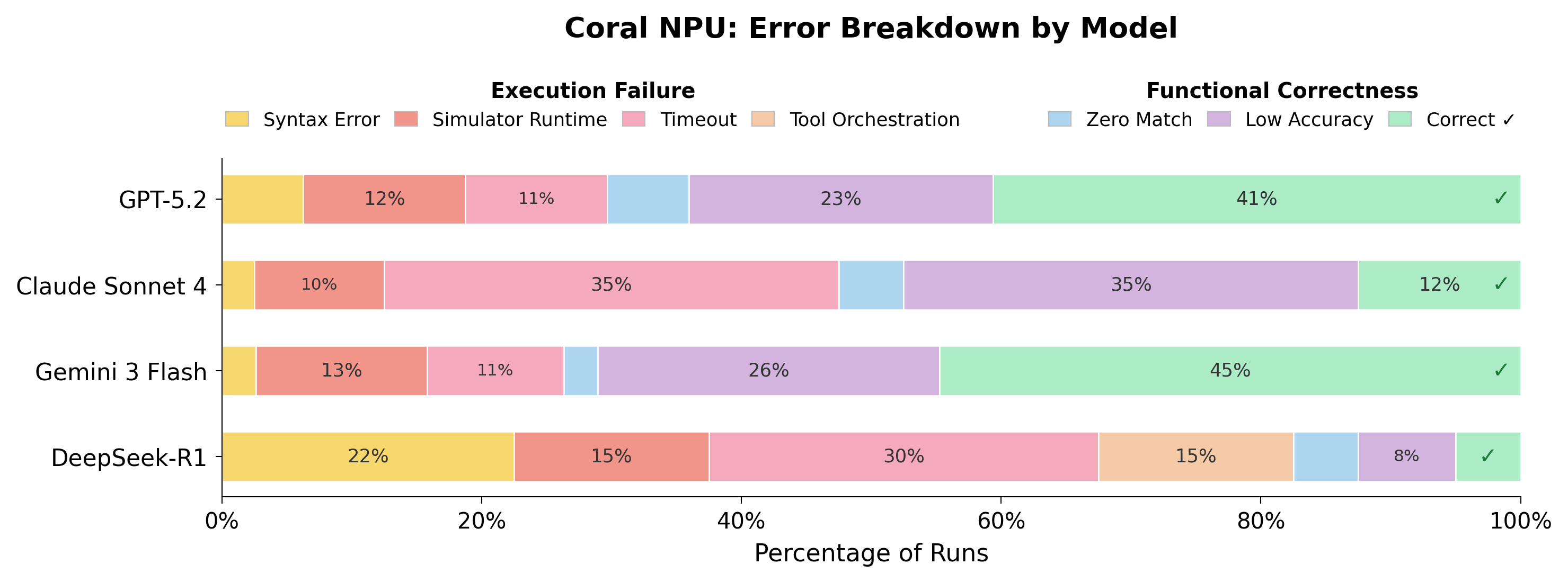}
    \caption{Coral NPU: Error Breakdown by Model.}
    \label{fig:coral-error-model}
\end{figure}

\begin{figure}[!h]
    \centering
    \includegraphics[width=0.8\linewidth]{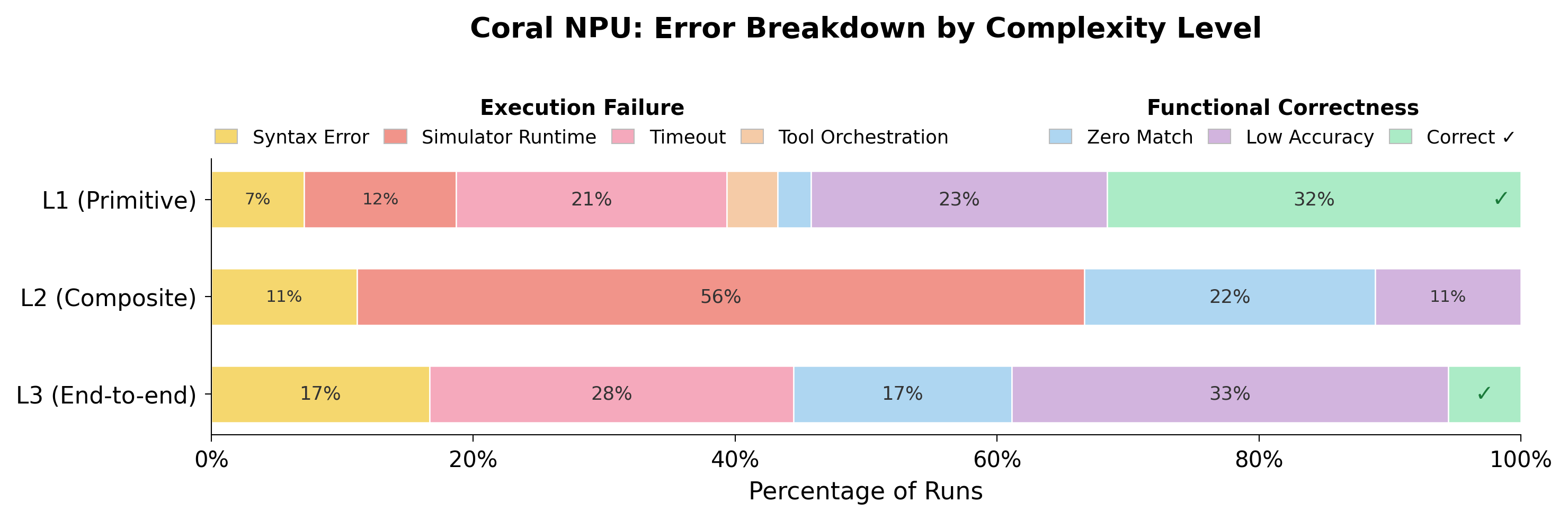}
    \caption{Coral NPU: Error Breakdown by Complexity Level.}
    \label{fig:coral-error-task-level}
\end{figure}

\begin{figure}[!h]
    \centering
    \includegraphics[width=0.8\linewidth]{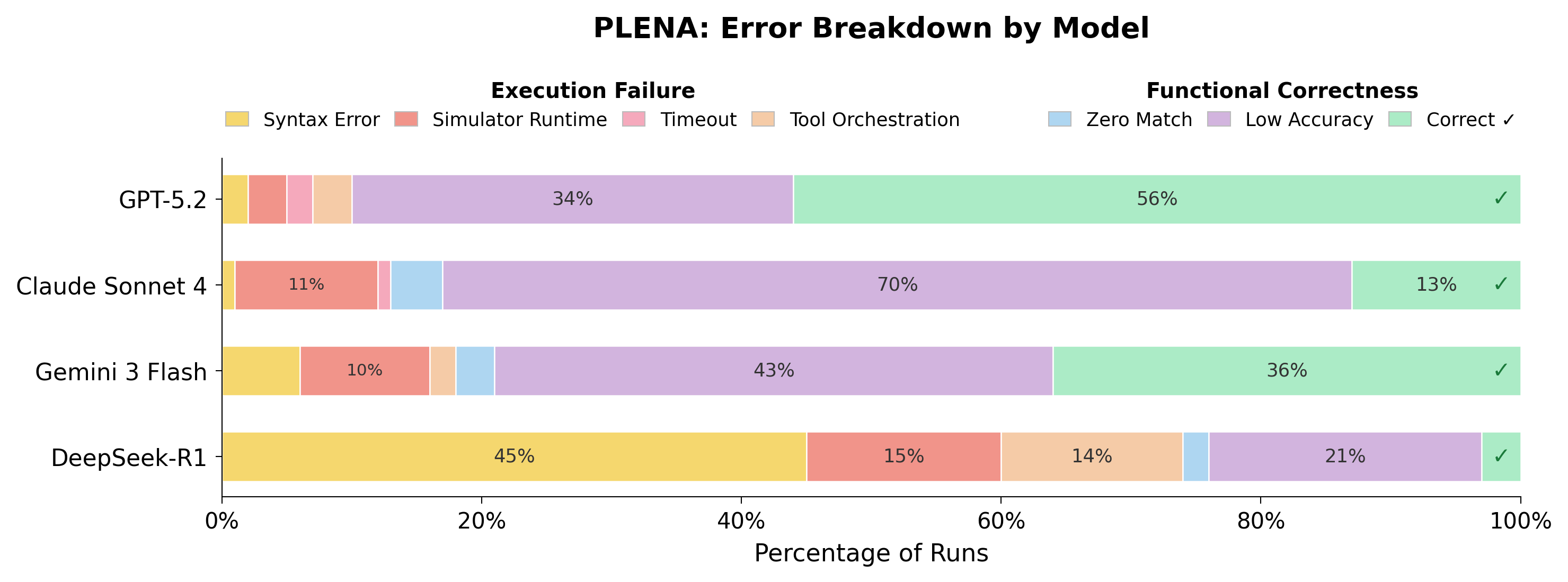}
    \caption{PLENA: Error Breakdown by Model.}
    \label{fig:plena-error-model}
\end{figure}

\begin{figure}[!h]
    \centering
    \includegraphics[width=0.8\linewidth]{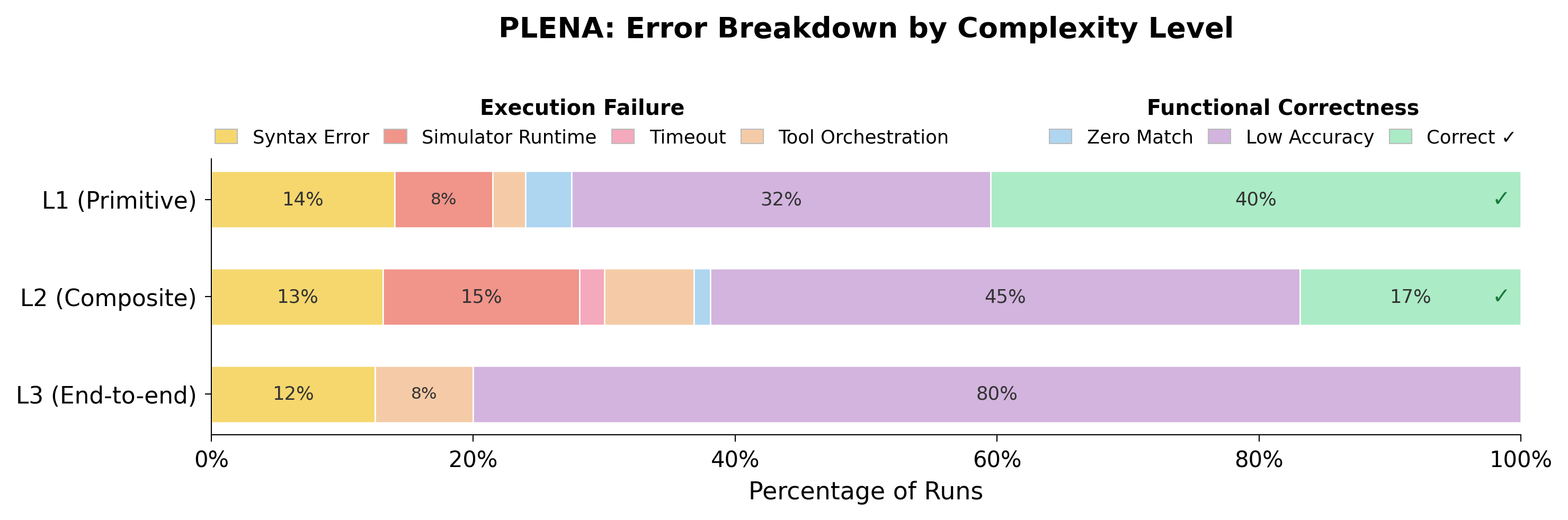}
    \caption{PLENA: Error Breakdown by Complexity Level.}
    \label{fig:plena-error-task-level}
\end{figure}

\subsection{Tool Usage Analysis}
\label{app:tool-usage-analysis}
We report how frequently the agent invokes each harness tool, broken down by
task-complexity level and, for the per-task view, by whether the run ultimately
succeeded.

\begin{figure}[!h]
    \centering
    \includegraphics[width=\linewidth]{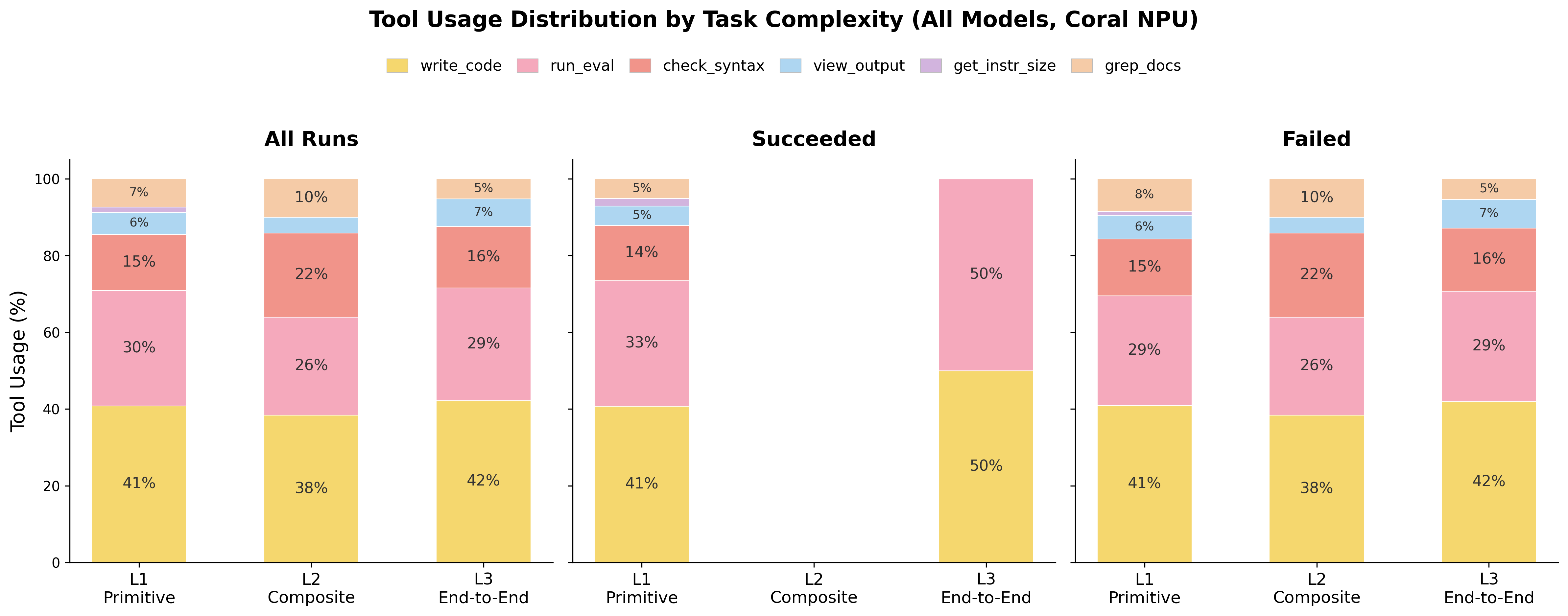}
    \caption{Tool Usage Distribution by Task Complexity (All Models, Coral NPU).}
    \label{fig:coral-tool-usage-by-level}
\end{figure}

\begin{figure}[!h]
    \centering
    \includegraphics[width=0.9\linewidth]{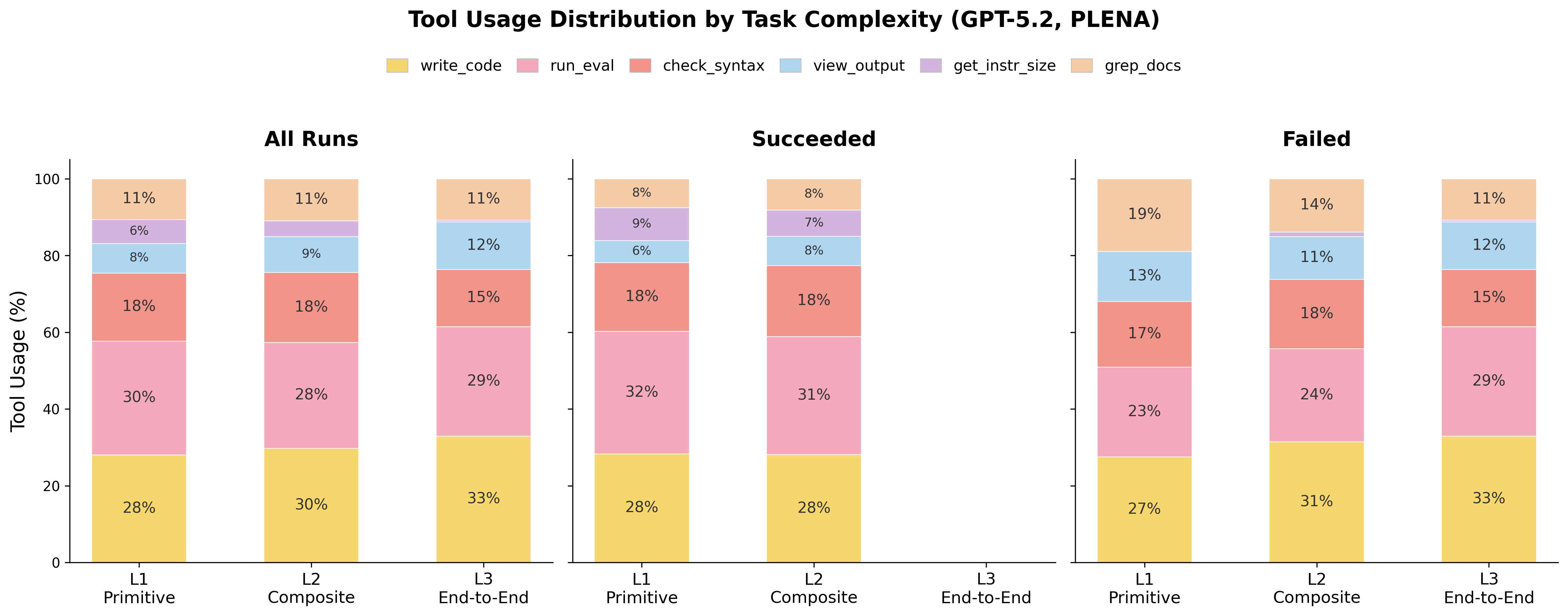}
    \caption{Tool Usage Distribution by Task Complexity (GPT-5.2, PLENA).}
    \label{fig:plena-gpt-tool-usage-by-level}
\end{figure}

\begin{figure}[!h]
    \centering
    \includegraphics[width=0.9\linewidth]{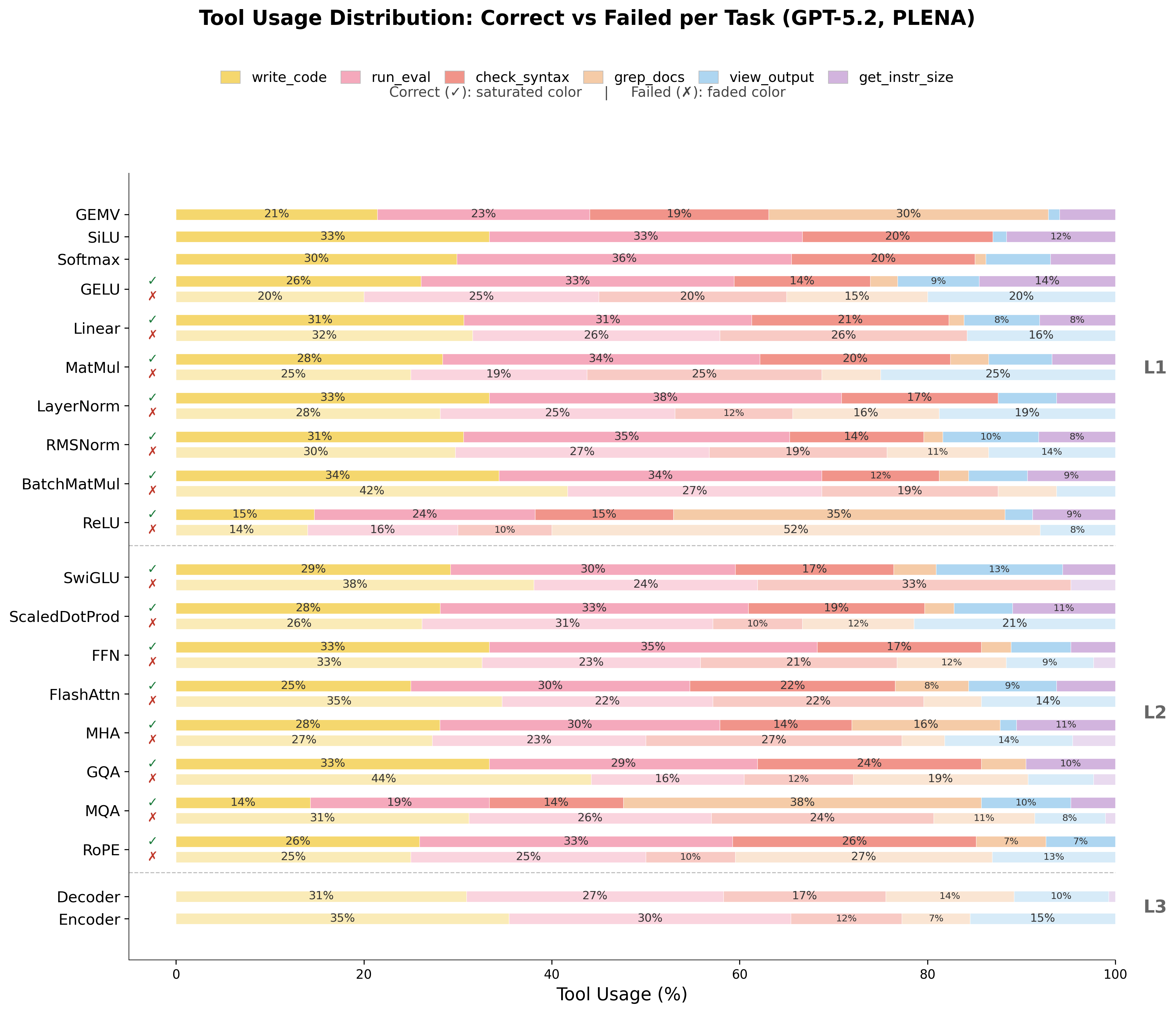}
    \caption{Tool Usage Distribution: Correct vs Failed per Task (GPT-5.2, PLENA).}
    \label{fig:plena-gpt-tool-use-by-task}
\end{figure}

\clearpage
\subsection{Scaling Analysis}
\label{app:scaling-analysis-figs}

\begin{figure}[!h]
    \centering
    \includegraphics[width=\linewidth]{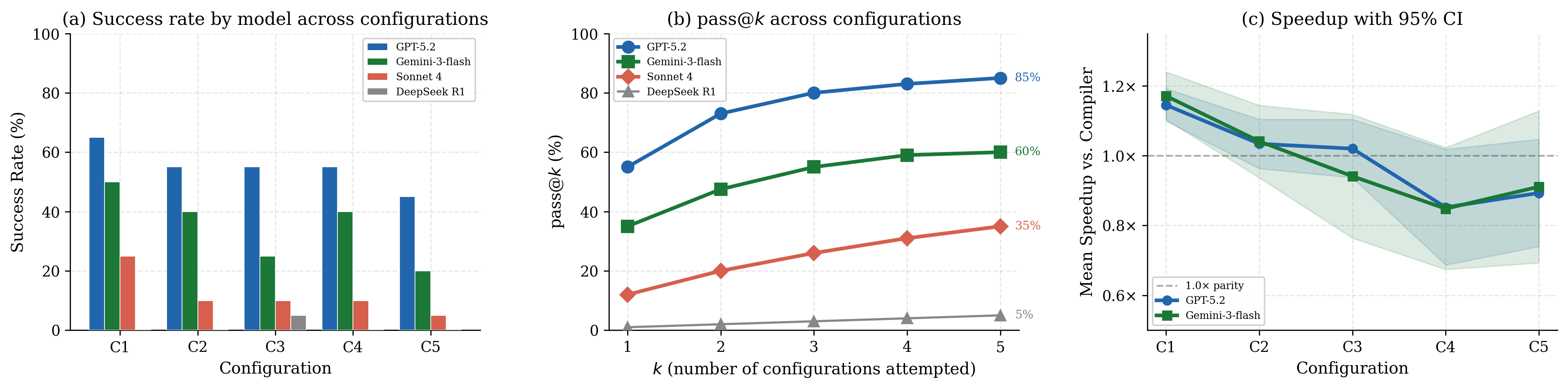}
    \caption{Scaling Analysis (PLENA).}
    \label{fig:scaling-analysis-plena}
\end{figure}

\begin{figure}[!h]
    \centering
    \includegraphics[width=\linewidth]{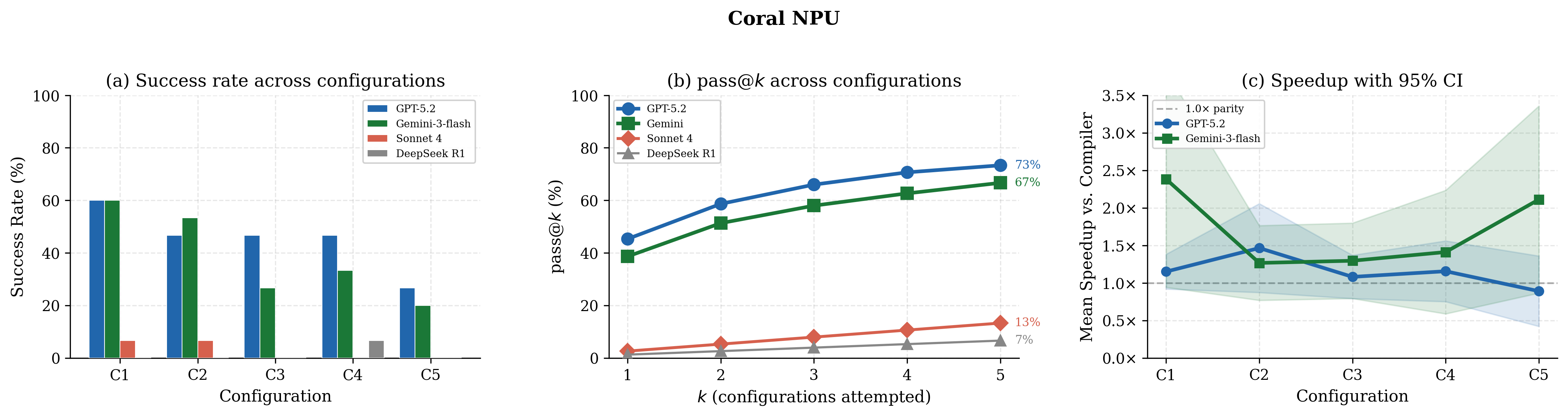}
    \caption{Scaling Analysis (Coral NPU).}
    \label{fig:scaling-analysis-coral}
\end{figure}

\begin{figure}[!h]
    \centering
    \includegraphics[width=\linewidth]{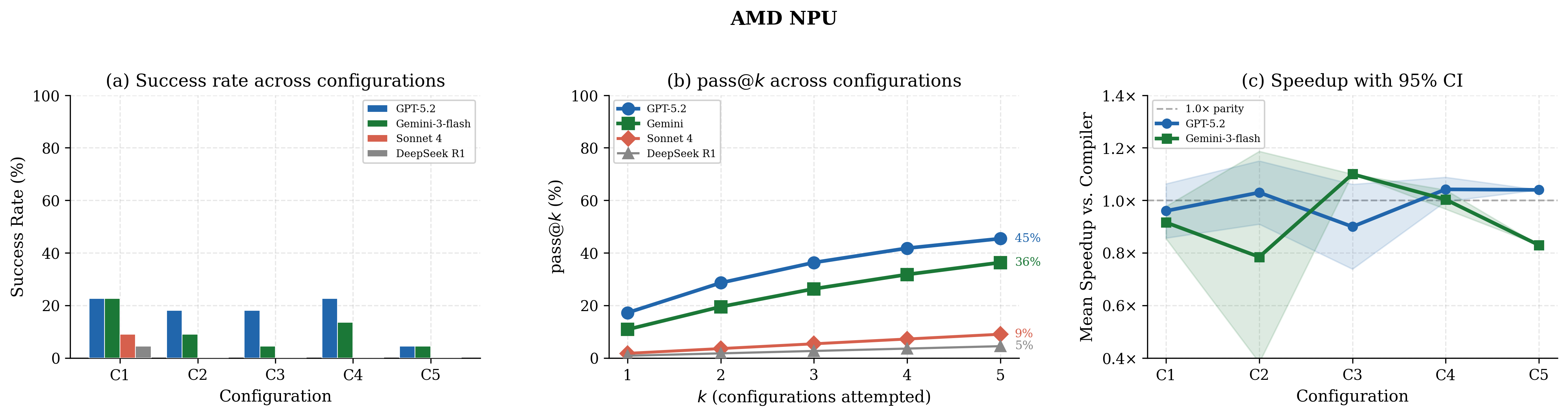}
    \caption{Scaling Analysis (AMD NPU).}
    \label{fig:scaling-analysis-amd}
\end{figure}

We provide scaling analysis on PLENA \Cref{fig:scaling-analysis-plena}, Coral NPU \Cref{fig:scaling-analysis-coral}, and AMD NPU \Cref{fig:scaling-analysis-amd}, plotting success rate, pass@k, and mean speedup with 95\% CI across configuration complexity (C1–C5) for all four models.

\subsection{Benchmark Design Analysis} 

To justify the completeness of our benchmark, we introduce an \emph{instruction coverage} metric that quantifies how comprehensively the selected workloads exercise ISA-level primitives. For each workload, the metric measures the fraction of each ISA extension invoked---spanning arithmetic, memory-access, vector, and control-flow instructions---together with the union across all workloads. \Cref{tab:coral-isa-coverage,tab:plena-isa-coverage} report this coverage for the RISC-V ISA in Coral NPU and the PLENA ISA, respectively. The selected tasks collectively exercise a broad spectrum of hardware-relevant operations, supporting the reliability and completeness of the benchmark.

\begin{table*}[!h]
\centering
\small
\caption{Instruction coverage percentage per ISA extension for each workload for RISC-V ISA used in CoralNPU.}
\label{tab:coral-isa-coverage}
\setlength{\tabcolsep}{4pt}
\renewcommand{\arraystretch}{1.05}
\begin{tabular}{lccccccc}
\toprule
\textbf{Workload} 
& \textbf{RV32I (\%)} 
& \textbf{RV32M (\%)} 
& \textbf{RV32F (\%)} 
& \textbf{Zbb (\%)} 
& \textbf{RVV (\%)} 
& \textbf{Pseudo (\%)} 
& \textbf{Total (\%)} \\
\midrule
ReLU                & 15.0 & 0.0  & 0.0  & 0.0  & 2.2 & 6  & 5.2 \\
SiLU                & 22.5 & 0.0  & 34.6 & 16.7 & 4.8 & 11 & 13.1 \\
GELU                & 22.5 & 0.0  & 34.6 & 16.7 & 5.2 & 10 & 13.1 \\
Softmax             & 22.5 & 0.0  & 42.3 & 22.2 & 6.6 & 13 & 15.9 \\
LayerNorm           & 32.5 & 12.5 & 46.2 & 16.7 & 8.3 & 16 & 19.5 \\
RMSNorm             & 30.0 & 0.0  & 34.6 & 11.1 & 4.8 & 14 & 14.6 \\
MatMul              & 25.0 & 0.0  & 0.0  & 0.0  & 3.9 & 9  & 8.5 \\
GEMV                & 42.5 & 37.5 & 0.0  & 5.6  & 3.9 & 9  & 11.9 \\
Batch MatMul        & 27.5 & 0.0  & 0.0  & 0.0  & 3.9 & 8  & 8.5 \\
Linear              & 25.0 & 0.0  & 0.0  & 0.0  & 7.4 & 10 & 11.3 \\
FFN (SwiGLU)        & 32.5 & 12.5 & 0.0  & 5.6  & 9.6 & 10 & 14.3 \\
Conv2D              & 32.5 & 0.0  & 0.0  & 0.0  & 7.4 & 10 & 12.2 \\
Depthwise Conv      & 22.5 & 0.0  & 0.0  & 0.0  & 8.3 & 13 & 12.5 \\
Conv Block          & 32.5 & 0.0  & 46.2 & 11.1 & 3.5 & 11 & 14.0 \\
Conv Block (scalar) & 32.5 & 12.5 & 46.2 & 11.1 & 1.3 & 10 & 12.5 \\
Attention           & 40.0 & 37.5 & 42.3 & 22.2 & 3.9 & 13 & 17.1 \\
\midrule
\textbf{Union (all)} 
& \textbf{60.0} 
& \textbf{50.0} 
& \textbf{53.8} 
& \textbf{33.3} 
& \textbf{19.2} 
& \textbf{25} 
& \textbf{35.7} \\
\midrule
ISA total & 40 & 8 & 26 & 18 & 229 & --- & 328 \\
\bottomrule
\end{tabular}
\end{table*}

\begin{table}[!h]
\centering
\small
\caption{Instruction coverage percentage per ISA extension for PLENA.}
\label{tab:plena-isa-coverage}
\setlength{\tabcolsep}{4pt}
\renewcommand{\arraystretch}{1.05}
\begin{tabular}{lcccccc}
\toprule
\textbf{Workload} 
& \textbf{Matrix (\%)} 
& \textbf{Vector (\%)} 
& \textbf{S.FP (\%)} 
& \textbf{S.INT (\%)} 
& \textbf{Mem/Ctrl (\%)} 
& \textbf{Total (\%)} \\
\midrule
BMM               & 16.7 & 0.0  & 10.0 & 28.6 & 60.0 & 22.4 \\
FFN               & 16.7 & 50.0 & 20.0 & 42.9 & 80.0 & 40.8 \\
FFN Intermediate  & 16.7 & 50.0 & 20.0 & 28.6 & 80.0 & 38.8 \\
FlashAttn Decode  & 66.7 & 60.0 & 70.0 & 42.9 & 80.0 & 65.3 \\
FlashAttn Prefill & 66.7 & 60.0 & 70.0 & 42.9 & 80.0 & 65.3 \\
FlashAttn QKT     & 0.0  & 0.0  & 10.0 & 28.6 & 70.0 & 20.4 \\
GELU              & 0.0  & 60.0 & 20.0 & 14.3 & 60.0 & 30.6 \\
LayerNorm         & 0.0  & 40.0 & 60.0 & 14.3 & 60.0 & 34.7 \\
Linear            & 16.7 & 0.0  & 10.0 & 28.6 & 80.0 & 26.5 \\
Linear Loop       & 16.7 & 0.0  & 10.0 & 28.6 & 80.0 & 26.5 \\
Loop              & 0.0  & 0.0  & 10.0 & 28.6 & 70.0 & 20.4 \\
RMSNorm           & 0.0  & 40.0 & 60.0 & 14.3 & 60.0 & 34.7 \\
SiLU              & 0.0  & 50.0 & 20.0 & 14.3 & 60.0 & 28.6 \\
\midrule
\textbf{Union (all)} 
& \textbf{66.7} 
& \textbf{90.0} 
& \textbf{80.0} 
& \textbf{42.9} 
& \textbf{80.0} 
& \textbf{73.5} \\
\midrule
ISA total & 12 & 10 & 10 & 7 & 10 & 49 \\
\bottomrule
\end{tabular}
\end{table}

\clearpage

%% file: text/appendix/template_case_study.tex
\clearpage

\section{Case-study: Improving Compiler Template (FFN) for PLENA}

\label{appendix:ffn-template}
\subsection{Original FFN Template}
\begin{lstlisting}[style=pythonstyle]
"""FFN (Feed-Forward Network) Assembly Template

Formula: Y = W_down @ silu(W_up @ X)
"""
import math
from typing import List

MXFP_RATIO = (8 * 8 + 8) / (8 * 8)
MLEN, BLEN, VLEN = 64, 4, 64
IMM2_BOUND = 2**18

def _mxfp_aligned(n: int) -> int:
    return ((int(n * MXFP_RATIO) + 63) // 64) * 64

def _load_imm(reg: int, value: int,
              temp_reg: int = None) -> List[str]:
    if value < IMM2_BOUND:
        return [f"S_ADDI_INT gp{reg}, gp0, {value}"]
    upper, lower = value >> 12, value & 0xFFF
    lines = [f"S_LUI_INT gp{reg}, {upper}"]
    if lower > 0:
        lines.append(
          f"S_ADDI_INT gp{reg}, gp{reg}, {lower}")
    return lines

def _projection(mlen, blen, batch, in_features,
                out_features, regs, w_hbm_reg,
                act_base, result_base) -> str:
    w_actual, w_temp, act_reg = regs[:3]
    out_reg, w_hbm_offset, result_reg = regs[3:6]

    lines = ["; Projection"]
    lines.append(f"; ({batch}, {in_features}) @ "
                 f"({in_features}, {out_features})")

    scale = in_features * out_features
    lines.extend(_load_imm(act_reg, scale, w_temp))
    lines.append(f"C_SET_SCALE_REG gp{act_reg}")
    lines.extend(_load_imm(act_reg, out_features))
    lines.append(f"C_SET_STRIDE_REG gp{act_reg}")
    lines.append(
      f"S_ADDI_INT gp{result_reg}, gp0, {result_base}")

    out_tiles = out_features // blen
    in_tiles = in_features // mlen
    tiles_per_mlen = mlen // blen

    for weight_row in range(out_tiles):
        if weight_row % tiles_per_mlen == 0:
            lines.append(
              f"S_ADDI_INT gp{w_actual}, gp0, 0")
            lines.append(
              f"S_ADDI_INT gp{w_hbm_offset}, gp0, "
              f"{weight_row * blen}")
            lines.append(
              f"S_ADDI_INT gp{out_reg}, "
              f"gp{result_reg}, 0")
            for _ in range(in_tiles):
                lines.append(
                  f"H_PREFETCH_M gp{w_actual}, "
                  f"gp{w_hbm_offset}, a{w_hbm_reg}, 1, 0")
                lines.append(
                  f"S_ADDI_INT gp{w_actual}, "
                  f"gp{w_actual}, {mlen * mlen}")
                lines.append(
                  f"S_ADDI_INT gp{w_hbm_offset}, "
                  f"gp{w_hbm_offset}, {mlen*out_features}")
            lines.append(
              f"S_ADDI_INT gp{w_actual}, gp0, 0")
        else:
            col_off = (weight_row % tiles_per_mlen) * blen
            lines.append(
              f"S_ADDI_INT gp{w_actual}, gp0, {col_off}")
            lines.append(
              f"S_ADDI_INT gp{out_reg}, "
              f"gp{result_reg}, {col_off}")

        for act_col in range(batch // blen):
            lines.append(
              f"S_ADDI_INT gp{act_reg}, gp0, "
              f"{act_base + act_col * mlen * blen}")
            lines.append(
              f"S_ADDI_INT gp{w_temp}, gp{w_actual}, 0")
            for _ in range(in_tiles):
                lines.append(
                  f"M_MM 0, gp{w_temp}, gp{act_reg}")
                lines.append(
                  f"S_ADDI_INT gp{w_temp}, "
                  f"gp{w_temp}, {mlen * mlen}")
                lines.append(
                  f"S_ADDI_INT gp{act_reg}, "
                  f"gp{act_reg}, {mlen * batch}")
            lines.append(
              f"M_MM_WO gp{out_reg}, gp0, 0")
            lines.append(
              f"S_ADDI_INT gp{out_reg}, "
              f"gp{out_reg}, {blen * mlen}")

        if ((weight_row + 1) % tiles_per_mlen == 0 
            and weight_row != out_tiles - 1):
            lines.append(
              f"S_ADDI_INT gp{result_reg}, "
              f"gp{result_reg}, {mlen * batch}")

    return "\n".join(lines)

def _silu(regs, act_base, scratch_base,
          vlen, batch, hidden_dim) -> str:
    act_addr, scratch_addr, loop_reg = regs
    num_vectors = (batch * hidden_dim) // vlen

    lines = ["; SiLU Activation: x * sigmoid(x)"]
    lines.append(
      f"S_ADDI_INT gp{act_addr}, gp0, {act_base}")
    lines.append(
      f"S_ADDI_INT gp{scratch_addr}, gp0, "
      f"{scratch_base}")
    lines.append("S_LD_FP f1, gp0, 1")

    lines.append(
      f"C_LOOP_START gp{loop_reg}, {num_vectors}")
    lines.append(
      f"V_SUB_VF gp{scratch_addr}, "
      f"gp{act_addr}, f0, 0, 1")
    lines.append(
      f"V_EXP_V gp{scratch_addr}, "
      f"gp{scratch_addr}, 0")
    lines.append(
      f"V_ADD_VF gp{scratch_addr}, "
      f"gp{scratch_addr}, f1, 0")
    lines.append(
      f"V_RECI_V gp{scratch_addr}, "
      f"gp{scratch_addr}, 0")
    lines.append(
      f"V_MUL_VV gp{act_addr}, "
      f"gp{scratch_addr}, gp{act_addr}, 0")
    lines.append(
      f"S_ADDI_INT gp{act_addr}, "
      f"gp{act_addr}, {vlen}")
    lines.append(f"C_LOOP_END gp{loop_reg}")

    return "\n".join(lines)
\end{lstlisting}

\subsection{KenrelCraft Agent generated optimized FFN Template (GPT-5.2)}
\begin{lstlisting}[style=pythonstyle]

"""FFN (Feed-Forward Network) Assembly Template for PLENA - Optimized v2
Formula: Y = W_down @ silu(W_up @ X)
Config targeted: batch=8, hidden=128, intermediate=256

Key optimizations:
- Chunk (64 columns) + slice (4 columns) hardware loops for GEMMs
- k=2 up-proj uses 4 precomputed activation pointers
- k=4 down-proj uses 4 precomputed bb0 activation pointers and computes bb1 via +256 temp,
  plus 4 weight-slice pointers (no +4096 ladder inside slice)
- SiLU uses a single scratch buffer to avoid overlap hazards
"""
from typing import List

MXFP_RATIO = 1.125
MLEN, BLEN, VLEN = 64, 4, 64
IMM2_BOUND = 2**18

def _mxfp_aligned(n: int) -> int:
    return ((int(n * MXFP_RATIO) + 63) // 64) * 64

def _load_imm(reg: int, value: int) -> List[str]:
    if value < IMM2_BOUND:
        return [f"S_ADDI_INT gp{reg}, gp0, {value}"]
    upper, lower = value >> 12, value & 0xFFF
    lines = [f"S_LUI_INT gp{reg}, {upper}"]
    if lower:
        lines.append(f"S_ADDI_INT gp{reg}, gp{reg}, {lower}")
    return lines

def _preload_addr_regs(x_size: int, wup_size: int) -> str:
    lines = ["; HBM base address regs"]
    lines.extend(_load_imm(1, 0))
    lines.append("C_SET_ADDR_REG a0, gp0, gp1")
    lines.extend(_load_imm(1, x_size))
    lines.append("C_SET_ADDR_REG a1, gp0, gp1")
    lines.extend(_load_imm(1, x_size + wup_size))
    lines.append("C_SET_ADDR_REG a2, gp0, gp1")
    return "\n".join(lines)

def _preload_x(batch: int, hidden: int) -> str:
    assert hidden % VLEN == 0 and batch % BLEN == 0
    tiles = hidden // VLEN
    bblks = batch // BLEN

    lines = ["; === Prefetch X: HBM -> VRAM ==="]
    lines.extend(_load_imm(2, batch * hidden))
    lines.append("C_SET_SCALE_REG gp2")
    lines.extend(_load_imm(3, hidden))
    lines.append("C_SET_STRIDE_REG gp3")

    # tile0, bb0
    lines.extend(_load_imm(4, 0))
    lines.extend(_load_imm(5, 0))
    lines.append("H_PREFETCH_V gp4, gp5, a0, 1, 0")
    # tile0, bb1
    if bblks > 1:
        lines.append(f"S_ADDI_INT gp4, gp4, {BLEN*VLEN}")
        lines.append(f"S_ADDI_INT gp5, gp5, {BLEN*hidden}")
        lines.append("H_PREFETCH_V gp4, gp5, a0, 1, 0")

    if tiles > 1:
        # tile1, bb0
        lines.extend(_load_imm(4, batch * VLEN))
        lines.extend(_load_imm(5, VLEN))
        lines.append("H_PREFETCH_V gp4, gp5, a0, 1, 0")
        if bblks > 1:
            lines.append(f"S_ADDI_INT gp4, gp4, {BLEN*VLEN}")
            lines.append(f"S_ADDI_INT gp5, gp5, {BLEN*hidden}")
            lines.append("H_PREFETCH_V gp4, gp5, a0, 1, 0")

    return "\n".join(lines)

def _proj_k2(batch, in_feat, out_feat, act_base, out_base, w_a) -> str:
    assert in_feat // MLEN == 2
    chunks = out_feat // MLEN

    k_hbm_step = MLEN * out_feat
    k_act_step = MLEN * batch

    lines = [f"; === Projection k=2: ({batch},{in_feat}) x ({in_feat},{out_feat}) ==="]

    lines.extend(_load_imm(12, in_feat * out_feat))
    lines.append("C_SET_SCALE_REG gp12")
    lines.extend(_load_imm(13, out_feat))
    lines.append("C_SET_STRIDE_REG gp13")

    # constant activation pointers: bb0 k0/k1, bb1 k0/k1
    lines.extend(_load_imm(8, act_base + 0))
    lines.extend(_load_imm(9, act_base + k_act_step))
    lines.extend(_load_imm(10, act_base + (BLEN * MLEN)))
    lines.extend(_load_imm(11, act_base + (BLEN * MLEN) + k_act_step))

    lines.extend(_load_imm(1, out_base))  # res_base
    lines.extend(_load_imm(2, 0))         # col

    lines.append(f"C_LOOP_START gp14, {chunks}")
    # prefetch weights: 2 k-tiles into MSRAM 0 and 4096
    lines.append("S_ADDI_INT gp3, gp2, 0")
    lines.append("S_ADDI_INT gp4, gp0, 0")
    lines.append(f"H_PREFETCH_M gp4, gp3, a{w_a}, 1, 0")
    lines.append(f"S_ADDI_INT gp4, gp4, {MLEN*MLEN}")
    lines.append(f"S_ADDI_INT gp3, gp3, {k_hbm_step}")
    lines.append(f"H_PREFETCH_M gp4, gp3, a{w_a}, 1, 0")

    # init slice pointers
    lines.append("S_ADDI_INT gp4, gp0, 0")
    lines.append(f"S_ADDI_INT gp5, gp0, {MLEN*MLEN}")
    lines.append("S_ADDI_INT gp6, gp1, 0")
    lines.append(f"S_ADDI_INT gp7, gp1, {BLEN*MLEN}")

    lines.append(f"C_LOOP_START gp15, {MLEN//BLEN}")
    # bb0
    lines.append("M_MM 0, gp4, gp8")
    lines.append("M_MM 0, gp5, gp9")
    lines.append("M_MM_WO gp6, gp0, 0")
    # bb1
    lines.append("M_MM 0, gp4, gp10")
    lines.append("M_MM 0, gp5, gp11")
    lines.append("M_MM_WO gp7, gp0, 0")
    # advance slice
    lines.append(f"S_ADDI_INT gp4, gp4, {BLEN}")
    lines.append(f"S_ADDI_INT gp5, gp5, {BLEN}")
    lines.append(f"S_ADDI_INT gp6, gp6, {BLEN}")
    lines.append(f"S_ADDI_INT gp7, gp7, {BLEN}")
    lines.append("C_LOOP_END gp15")

    # next chunk
    lines.append(f"S_ADDI_INT gp1, gp1, {MLEN*batch}")
    lines.append(f"S_ADDI_INT gp2, gp2, {MLEN}")
    lines.append("C_LOOP_END gp14")

    return "\n".join(lines)

def _proj_k4(batch, in_feat, out_feat, act_base, out_base, w_a) -> str:
    assert in_feat // MLEN == 4
    chunks = out_feat // MLEN

    k_hbm_step = MLEN * out_feat
    k_act_step = MLEN * batch

    lines = [f"; === Projection k=4 (optimized): ({batch},{in_feat}) x ({in_feat},{out_feat}) ==="]

    lines.extend(_load_imm(12, in_feat * out_feat))
    lines.append("C_SET_SCALE_REG gp12")
    lines.extend(_load_imm(13, out_feat))
    lines.append("C_SET_STRIDE_REG gp13")

    # Precompute bb0 activation pointers for k=0..3 (independent of chunk/slice)
    lines.extend(_load_imm(10, act_base + 0 * k_act_step))
    lines.extend(_load_imm(11, act_base + 1 * k_act_step))
    lines.extend(_load_imm(12, act_base + 2 * k_act_step))
    lines.extend(_load_imm(13, act_base + 3 * k_act_step))

    lines.extend(_load_imm(1, out_base))
    lines.extend(_load_imm(2, 0))

    lines.append(f"C_LOOP_START gp14, {chunks}")
    # Prefetch 4 k tiles into MSRAM at 0,4096,8192,12288
    lines.append("S_ADDI_INT gp3, gp2, 0")
    lines.append("S_ADDI_INT gp4, gp0, 0")
    lines.append(f"H_PREFETCH_M gp4, gp3, a{w_a}, 1, 0")
    lines.append(f"S_ADDI_INT gp4, gp4, {MLEN*MLEN}")
    lines.append(f"S_ADDI_INT gp3, gp3, {k_hbm_step}")
    lines.append(f"H_PREFETCH_M gp4, gp3, a{w_a}, 1, 0")
    lines.append(f"S_ADDI_INT gp4, gp4, {MLEN*MLEN}")
    lines.append(f"S_ADDI_INT gp3, gp3, {k_hbm_step}")
    lines.append(f"H_PREFETCH_M gp4, gp3, a{w_a}, 1, 0")
    lines.append(f"S_ADDI_INT gp4, gp4, {MLEN*MLEN}")
    lines.append(f"S_ADDI_INT gp3, gp3, {k_hbm_step}")
    lines.append(f"H_PREFETCH_M gp4, gp3, a{w_a}, 1, 0")

    # slice pointers: w0..w3, out0/out1
    lines.append("S_ADDI_INT gp4, gp0, 0")
    lines.append(f"S_ADDI_INT gp5, gp0, {MLEN*MLEN}")
    lines.append(f"S_ADDI_INT gp6, gp0, {2*MLEN*MLEN}")
    lines.append(f"S_ADDI_INT gp7, gp0, {3*MLEN*MLEN}")
    lines.append("S_ADDI_INT gp8, gp1, 0")
    lines.append(f"S_ADDI_INT gp9, gp1, {BLEN*MLEN}")

    lines.append(f"C_LOOP_START gp15, {MLEN//BLEN}")
    # bb0: use precomputed act pointers gp10..gp13
    lines.append("M_MM 0, gp4, gp10")
    lines.append("M_MM 0, gp5, gp11")
    lines.append("M_MM 0, gp6, gp12")
    lines.append("M_MM 0, gp7, gp13")
    lines.append("M_MM_WO gp8, gp0, 0")

    # bb1: compute temp act pointers = bb0 + 256 (BLEN*MLEN)
    lines.append(f"S_ADDI_INT gp3, gp10, {BLEN*MLEN}")
    lines.append("M_MM 0, gp4, gp3")
    lines.append(f"S_ADDI_INT gp3, gp11, {BLEN*MLEN}")
    lines.append("M_MM 0, gp5, gp3")
    lines.append(f"S_ADDI_INT gp3, gp12, {BLEN*MLEN}")
    lines.append("M_MM 0, gp6, gp3")
    lines.append(f"S_ADDI_INT gp3, gp13, {BLEN*MLEN}")
    lines.append("M_MM 0, gp7, gp3")
    lines.append("M_MM_WO gp9, gp0, 0")

    # advance slice pointers
    lines.append(f"S_ADDI_INT gp4, gp4, {BLEN}")
    lines.append(f"S_ADDI_INT gp5, gp5, {BLEN}")
    lines.append(f"S_ADDI_INT gp6, gp6, {BLEN}")
    lines.append(f"S_ADDI_INT gp7, gp7, {BLEN}")
    lines.append(f"S_ADDI_INT gp8, gp8, {BLEN}")
    lines.append(f"S_ADDI_INT gp9, gp9, {BLEN}")
    lines.append("C_LOOP_END gp15")

    # next chunk
    lines.append(f"S_ADDI_INT gp1, gp1, {MLEN*batch}")
    lines.append(f"S_ADDI_INT gp2, gp2, {MLEN}")
    lines.append("C_LOOP_END gp14")

    return "\n".join(lines)

def _silu(act_base: int, scratch_base: int, batch: int, dim: int) -> str:
    num_vec = (batch * dim) // VLEN
    lines = ["; === SiLU: in-place on act_base (single scratch buffer) ==="]
    lines.extend(_load_imm(1, act_base))
    lines.extend(_load_imm(2, scratch_base))
    lines.append("S_LD_FP f1, gp0, 1")
    lines.append(f"C_LOOP_START gp3, {num_vec}")
    lines.append("V_SUB_VF gp2, gp1, f0, 0, 1")
    lines.append("V_EXP_V gp2, gp2, 0")
    lines.append("V_ADD_VF gp2, gp2, f1, 0")
    lines.append("V_RECI_V gp2, gp2, 0")
    lines.append("V_MUL_VV gp1, gp2, gp1, 0")
    lines.append(f"S_ADDI_INT gp1, gp1, {VLEN}")
    lines.append("C_LOOP_END gp3")
    return "\n".join(lines)

def ffn_asm(batch: int, hidden_size: int, intermediate_size: int) -> str:
    assert batch == 8 and hidden_size == 128 and intermediate_size == 256

    x_hbm_size = _mxfp_aligned(batch * hidden_size)
    wup_hbm_size = _mxfp_aligned(hidden_size * intermediate_size)

    input_vram = 0
    up_out_vram = batch * hidden_size  # 1024
    output_vram = 0

    lines = [
        "; FFN optimized kernel v2",
        f"; batch={batch}, hidden={hidden_size}, intermediate={intermediate_size}",
        ""
    ]

    lines.append(_preload_addr_regs(x_hbm_size, wup_hbm_size))
    lines.append("")

    lines.append(_preload_x(batch, hidden_size))
    lines.append("")

    # Up projection: X @ W_up.T -> up_out
    lines.append(_proj_k2(batch, hidden_size, intermediate_size, input_vram, up_out_vram, w_a=1))
    lines.append("")

    # SiLU on up_out, scratch uses VRAM[0]
    lines.append(_silu(up_out_vram, input_vram, batch, intermediate_size))
    lines.append("")

    # Down projection: silu_out @ W_down.T -> output in-place
    lines.append(_proj_k4(batch, intermediate_size, hidden_size, up_out_vram, output_vram, w_a=2))

    return "\n".join(lines)
\end{lstlisting}

%% file: text/appendix/co_design_dllm.tex
\section{KernelCraft Case-study: Co-Designing ISA for Emerging ML Workloads (dLLM)}
\label{appendix:dllm}

This section presents the agent trace for the ISA co-design case study described in \Cref{sec:casestudy-isa}. The agent is tasked with implementing a \texttt{get\_transfer\_index} kernel for diffusion language model (dLLM) sampling on the PLENA accelerator. We include the task specification and show how the agent identifies ISA gaps, proposes new instructions, and iteratively implements the kernel once the human expert provides the final instruction designs. The final kernel achieved 100\% correctness at 14,201\,ns latency. We use \texttt{Gemini-3-Pro-Preview} with thinking level set to high.

\begin{tcolorbox}[colback=gray!10, colframe=gray!50, title=\textbf{Task Specification: Get Transfer Index (dLLM Sampling)}]
\small

\textbf{Formula:}
\vspace{-0.3em}
\begin{flushleft}
$\texttt{x0} = \text{argmax}(\texttt{logits}, \text{dim}=-1)$ \\
$\texttt{confidence} = \text{softmax}(\texttt{logits})[\texttt{x0}]$ \\
$\texttt{x0} = \text{where}(\texttt{mask}, \texttt{x0}, \texttt{x})$ \\
$\texttt{confidence} = \text{where}(\texttt{mask}, \texttt{confidence}, -\infty)$ \\
$\texttt{transfer\_index} = \text{top\_k}(\texttt{confidence}, k) \;\&\; \texttt{mask}$ \\
$\texttt{x}[\texttt{transfer\_index}] = \texttt{x0}[\texttt{transfer\_index}]$
\end{flushleft}

\vspace{-0.5em}
\textbf{Test Configuration:}
batch $= 2$, gen\_length $= 64$, vocab\_size $= 256$, $k = 8$, vlen $= 64$, dtype $=$ bfloat16

\vspace{0.5em}
\textbf{Shapes:}
$Q \in \mathbb{R}^{2 \times 16384}$ (logits), mask $\in \mathbb{R}^{2 \times 64}$, $\texttt{x}, \texttt{x0} \in \mathbb{Z}^{2 \times 64}$

\vspace{0.5em}
\textbf{PLENA Memory Layout:}
\begin{itemize}[nosep, leftmargin=*]
    \item HBM: $[\texttt{logits}, \texttt{mask}]$ sequential
    \item INT\_MEM: $[\texttt{x}, \texttt{x0}]$ sequential
    \item FP\_MEM: $[0] = 0.0$
\end{itemize}
\end{tcolorbox}

\subsection*{Agent Identifying ISA Gaps}

\begin{agentthinking}
I'm tasked with generating assembly for a ``Get Transfer Index'' kernel on PLENA. This means finding the \texttt{argmax} of some logits, calculating a confidence score via softmax, applying a mask, selecting the top-k confidence values, and updating tokens. [\ldots] The instruction set looks somewhat restrictive, and it looks like it lacks critical instructions for this particular workload.

The heart of the challenge is in the logic. I need to iterate over tokens, compute the index of the max logit (that's \texttt{argmax}), calculate the confidence score, apply the mask, and select the top-k indices. Here's where the architectural limitations start to bite. The instruction set seems to lack instructions for finding the argmax, particularly in the context of vector operations. Finding the argmax without a vector comparison instruction (e.g., \texttt{V\_EQ}, \texttt{V\_GT}) presents a big hurdle, as I cannot extract index value to GP reg. Also, I cannot use INT\_MEM directly to index into VRAM as there is no way to do it.

[\ldots] I am essentially blocked by a lack of a mechanism to ``read back'' vector data into scalar registers to drive control flow or update memory. I can't directly read from VRAM or FP register to make decisions or index into the \texttt{INT\_MEM}, which is how the tokens \texttt{x} and \texttt{x0} are stored.
\end{agentthinking}

\begin{toolcall}[title={\textbf{$\blacktriangleright$} Tool Calls: \texttt{grep\_docs}}]
\textbf{Query 1:} \texttt{"argmax"} $\;\rightarrow\;$ \textit{No matches} \\
\textbf{Query 2:} \texttt{"compare"} $\;\rightarrow\;$ \textit{No matches} \\
\textbf{Query 3:} \texttt{"move vector scalar"} $\;\rightarrow\;$ \textit{No matches}
\end{toolcall}

\begin{agentthinking}
Confirmed: the architecture does not provide a straightforward way to move data between VRAM and scalar registers or to select/mask elements in a data-dependent manner. The instruction set makes argmax computation and top-k selection quite difficult.
\end{agentthinking}

\subsection*{Conversation Mode: User Prompting ISA Proposals}

\begin{tcolorbox}[colback=white, colframe=black!60, title=\textbf{User Message}]
\small
We are working on a new workload which like you have perhaps identified requires more efficient operations that don't exist yet in the current ISA (argmax with index tracking, top-k selection, conditional element-wise selection etc) as indicated by the workload specific prompts. Therefore you could just try to propose the new instructions you think we will need before any tool calls --- define their format, operands, and behavior just like the ones you see in the provided instruction sets.
\end{tcolorbox}

\subsection*{Agent-Proposed ISA Extensions}

\begin{agentthinking}
The current ISA requires enhancements. I propose:

\textbf{1. \texttt{V\_CMP\_GT\_VV\_MASK gp\_rd, v\_rs1, v\_rs2}} --- Compare vector \texttt{rs1} with \texttt{rs2} element-wise. If \texttt{rs1[i] > rs2[i]}, set bit \texttt{i} in \texttt{gp\_rd} to 1, otherwise 0. \textit{Use case: updating max\_val and argmax\_idx in the argmax loop.}

\textbf{2. \texttt{V\_TOPK\_MASK v\_rd, v\_rs1, imm\_k}} --- Identifies the \texttt{k} largest elements in vector \texttt{rs1}. Writes 1.0 to \texttt{v\_rd} at top-k positions, 0.0 elsewhere. \textit{Use case: \texttt{transfer\_index = top\_k(confidence, k)}.}

\textbf{3. \texttt{V\_LD\_INT v\_rd, gp\_rs1, imm}} --- Load VLEN integers from INT\_MEM, convert to BF16, store in VRAM. \textit{Use case: loading \texttt{x} and \texttt{x0} tokens into VRAM for processing.}

\textbf{4. \texttt{V\_ST\_INT v\_rd, gp\_rs1, imm}} --- Store BF16 values from VRAM to INT\_MEM as integers. \textit{Use case: storing updated \texttt{x} tokens back to INT\_MEM.}

\textbf{5. \texttt{V\_AND\_VV rd, rs1, rs2}} --- Bitwise AND on two vectors. \textit{Use case: combining masks.}

\textbf{Planned implementation:} Loop 256 times (vocab). Use \texttt{V\_CMP\_GT\_VV\_MASK} to compare with running max, update both \texttt{max\_val} and \texttt{argmax\_idx}. Then softmax denominator loop, \texttt{V\_TOPK\_MASK} on confidence, and final update via masked merge.
\end{agentthinking}

\vspace{-0.8em}
\subsection*{Conversation Mode: Providing Held-out Instruction Specifications}
After reviewing the agent's proposals, the human expert appended the held-out instruction specifications to the conversation context. These instructions were designed independently by a hardware engineer and intentionally withheld during the initial agent interaction to test whether the agent could identify the architectural gaps on its own.

\begin{tcolorbox}[colback=gray!10, colframe=gray!50, title=\textbf{Expert-Designed Instruction: \texttt{V\_RED\_MAX\_IDX}}]
\small
\textbf{Format:} \texttt{V\_RED\_MAX\_IDX rd, rs1, rs2, rs3}

\textbf{Operation:} Find maximum value and its global index across multiple vector chunks, updating both \texttt{gp\_reg<rd>} (index) and \texttt{fp\_reg<rs3>} (max value).

\textbf{Operands:}
\begin{itemize}[nosep, leftmargin=*]
    \item \texttt{rd}: GP register storing the global maximum index (updated on each call)
    \item \texttt{rs1}: GP register containing address of current vector chunk in Vector SRAM
    \item \texttt{rs2}: GP register containing offset value (for global index calculation)
    \item \texttt{rs3}: FP register storing/updating the maximum value
\end{itemize}

\textbf{Behavior:} Finds max value and local index in the current chunk, computes global index $=$ offset $+$ local\_index, and updates \texttt{rd} and \texttt{rs3} only if the current max exceeds the previous max.
\end{tcolorbox}

\begin{tcolorbox}[colback=gray!10, colframe=gray!50, title=\textbf{Expert-Designed Instruction: \texttt{V\_TOPK\_MASK}}]
\small
\textbf{Format:} \texttt{V\_TOPK\_MASK rd, rs1, rs2, k\_scalar, len\_reg}

\textbf{Operation:} Generate a mask vector selecting the top-$k$ elements based on confidence scores.

\textbf{Operands:}
\begin{itemize}[nosep, leftmargin=*]
    \item \texttt{rd}: GP register containing output mask address in Vector SRAM
    \item \texttt{rs1}: GP register containing confidence values address in Vector SRAM
    \item \texttt{rs2}: GP register containing input mask address in Vector SRAM
    \item \texttt{k\_scalar}: GP register containing $k$ value
    \item \texttt{len\_reg}: GP register containing vector length
\end{itemize}

\textbf{Behavior:} Sets non-masked positions to $-\infty$, finds top-$k$ indices, creates output mask (1.0 at selected positions), ANDs with input mask.
\end{tcolorbox}

\begin{tcolorbox}[colback=gray!10, colframe=gray!50, title=\textbf{Expert-Designed Instruction: \texttt{S\_SELECT\_INT}}]
\small
\textbf{Format:} \texttt{S\_SELECT\_INT rd, rs1, rs2, rs3, len\_reg}

\textbf{Operation:} Element-wise conditional selection on INT SRAM using mask from Vector SRAM.

\textbf{Operands:}
\begin{itemize}[nosep, leftmargin=*]
    \item \texttt{rd}: GP register containing output base address in INT SRAM
    \item \texttt{rs1}: GP register containing src1 base address (selected when mask $\neq 0$)
    \item \texttt{rs2}: GP register containing src2 base address (selected when mask $= 0$)
    \item \texttt{rs3}: GP register containing mask vector address in Vector SRAM
    \item \texttt{len\_reg}: GP register containing number of elements
\end{itemize}

\textbf{Behavior:} For each $i$: if mask[$i$] $\neq 0.0$, output[$i$] $=$ src1[$i$]; else output[$i$] $=$ src2[$i$].
\end{tcolorbox}

\begin{tcolorbox}[colback=gray!10, colframe=gray!50, title=\textbf{Expert-Designed Instruction: \texttt{S\_MAP\_V\_FP}}]
\small
\textbf{Format:} \texttt{S\_MAP\_V\_FP rd, rs1, imm, len\_reg}

\textbf{Operation:} Copy a vector of length \texttt{len\_reg} from FP\_MEM to Vector SRAM.

\textbf{Operands:}
\begin{itemize}[nosep, leftmargin=*]
    \item \texttt{rd}: Register containing destination address in Vector SRAM
    \item \texttt{rs1}: Register containing base address in FP\_MEM
    \item \texttt{imm}: Immediate offset added to \texttt{rs1}
    \item \texttt{len\_reg}: Register containing the vector length
\end{itemize}
\end{tcolorbox}

\subsection*{Iteration 1: Initial Implementation}

After receiving the expert-designed instruction specifications, the agent implemented the full kernel.

\begin{toolcall}[title={\textbf{$\blacktriangleright$} Tool Call: \texttt{write\_code}}]
\textbf{Input:} \texttt{<assembly\_code>} (164 lines)
\end{toolcall}

\begin{toolresult}
\texttt{success: true, lines: 164}
\end{toolresult}

\begin{toolcall}[title={\textbf{$\blacktriangleright$} Tool Call: \texttt{run\_evaluation}}]
\textbf{Workload:} \texttt{get\_transfer\_index}
\end{toolcall}

\begin{toolresult}
\texttt{success: false} \\
\texttt{error: Loop at PC 15 exceeded max instructions limit (10000000)}
\end{toolresult}

\subsection*{Iteration 2: Loop Counter Bug Fix}

\begin{agentthinking}
The issue is with the inner loop using \texttt{C\_LOOP\_START}. The specified register is used by the \emph{hardware} to track iterations. My manual increment of \texttt{gp2} is fighting the hardware's internal decrement---the register is being modified by two independent mechanisms, leading to the infinite loop. The fix is to use separate registers for hardware loop counters (\texttt{gp11}, \texttt{gp12}) and my manual indices (\texttt{gp1}, \texttt{gp2}).
\end{agentthinking}

\begin{toolcall}[title={\textbf{$\blacktriangleright$} Tool Call: \texttt{write\_code}}]
\textbf{Input:} \texttt{<assembly\_code>} (125 lines, separate HW loop counters)
\end{toolcall}

\begin{toolresult}
\texttt{success: true, lines: 125}
\end{toolresult}

\begin{toolcall}[title={\textbf{$\blacktriangleright$} Tool Call: \texttt{run\_evaluation}}]
\textbf{Workload:} \texttt{get\_transfer\_index}
\end{toolcall}

\begin{toolresult}
\texttt{success: true, latency\_ns: 14897.0, mse: 2123.9, match\_rate: 90.6\%}
\end{toolresult}

\subsection*{Iteration 3: Memory Address Correction}

\begin{agentthinking}
The critical error was in the HBM address for the mask data. Logits reside at address 0, followed by their MXFP scales. I was pointing at the logits \emph{scale} region, not the mask data. Corrected: Mask Base $= 32768 + 4096 = 36864$.
\end{agentthinking}

\begin{toolcall}[title={\textbf{$\blacktriangleright$} Tool Call: \texttt{write\_code}}]
\textbf{Input:} \texttt{<assembly\_code>} (124 lines, corrected mask HBM offset)
\end{toolcall}

\begin{toolresult}
\texttt{success: true, lines: 124}
\end{toolresult}

\begin{toolcall}[title={\textbf{$\blacktriangleright$} Tool Call: \texttt{run\_evaluation}}]
\textbf{Workload:} \texttt{get\_transfer\_index}
\end{toolcall}

\begin{toolresult}
\texttt{success: true, latency\_ns: 14897.0, mse: 0.0, match\_rate: 100.0\%} \textcolor{green!60!black}{\textbf{$\checkmark$}}
\end{toolresult}

\subsection*{Iteration 4: Performance Optimization}

\begin{agentthinking}
With 100\% correctness, I'll optimize. Moving \texttt{C\_SET\_SCALE\_REG} outside the inner loop and applying 4$\times$ loop unrolling with a single buffer to reduce loop overhead.
\end{agentthinking}

\begin{toolcall}[title={\textbf{$\blacktriangleright$} Tool Call: \texttt{write\_code}}]
\textbf{Input:} \texttt{<assembly\_code>} (232 lines, 4$\times$ unrolled inner loop)
\end{toolcall}

\begin{toolresult}
\texttt{success: true, lines: 232}
\end{toolresult}

\begin{toolcall}[title={\textbf{$\blacktriangleright$} Tool Call: \texttt{run\_evaluation}}]
\textbf{Workload:} \texttt{get\_transfer\_index}
\end{toolcall}

\begin{toolresult}
\texttt{success: true, latency\_ns: 14201.0, mse: 0.0, match\_rate: 100.0\%} \textcolor{green!60!black}{\textbf{$\checkmark$}} \\
\textit{4.7\% latency reduction over initial correct implementation.}
\end{toolresult}

%% file: text/appendix/system_prompt.tex
\section{KernelCraft System Prompt}

\label{appendix:agent-prompt}

The KernelCraft agent system prompt follows a modular template structure composed of four components: (1) a shared base prompt across accelerators defining the agent's role and workflow, (2) a tool specification appended inside the system prompt to emphasize available tools, (3) accelerator-specific content containing ISA and memory documentation, and (4) hardware-specific guidance with debugging strategies and common pitfalls. Components (1) and (2) are shared across all accelerators, while (3) and (4) are instantiated per backend.

\subsection*{Part 1: Shared Base Prompt}

The shared prompt establishes the agent as an expert assembly code generator operating in an autonomous multi-turn tool-calling loop. It enforces a two-phase workflow: first achieving functional correctness, then optimizing for performance.

\begin{promptbox}[title={\texttt{\_SHARED\_PROMPT}}]
\begin{lstlisting}
You are KernelCraft, an expert assembly code generator and optimizer 
for custom accelerators and emerging hardware.
You operate inside a multi-turn automated tool-calling loop. 
There is **no human in the loop** after the first message.

Your goal is to produce **correct AND high-performance {accelerator_name} 
assembly kernels**, using tools strategically and iteratively.
...

===============================================================
WORKFLOW (TWO PHASES)
===============================================================
**PHASE 1: CORRECTNESS** (Target: match_rate == 100%)
1. Plan the kernel: compute tiling, memory layout, loop structure
2. Write complete assembly (not incremental snippets)
3. Save code with write_code(assembly_code)
4. Run evaluation with run_evaluation(workload_type)
5. If match_rate is low, use view_output() to diagnose, then fix
6. Iterate until match_rate == 100%

**PHASE 2: PERFORMANCE OPTIMIZATION** (Target: minimize latency)
7. Note the baseline latency from the passing run
8. Apply optimization techniques to reduce latency
9. Re-run run_evaluation() to verify correctness AND measure latency
...
\end{lstlisting}
\end{promptbox}

\subsection*{Part 2: Tool Specification}

The agent interacts with a file-based tool interface. All tools operate on a shared assembly file written by \texttt{write\_code()}.

\begin{promptbox}[title={\texttt{\_TOOLS\_DESCRIPTION}}]
\begin{lstlisting}
Tools read from a shared file - call write_code() first, then other tools.

- write_code(assembly_code)   : Save code to file for other tools
- run_evaluation(workload)    : Evaluate correctness + performance
- check_syntax()              : Compile and check for syntax errors
- view_output()               : Compare actual vs expected output
- grep_docs(query)            : Search ISA and hardware documentation

TYPICAL FLOW:
  Phase 1: write_code -> check_syntax -> run_evaluation -> view_output
  Phase 2: optimize -> write_code -> run_evaluation -> compare latency
  ...
\end{lstlisting}
\end{promptbox}

\subsection*{Part 3: Accelerator-Specific Content}

Each accelerator backend instantiates three template placeholders with domain-specific documentation. We show abbreviated examples from two backends.

\subsubsection*{PLENA}

\begin{promptbox}[title={\texttt{\{hardware\_config\}}}]
\begin{lstlisting}
MLEN=64      ; Matrix tile dimension
VLEN=64      ; Vector register length  
BLEN=4       ; Block size for writeout
HLEN=16      ; Half-precision tile dimension
...
\end{lstlisting}
\end{promptbox}

\begin{promptbox}[title={\texttt{\{memory\_layout\}}}]
\begin{lstlisting}
SRAM Layout:
  Matrix SRAM (MSRAM):  0x0000 - 0x3FFF   ; Weight tiles
  Vector SRAM (VRAM):   0x4000 - 0x5FFF   ; Activations & outputs
  Scalar Registers:     gp0-gp15, f0-f7, a0-a7
...
\end{lstlisting}
\end{promptbox}

\begin{promptbox}[title={\texttt{\{isa\_spec\}}}]
\begin{lstlisting}
IMPLEMENTED INSTRUCTIONS:
- Matrix:  M_MM, M_TMM, M_BMM, M_MM_WO, M_BMM_WO, M_MV, ...
- Vector:  V_ADD_VV, V_MUL_VF, V_EXP_V, V_RED_SUM, V_RED_MAX, ...
- Scalar:  S_ADD_INT, S_ADDI_INT, S_MUL_INT, S_LD_FP, S_EXP_FP, ...
- Memory:  H_PREFETCH_M, H_PREFETCH_V, H_STORE_V
- Control: C_SET_ADDR_REG, C_SET_STRIDE_REG, C_LOOP_START, ...
\end{lstlisting}
\end{promptbox}

\subsubsection*{Coral NPU}

\begin{promptbox}[title={\texttt{\{hardware\_config\}}}]
\begin{lstlisting}
ISA: rv32imf_zve32x (RISC-V with RVV vector extension)
VLEN=128 bits    ; Vector register width
XLEN=32 bits     ; Scalar register width (RV32)
ELEN=32 bits     ; Maximum element width

Vector Registers: v0-v31 (128 bits each)
  - e8:  16 x int8  per register
  - e16: 8 x int16  per register
  - e32: 4 x int32  per register

LMUL (Register Grouping):
  m1: 1 reg  | m2: 2 regs | m4: 4 regs | m8: 8 regs
...
\end{lstlisting}
\end{promptbox}

\begin{promptbox}[title={\texttt{\{memory\_layout\}}}]
\begin{lstlisting}
Memory Regions:
  ITCM:     0x00000000  (8 KB)   ; Instruction memory
  DTCM:     0x00010000  (32 KB)  ; Data memory, single-cycle
  External: 0x20000000  (4 MB)   ; External memory via AXI4

Test Harness Memory Map:
  input_a:  0x20000000           ; First input array
  input_b:  0x20000000 + sizeof(input_a)
  output:   Dynamic (4KB aligned after inputs)
...
\end{lstlisting}
\end{promptbox}

\begin{promptbox}[title={\texttt{\{isa\_spec\}}}]
\begin{lstlisting}
RVV 1.0 VECTOR INSTRUCTIONS:
- Config:    vsetvli, vsetivli
- Arith:     vadd, vsub, vmul, vdiv, vrem, ...
- Widening:  vwadd, vwsub, vwmul, vwmacc, ...
- Saturate:  vsadd, vssub, vsmul, vssra, ...
- Narrow:    vnsrl, vnsra, vnclip, vnclipu
- Memory:    vle8/16/32, vse8/16/32, vlse, vsse, ...
- Reduce:    vredsum, vredmax, vredmin, ...

...
\end{lstlisting}
\end{promptbox}

\subsection*{Part 4: Hardware-Specific Guidance}

The prompt includes detailed guidance to help agents understand hardware constraints and debug low match rates.

\subsubsection*{PLENA}

\begin{promptbox}[title={Address Formulas}]
\begin{lstlisting}
1. Weight HBM offset (for H_PREFETCH_M):
   k_tile * (MLEN * out_features) + out_tile * MLEN

2. STRIDE_REG must match weight matrix layout:
   For W[in_features, out_features]: STRIDE = out_features
   ...
\end{lstlisting}
\end{promptbox}

\begin{promptbox}[title={Debugging Checklist}]
\begin{lstlisting}
If match_rate is low, verify:
1. M_MM vs M_TMM: Use M_MM when weights are pre-transposed
2. STRIDE_REG: Must equal number of COLUMNS, not rows
3. MSRAM offset: col_block * BLEN, NOT col_block * MLEN
   ...
\end{lstlisting}
\end{promptbox}

\begin{promptbox}[title={Common Pitfalls}]
\begin{lstlisting}
- S_MUL_INT takes ONLY registers (no immediates):
    WRONG:  S_MUL_INT gp1, gp2, 64
    RIGHT:  S_ADDI_INT gp3, gp0, 64
            S_MUL_INT gp1, gp2, gp3
- HBM addresses must be 64-element aligned
  ...
\end{lstlisting}
\end{promptbox}

\subsubsection*{Coral NPU}

\begin{promptbox}[title={Memory Access Patterns}]
\begin{lstlisting}
CRITICAL: Use RVV vector instructions for ALL data processing.
Scalar instructions only for loop control and address setup.

  BAD:  lh t0, 0(a0); sh t0, 0(a1)    <- Scalar (1 element)
  GOOD: vle16.v v0, (a0); vse16.v v0, (a1)  <- Vector (multiple)
  ...
\end{lstlisting}
\end{promptbox}

\begin{promptbox}[title={Debugging Checklist}]
\begin{lstlisting}
If match_rate is low, verify:
1. Missing vsetvli before vector operations
2. Using t7 (doesn't exist! only t0-t6)
3. Output to DTCM instead of External Memory (0x20000000+)
4. LMUL register overlap: With m4, use v0/v4/v8/v12...
   ...
\end{lstlisting}
\end{promptbox}

\begin{promptbox}[title={Common Pitfalls}]
\begin{lstlisting}
- .vi immediate range: -16 to +15 only
    WRONG:  vadd.vi v4, v4, 128
    RIGHT:  li t0, 128; vadd.vx v4, v4, t0

- vsext.vfN requires DESTINATION SEW set BEFORE instruction:
    WRONG:  vsetvli e8; vsext.vf2   <- extends 4-bit to 8-bit!
    RIGHT:  vsetvli e16; vsext.vf2  <- extends 8-bit to 16-bit

- Non-existent instructions: vwcvt, vcvt, vfadd, vfmul (no vec float)
  ...
\end{lstlisting}
\end{promptbox}

%% file: text/appendix/example_trace.tex
\section{KernelCraft Example Entire Agent Trace: Scaled Dot-Product Attention Kernel Generation for PLENA}
\label{appendix:agent-trace}

This section provides a complete reasoning trace of an LLM agent generating an optimized Scaled Dot-Product Attention kernel for the PLENA accelerator. The trace illustrates the iterative refinement process, including debugging, correctness verification, and performance optimization. We include the workload specification in natural language and conclude with the final optimized kernel that achieved 100\% correctness with an 11.3\% latency reduction over its initial correct implementation.

\begin{tcolorbox}[colback=gray!10, colframe=gray!50, title=\textbf{Task Specification: Scaled Dot-Product Attention}]
\small

\textbf{Formula:} $Y = \text{softmax}(QK^\top / \sqrt{d}) \cdot V$

\vspace{0.5em}
\textbf{Test Configuration:}
batch $= 1$, seq\_len $= 64$, head\_dim $= 64$, input\_dtype $=$ bfloat16, qk\_scale $= 0.125$

\vspace{0.5em}
\textbf{Shapes:}
$Q, K, V, Y \in \mathbb{R}^{1 \times 64 \times 64}$

\vspace{0.5em}
\textbf{Computation Stages:}
\begin{enumerate}[nosep, leftmargin=*]
    \item $S = QK^\top \rightarrow [1, 64, 64]$
    \item $S = S / \sqrt{d}$
    \item $A = \text{softmax}(S, \text{dim}=-1)$
    \item $Y = A \cdot V \rightarrow [1, 64, 64]$
\end{enumerate}

\vspace{0.5em}
\textbf{PLENA Specific Workload Memory Layout:}
\begin{itemize}[nosep, leftmargin=*]
    \item HBM: $[Q, K, V]$ sequential
    \item Scores: Vector SRAM scratch
    \item Output: Vector SRAM at address 0
    \item FP\_MEM: $[0]=0.0$, $[1]=1/\sqrt{d}$, $[2]=-\infty$
\end{itemize}
\end{tcolorbox}

\vspace{0.5em}


\subsection*{Iteration 1: Initial Implementation}

\begin{agentthinking}
\textbf{Okay, let's break down this Scaled Dot-Product Attention implementation.} My goal is to get this running with high efficiency, targeting over 99\% match rate, and minimize latency.

First, I see the core operation: $Y = \text{softmax}(\frac{Q K^T}{\sqrt{d}}) V$. The parameters are set, including the scaling factor and the dimensions. The data ($Q$, $K$, $V$, and the final output $Y$) is moving between HBM and Vector/Matrix SRAM. It looks like I need to pay close attention to the HBM layout to ensure optimal memory access patterns. The sizes are calculated and the memory offsets in HBM are set.

Next, I need to compute the Attention Scores, which is the $Q @ K^T$ operation. $Q$ and $K$ are loaded from HBM into Vector SRAM and Matrix SRAM, respectively. I need to make use of the \texttt{M\_TMM} which can calculate the matrix multiplication. The strategy will involve tiling to get the right output. The key is going to be the loops and how data is transferred for efficient use of the Systolic array. The inner loop will need to consider the \texttt{MLEN} and \texttt{BLEN} parameters to configure the address within MSRAM.

Once I have the scores, I will apply softmax to normalize them. I'll need to calculate the max value for each row, compute the exponentials, the normalizers, and the actual attention weights. I'll put some thoughts into the order of calculations to keep the code efficient.

The last step is the $A @ V$ multiplication, which computes the final output Y. I will load V into the Matrix...
\end{agentthinking}

\begin{toolcall}[title={\textbf{$\blacktriangleright$} Tool Call: \texttt{write\_code}}]
\textbf{Input:} \texttt{<assembly\_code>} (127 lines)
\end{toolcall}

\begin{toolresult}
\texttt{success: true, lines: 127}
\end{toolresult}


\subsection*{Iteration 2: Syntax Verification}

\begin{agentthinking}
\textbf{Deep Dive: Code Verification and Optimization}

Alright, the code is ready, time to thoroughly examine it. First, I need to meticulously check the syntax and, of course, the evaluation procedures. Let me just pause here and ensure that the memory addresses for \texttt{M\_TMM} and \texttt{M\_MM} are spot on. For \texttt{M\_TMM}, the source registers are \texttt{rs1=VRAM} and \texttt{rs2=MSRAM} -- perfect. The MSRAM address stride is 256, again, correct. Similarly, for \texttt{M\_MM}, \texttt{rs1=MSRAM} and \texttt{rs2=VRAM} are accurate. The MSRAM address stride is 4. Yep, all good. The \texttt{M\_MM\_WO} address calculation, \texttt{row\_blk*256 + col\_blk*4} is right.

Now, a small, but crucial detail: \texttt{M\_TMM} accumulates. So, I must verify the accumulator is cleared before each block's first \texttt{M\_TMM}. Oh, wait! \texttt{M\_MM\_WO} clears the accumulator, so each \texttt{col\_blk} loop iteration starts with a clean slate.

Wait, did I forget to include the inner \texttt{k\_blk} loop? My code doesn't have it. I need to be sure. Ah, that's because $d=64$, and \texttt{M\_MM} and \texttt{M\_TMM} work on $MLEN=64$ elements at a time. The code's design processes a full summation over $k=0 \dots 63$ with just one \texttt{M\_MM} call. But wait, is that true? Let's confirm by reading the workload spec carefully. ``Fetch a (BLEN, MLEN) tile from Vector SRAM ... and a (MLEN, BLEN) tile from Matrix SRAM''. Yes, $BLEN=4$ and $MLEN=64$. So, it computes $(4, 64) @ (64, 4) = (4, 4)$. Exactly what is needed for one block of the score matrix! So, the loops over row and col blocks are set up correctly, and the $k$ loop...
\end{agentthinking}

\begin{toolcall}[title={\textbf{$\blacktriangleright$} Tool Call: \texttt{check\_syntax}}]
\textbf{Input:} \texttt{generated\_kernel.asm}
\end{toolcall}

\begin{toolresult}
\texttt{success: true, errors: []}
\end{toolresult}


\subsection*{Iteration 3: First Evaluation}

\begin{agentthinking}
Okay, here's the summary, rendered as a thought process from a field expert:

\textbf{Assessment and Execution}

Alright, I've meticulously checked the syntax. It's clean, no glaring errors are visible from my review. Now, it's time to run the evaluation. The code seems to be formatted as expected, so the next step is to initiate the test and verify how it behaves. I will run the evaluation function and see if the output is as expected or if there is something I have missed.
\end{agentthinking}

\begin{toolcall}[title={\textbf{$\blacktriangleright$} Tool Call: \texttt{run\_evaluation}}]
\textbf{Input:} \texttt{workload: attention}
\end{toolcall}

\begin{toolresult}
\begin{tabular}{ll}
\texttt{success:} & \texttt{true} \\
\texttt{latency\_ns:} & \texttt{41845.0} \\
\texttt{match\_rate:} & \textcolor{red}{\textbf{0.46\%}} \\
\end{tabular}
\end{toolresult}


\subsection*{Iteration 4: Investigating Low Match Rate}

\begin{agentthinking}
Here's my analysis of the situation:

\textbf{Low Match Rate Investigation}

Okay, this match rate of 0.46\% is concerning. That's practically a non-starter. In my experience, a rate this low almost always points to a deep-seated problem with either the address mapping itself -- something fundamentally wrong in how we're locating data -- or a flaw in the core logic of the process. Time to dig in. Given the severity of this, I need to take a look at the output to understand what is happening. Let's use \texttt{view\_output} and see what the data is showing us.
\end{agentthinking}

\begin{toolcall}[title={\textbf{$\blacktriangleright$} Tool Call: \texttt{view\_output}}]
\textbf{Input:} \texttt{num\_rows: 10}
\end{toolcall}

\begin{toolresult}
\begin{lstlisting}
row_analysis: [
  {row: 0, sim_nonzero: 64, golden_min: -0.41, golden_max: 0.31},
  {row: 1, sim_nonzero: 64, golden_min: -0.62, ...}
]
\end{lstlisting}
\end{toolresult}


\subsection*{Iteration 5: Root Cause Analysis and Fix}

\begin{agentthinking}
\textbf{Diagnosing Nan and Large Values in Attention Mechanism Simulation}

Okay, here's what I've got to sort through. The simulated output is riddled with \texttt{nan} values and massive, incorrect numbers. This immediately points to a potential overflow issue with the \texttt{exp} function \textit{before} the softmax or a bug in my softmax implementation or incorrect matrix multiplication offsets.

First, let's trace the scores computation. The dimensions are right: $Q$ and $K$ are both [64, 64], so the matrix multiplication should result in $S = Q K^T$. The code uses \texttt{M\_TMM} which computes $V @ M^T$. I have $Q$ as a $(4, 64)$ tile and $K$ as a $(4, 64)$ tile, resulting in a $(4, 4)$ output block. That's good.

I did, however, find a potential error in the memory address calculation for the \texttt{M\_MM\_WO} instruction. For the write address of \texttt{M\_MM\_WO}, the ISA says \texttt{vec\_base + i * MLEN} where \texttt{i} goes from 0 to \texttt{BLEN}. My destination for \texttt{M\_MM\_WO} was \texttt{4096 + row\_blk*256 + col\_blk*4}. Looking at the \texttt{MLEN} (64) and \texttt{BLEN} (4) values, my write address calculation is correct in my row-major layout, since I am writing a block of 4 elements at addresses offset by 64.

But, back to the softmax. I'm using the correct register operands in \texttt{V\_RED\_MAX} and \texttt{V\_RED\_SUM}. My row size is also correct, using the full \texttt{VLEN} of 64 elements, which looks good. But, where's the \texttt{nan} coming from? It's possible I need to apply \texttt{qk\_scale} (0.125) before the max for numerical stability. My code calculates $S = QK^T$, then...
\end{agentthinking}

\begin{toolcall}[title={\textbf{$\blacktriangleright$} Tool Call: \texttt{write\_code}}]
\textbf{Input:} \texttt{<assembly\_code>} (107 lines)
\end{toolcall}

\begin{toolresult}
\texttt{success: true, lines: 107}
\end{toolresult}


\subsection*{Iteration 6: Verification After Bug Fix}

\begin{agentthinking}
\textbf{My Line of Reasoning for M\_MM\_WO Address Calculation}

Okay, wait a minute! I think I spotted something in my \texttt{M\_MM\_WO} address calculation. Let me step through this...

I'm starting with this code:

\begin{lstlisting}
S_ADDI_INT gp1, gp0, 64
S_MUL_INT gp13, gp7, gp1
S_ADDI_INT gp1, gp0, 4
S_MUL_INT gp13, gp13, gp1  ; row_blk * 256
\end{lstlisting}

Right, I'm multiplying \texttt{row\_blk} by 64, then multiplying that result by 4. So \texttt{gp13} should eventually equal \texttt{row\_blk * 256}. Now, $row\_blk$ ranges from 0 to 15, meaning the maximum value is $15 \times 256 = 3840$. Okay, that's within the expected bounds.

I was momentarily thrown off because \texttt{S\_MUL\_INT} doesn't take immediate values directly, so I had to use \texttt{S\_ADDI\_INT} to load the constants.

Let's break it down:
\texttt{gp13 = row\_blk * 64}
\texttt{gp13 = gp13 * 4} which does indeed give me \texttt{row\_blk * 256}.

Is \texttt{row\_blk * 256} the correct offset for this? Each row block is 4 rows, and each row has 64 elements. So, 4 rows * 64 elements/row gives me 256 elements, exactly. Therefore, \texttt{row\_blk * 256} correctly gives me the starting offset of that block. That's good.

Next, I need to check the column offset:

\begin{lstlisting}
S_ADDI_INT gp1, gp0, 4
S_MUL_INT gp14, gp9, gp1
S_ADD_INT gp13, gp13, gp14
\end{lstlisting}

Here, $col\_blk$ goes from 0 to 15. The \texttt{gp14} register becomes \texttt{col\_blk * 4}. This seems right. This adds an offset \textit{within} the 4-row block. Perfect.

Now, a final check... the scale registers. I've been a little paranoid about these lately.

$Q\_offset = 0$, $Q\_$...
\end{agentthinking}

\begin{toolcall}[title={\textbf{$\blacktriangleright$} Tool Call: \texttt{run\_evaluation}}]
\textbf{Input:} \texttt{workload: attention}
\end{toolcall}

\begin{toolresult}
\begin{tabular}{ll}
\texttt{success:} & \texttt{true} \\
\texttt{latency\_ns:} & \texttt{42376.0} \\
\texttt{match\_rate:} & \textcolor{green!60!black}{\textbf{100.0\%}} \\
\end{tabular}

\vspace{0.5em}
$\rightarrow$ \textit{Correctness achieved!}
\end{toolresult}


\subsection*{Iteration 7: Performance Optimization - Phase 1}

\begin{agentthinking}
\textbf{Focusing on Performance After Perfect Match}

Okay, 100\% match, that's excellent. Now the real work begins: optimization! Let's see... current latency is sitting at 42376 nanoseconds. Time to get this number down.

My first thought is always instruction reduction. Let's look at those loops. I see some immediate opportunities. Those multiplications:

\begin{lstlisting}
S_ADDI_INT gp1, gp0, 256
S_MUL_INT gp11, gp7, gp1
\end{lstlisting}

can be done with incrementing. We'll precalculate that outside the loop.

Next, prefetching! Can we get ahead of the game with loading data? Specifically, can we prefetch the $V$ matrix? The memory model is usually \texttt{MLEN $\times$ MLEN $\times$ 4}, with \texttt{MLEN} being 64, so that's 4096 elements per tile. The ISA tells us how the matrix tiles are addressed sequentially. Let's see... the matrix SRAM can hold at least 4 tiles, with the layout specified. $K$ is currently in tile zero, so we can prefetch $V$ to address 4096.

Finally, loop overhead. We could try combining the loops, but the 2D tiling is already pretty efficient, so I am not too worried about this. The priority is definitely in optimizing the address calculations, I will start there.
\end{agentthinking}

\begin{toolcall}[title={\textbf{$\blacktriangleright$} Tool Call: \texttt{write\_code}}]
\textbf{Input:} \texttt{<assembly\_code>} (85 lines)
\end{toolcall}

\begin{toolresult}
\texttt{success: true, lines: 85}
\end{toolresult}


\subsection*{Iteration 8: Evaluation of Optimized Code}

\begin{agentthinking}
Okay, here's my summary, taking the role of an expert:

\textbf{Assessing the Optimized Version's Performance}

Alright, time to get down to brass tacks. We've poured a lot of effort into this optimization, so the key now is to see if it's actually paying off. I need to get a concrete handle on the performance gains. Specifically, I'm thinking about latency. This is where the rubber meets the road. No matter how elegant the code, if the latency isn't where we need it to be, all that refactoring is just intellectual exercise. So, let's fire up those profiling tools and see what kind of impact we've made. Let's see if the numbers are speaking to the effort!
\end{agentthinking}

\begin{toolcall}[title={\textbf{$\blacktriangleright$} Tool Call: \texttt{run\_evaluation}}]
\textbf{Input:} \texttt{workload: attention}
\end{toolcall}

\begin{toolresult}
\begin{lstlisting}
success: false
errors: ["Invalid operand '13_r': PLENA doesn't support labels/symbols. 
         Use registers (gp0-gp15, f0-f7, a0-a7) or integer immediates only."]
\end{lstlisting}

\vspace{0.5em}
$\rightarrow$ \textit{Syntax error: invalid register naming}
\end{toolresult}


\subsection*{Iteration 9: Fixing Register Naming Error}

\begin{agentthinking}
\textbf{Re-thinking Register Allocation}

Okay, so I messed up a bit. I got a little carried away and used these placeholder names like \texttt{gp13\_r}. No, no, no. I need to be precise. I'm working with actual register names here, \texttt{gp0} through \texttt{gp15}. Time to sort this out properly.

Let's start by defining my registers. I need to get this straight.

\begin{itemize}[nosep]
\item \texttt{gp0}: Okay, start with zero. Let's make that a safe default.
\item \texttt{gp1}: Let's make that our general-purpose ``temp'' register.
\item \texttt{gp2}: That can be used for ``scale'' and another temp register.
\item \texttt{gp3}: Another temp register, potentially the ``sram\_dest''.
\item \texttt{gp4}: Another temp register, maybe the ``hbm\_offset''.
\item \texttt{gp5}: Let's use this one for \texttt{loop\_cnt1}.
\item \texttt{gp6}: We'll use this for \texttt{loop\_cnt2}.
\item \texttt{gp7}: And \texttt{loop\_cnt3}, in case we need them.
\item \texttt{gp8}: The register \texttt{q\_addr} will be used to track the address for $q$.
\item \texttt{gp9}: The register \texttt{k\_addr} will be used to track the address for $k$.
\item \texttt{gp10}: I'll use \texttt{s\_addr\_r} as the base address for $s$.
\item \texttt{gp11}: And I'll use \texttt{s\_addr} as the full address for $s$.
\item \texttt{gp12}: This is our \texttt{row\_ptr}.
\item \texttt{gp13}: This is our register for \texttt{v\_addr}.
\item \texttt{gp14}: This is the base address for \texttt{a\_addr\_r}.
\item \texttt{gp15}: And this is the base address for \texttt{y\_addr\_r}.
\end{itemize}

Okay, 16 registers. Got it. Now let's think about how I'm going to allocate these registers across the steps.

In \textbf{Step 2 (Scores)}, for the key loops, these allocations are what I see:

\texttt{gp5}: Now this one's the outer lo...
\end{agentthinking}

\begin{toolcall}[title={\textbf{$\blacktriangleright$} Tool Call: \texttt{write\_code}}]
\textbf{Input:} \texttt{<assembly\_code>} (86 lines)
\end{toolcall}

\begin{toolresult}
\texttt{success: true, lines: 86}
\end{toolresult}


\subsection*{Iteration 10: Evaluation After Register Fix}

\begin{agentthinking}
Okay, here's the summary, rendered as a first-person account from an expert:

\textbf{Latency Analysis}

Alright, time to dig in. I've put in the work, optimized this code, and now it's crucial to understand the impact. My primary focus is the latency. Let's see what the numbers are saying. I need to get a handle on the end-to-end response time. Is this optimized version \textit{actually} performing better? Or have I introduced a subtle bottleneck somewhere I didn't anticipate? I need to analyze the data, looking at the distribution, the mean, and the tail latency, especially. Are there any outliers that are skewing things? If there are, I need to know why. And the tools? Are they working as they should be? The profiling tools I've set up are important in the process. It's time to crunch the numbers and validate the improvement, or identify the problem.
\end{agentthinking}

\begin{toolcall}[title={\textbf{$\blacktriangleright$} Tool Call: \texttt{run\_evaluation}}]
\textbf{Input:} \texttt{workload: attention}
\end{toolcall}

\begin{toolresult}
\begin{tabular}{ll}
\texttt{success:} & \texttt{true} \\
\texttt{latency\_ns:} & \textbf{37573.0} $\leftarrow$ \textit{11\% improvement!} \\
\texttt{match\_rate:} & \texttt{100.0\%} \\
\end{tabular}
\end{toolresult}


\subsection*{Iteration 11: Further Optimization - Prefetching V}

\begin{agentthinking}
\textbf{Optimization Musings: Dropping Latency}

Okay, so I've managed to shave off some latency, bringing it down from 42376 ns to 37573 ns -- roughly an 11\% improvement. That's a good start, but as always, the question is: can we push it further? My mind immediately jumps to prefetching, specifically targeting that bottleneck around the V matrix.

I'm thinking, if I can get V loaded \textit{before} that softmax loop kicks off, we might see another boost. Now, the matrix SRAM has K at address 0, perfect. I'll prefetch V to 4096.

The plan is to use that \texttt{H\_PREFETCH\_M} instruction. The beauty of this is that it should run concurrently with the scalar and vector operations in Step 3. Let's get to work, time to modify the implementation.
\end{agentthinking}

\begin{toolcall}[title={\textbf{$\blacktriangleright$} Tool Call: \texttt{write\_code}}]
\textbf{Input:} \texttt{<assembly\_code>} (74 lines)
\end{toolcall}

\begin{toolresult}
\texttt{success: true, lines: 74}
\end{toolresult}


\subsection*{Iteration 12: Evaluation with Prefetching}

\begin{agentthinking}
Okay, here's the summary, rendered as a first-person account from an expert:

\textbf{Latency Check: Time to Assess Performance}

Alright, let's get down to brass tacks. We've been working on this system for a while, made some significant adjustments. Now it's the moment of truth. Time to dive into the nitty-gritty and measure that all-important factor: latency. This isn't just a casual glance; we need solid data. I want to see if our recent changes have meaningfully impacted response times. We need the numbers. I'll be looking for any bottlenecks or unexpected delays. We're looking for performance, and the numbers never lie.
\end{agentthinking}

\begin{toolcall}[title={\textbf{$\blacktriangleright$} Tool Call: \texttt{run\_evaluation}}]
\textbf{Input:} \texttt{workload: attention}
\end{toolcall}

\begin{toolresult}
\begin{tabular}{ll}
\texttt{success:} & \texttt{true} \\
\texttt{latency\_ns:} & \texttt{37573.0} \\
\texttt{match\_rate:} & \texttt{100.0\%} \\
\end{tabular}

\vspace{0.5em}
$\rightarrow$ \textit{Latency unchanged --- prefetch already masked by compute}
\end{toolresult}


\subsection*{Iteration 13: Analysis and Final Verification}

\begin{agentthinking}
\textbf{Reflections on Optimization Results}

Okay, so the latency has stayed exactly the same at 37573 ns. Interesting. This means one of two things is happening. Either the \texttt{H\_PREFETCH\_M} operation for $V$ is taking significantly less time than the softmax computation, so its latency was already masked by the bigger operation, or the prefetch engine and the compute engine in the simulator aren't perfectly overlapping in \textit{this} particular instance.

However, a 100\% match rate and an improvement in latency compared to the baseline is a win, so I'm happy with that outcome. But, as always, my mind is still ticking, wondering if there are any further gains to be made.

Let's see... within the softmax loop, could we unroll it a little bit? But I should pause there for a moment. Given that the vector units are already being fully utilized with a VLEN of 64, this might not be possible. The current code \textit{is} quite clean and efficient. I don't think I can improve the performance by optimizing the softmax loop.
\end{agentthinking}

\begin{toolcall}[title={\textbf{$\blacktriangleright$} Tool Call: \texttt{run\_evaluation}}]
\textbf{Input:} \texttt{workload: attention}
\end{toolcall}

\begin{toolresult}
\begin{tabular}{ll}
\texttt{success:} & \texttt{true} \\
\texttt{latency\_ns:} & \textbf{37573.0} \\
\texttt{match\_rate:} & \textbf{100.0\%} \\
\end{tabular}

\vspace{0.5em}
\textbf{[COMPLETED]} Finished in 14 iterations (including iteration 14 confirmation run)
\end{toolresult}

\subsection*{Summary}
\begin{tcolorbox}[colback=white, colframe=black, title=\textbf{Analysing Agent Behavior}]
\small
\begin{itemize}[nosep, leftmargin=*]
    \item \textbf{Total Iterations:} 14 (13 shown + 1 final confirmation)
    \item \textbf{Key Debugging Insight:} Identified numerical overflow in softmax due to missing scaling before exponentiation (Iteration 5)
    \item \textbf{Optimization Techniques Applied:}
    \begin{enumerate}[nosep]
        \item Direct register-based pointer increments (eliminated redundant multiplications)
        \item Prefetch overlapping (V matrix prefetch during softmax)
        \item Tiling strategy exploiting $(BLEN, MLEN)$ systolic array geometry
    \end{enumerate}
    \item \textbf{Performance Progression:}
\end{itemize}

\vspace{0.5em}
\centering
\begin{tabular}{@{}lccc@{}}
\toprule
\textbf{Stage} & \textbf{Iteration} & \textbf{Match Rate} & \textbf{Latency (ns)} \\
\midrule
Initial implementation & 3 & 0.46\% & 41,845 \\
After bug fix & 6 & 100\% & 42,376 \\
After optimization & 10 & 100\% & 37,573 \\
With prefetching & 12--13 & 100\% & 37,573 \\
\bottomrule
\end{tabular}

\vspace{0.5em}
\raggedright
\textbf{Total Improvement:} 42,376 ns $\rightarrow$ 37,573 ns (\textbf{11.3\% reduction})
\end{tcolorbox}

\begin{tcolorbox}[colback=gray!10, colframe=gray!50, title=\textbf{Final Optimized Kernel}]
\small
\textbf{Match Rate:} 100\% \quad \textbf{Latency:} 37,573 ns

\begin{lstlisting}[language={[x86masm]Assembler}, basicstyle=\ttfamily\scriptsize, breaklines=true]
; Scaled Dot-Product Attention: Y = softmax(Q @ K^T / sqrt(d)) @ V
; Q, K, V are [1, 64, 64], MLEN=64, VLEN=64, BLEN=4
; --- Step 1: Prefetch Q and K ---
S_ADDI_INT gp1, gp0, 64
C_SET_STRIDE_REG gp1
S_ADDI_INT gp2, gp0, 4096
C_SET_SCALE_REG gp2
S_ADDI_INT gp3, gp0, 0             ; VRAM Q dest
S_ADDI_INT gp4, gp0, 0             ; HBM Q offset
C_LOOP_START gp5, 16
    H_PREFETCH_V gp3, gp4, a0, 1, 0
    S_ADDI_INT gp3, gp3, 256
    S_ADDI_INT gp4, gp4, 256
C_LOOP_END gp5
S_ADDI_INT gp2, gp0, 8128
C_SET_SCALE_REG gp2
S_ADDI_INT gp4, gp0, 4608
S_ADDI_INT gp3, gp0, 0             ; MSRAM K dest
H_PREFETCH_M gp3, gp4, a0, 1, 0
; --- Step 2: Compute Scores S = Q @ K^T ---
S_ADDI_INT gp7, gp0, 0             ; q_addr
S_ADDI_INT gp9, gp0, 4096          ; s_addr_r
C_LOOP_START gp5, 16
    S_ADDI_INT gp8, gp0, 0         ; k_addr
    S_ADD_INT gp10, gp0, gp9       ; s_addr
    C_LOOP_START gp6, 16
        M_TMM 0, gp7, gp8
        M_MM_WO gp10, gp0, 0
        S_ADDI_INT gp8, gp8, 256
        S_ADDI_INT gp10, gp10, 4
    C_LOOP_END gp6
    S_ADDI_INT gp7, gp7, 256
    S_ADDI_INT gp9, gp9, 256
C_LOOP_END gp5
; --- Step 3: Prefetch V to MSRAM[4096] while doing Softmax ---
S_ADDI_INT gp2, gp0, 12160
C_SET_SCALE_REG gp2
S_ADDI_INT gp4, gp0, 9216
S_ADDI_INT gp3, gp0, 4096          ; MSRAM V dest at 4096
H_PREFETCH_M gp3, gp4, a0, 1, 0
S_LD_FP f1, gp0, 1                 ; Load 1/sqrt(d)
S_ADDI_INT gp10, gp0, 4096
C_LOOP_START gp11, 64
    V_MUL_VF gp10, gp10, f1, 0
    S_ADDI_INT gp10, gp10, 64
C_LOOP_END gp11
S_LD_FP f2, gp0, 2                 ; Load -inf
S_ADDI_INT gp10, gp0, 4096         ; row_ptr
C_LOOP_START gp11, 64
    S_ADD_FP f3, f2, f0            ; Initialize max = -inf
    V_RED_MAX f3, gp10, 0
    V_SUB_VF gp10, gp10, f3, 0, 0
    V_EXP_V gp10, gp10, 0
    S_ADD_FP f4, f0, f0            ; Initialize sum = 0
    V_RED_SUM f4, gp10
    S_RECI_FP f4, f4
    V_MUL_VF gp10, gp10, f4, 0
    S_ADDI_INT gp10, gp10, 64
C_LOOP_END gp11
; --- Step 4: Compute Y = A @ V ---
S_ADDI_INT gp7, gp0, 4096          ; a_addr_r
S_ADDI_INT gp9, gp0, 0             ; y_addr_r
C_LOOP_START gp5, 16
    S_ADDI_INT gp8, gp0, 4096      ; v_addr at MSRAM[4096]
    S_ADD_INT gp10, gp0, gp9       ; y_addr
    C_LOOP_START gp6, 16
        M_MM 0, gp8, gp7
        M_MM_WO gp10, gp0, 0
        S_ADDI_INT gp8, gp8, 4
        S_ADDI_INT gp10, gp10, 4
    C_LOOP_END gp6
    S_ADDI_INT gp7, gp7, 256
    S_ADDI_INT gp9, gp9, 256
C_LOOP_END gp5
\end{lstlisting}
\end{tcolorbox}